\documentclass[12pt]{article}
\usepackage{jheppub}
\usepackage{epstopdf}

\usepackage{mathrsfs}
\usepackage{psfrag}
\usepackage{color}
\usepackage{hyperref}

\usepackage{slashed}
\usepackage{feynmp-auto}
\usepackage{simplewick}
\usepackage{cancel}

\usepackage[table]{xcolor}

\usepackage{amsmath,bbm,array,amsfonts,graphicx,wrapfig,arydshln,lscape,float,multirow,longtable,rotating,makecell}
\usepackage{url}

\newcommand{\la}{\langle}
\newcommand{\ra}{\rangle}
\newcommand{\rotp}[1]{\rotatebox{90}{#1}}

\let\ifpdf\relax
\usepackage[T1]{fontenc}
\usepackage[utf8]{inputenc}
\usepackage{ifpdf,tocloft,rotating}
\usepackage{hyperref}
\let\normalcolor\relax
\hypersetup{pdftitle={},pdfcreator={},linkcolor=[rgb]{0.15,0.35,0.75},colorlinks=true,citecolor=[rgb]{0.675,0,0.2},urlcolor=[rgb]{0.15,0.35,0.65}}
\let\olditemize\itemize\renewcommand{\itemize}{\vspace{-2pt}\olditemize\setlength{\itemsep}{1pt}\setlength{\parskip}{0pt}\setlength{\parsep}{-0pt}}
\let\oldenumerate\enumerate\renewcommand{\enumerate}{\vspace{-4pt}\oldenumerate\setlength{\itemsep}{1pt}\setlength{\parskip}{0pt}\setlength{\parsep}{0pt}}

\renewcommand{\bar}{\overline}

\renewcommand{\tilde}{\widetilde}
\newcommand{\ab}[1]{\langle #1\rangle}
\newcommand{\spb}[1]{[ #1]}

\definecolor{mhvBlue}{rgb}{0.3,0.2,0.75}
\definecolor{fRed}{rgb}{0.48,0.02824,0.18824}
\definecolor{cut2}{rgb}{0.18824,0.18824,0.48}
\definecolor{cut1}{rgb}{0.48,0.02824,0.18824}

\newcommand{\M}{\mathcal{M}}
\renewcommand{\O}{\mathcal{O}}

\renewcommand{\L}{\mathcal{L}}

\newcommand{\C}{\mathcal{C}}
\newcommand{\e}{\text{ e}}

\newcommand{\tblue}[1]{\textcolor{blue}{#1}}

\newcommand{\cR}{\cellcolor{LightRed}}
\newcommand{\cRed}{\cellcolor{MediumRed}}
\newcommand{\cB}{\cellcolor{LightBlue}}
\newcommand{\cT}{\cellcolor{testcolor}}
\definecolor{LightGray}{gray}{0.8}
\definecolor{LightBlue}{rgb}{0.8,0.8,1}
\definecolor{LightRed}{rgb}{1,0.8,0.8}
\definecolor{MediumRed}{rgb}{1,0.5,0.5}
\definecolor{testcolor}{rgb}{1,.7,0.5}

\newcommand{\boldS}[1]{\boldsymbol{#1}}

\def\Tr{\mathop{\rm Tr}}

\def\d{{\rm d}}

\def\eps{\epsilon}

\def\Ff#1{F^{4}_{#1}}
\def\Ft#1{(F^{2}_{#1})^2}
\usepackage{mathtools}

\newcommand{\opcoef}{c}
\newcommand{\dLIPStwo}{d\text{LIPS}_2}
\newcommand{\dLIPSthree}{d\text{LIPS}_3}

\newcommand{\re}{\text{Re}}

\newcommand{\phim}[1]{#1{}_{\vphantom{\overline \phi}\varphi}^{\vphantom{+}}}
\newcommand{\phibarm}[1]{#1{}_{\,\vphantom{\overline \phi}\overline \varphi}^{\vphantom{+}}\,}

\newcommand{\psim}[2]{{#1{}_{\vphantom{\overline \psi}\psi}^{#2}}}
\newcommand{\psibarm}[2]{{#1{}_{\,\overline \psi}^{#2}\,}}

\newcommand{\varphibar}{\overline \varphi}

\newcommand{\sect}[1]{Section~\ref{#1}}

\newcommand{\FCube}{{F^3}}
\newcommand{\PhiSquareFSquareA}{{(\varphi^2F^2)_1}}
\newcommand{\PhiSquareFSquareB}{{(\varphi^2F^2)_2}}
\newcommand{\DSquarePhiFourthA}{{(D^2\varphi^4)_1}}
\newcommand{\DSquarePhiFourthB}{{(D^2\varphi^4)_2}}
\newcommand{\PhiSixth}{{\varphi^6}}
\newcommand{\DPhiSquarePsiSquareA}{{(D\varphi^2\psi^2)_1}}
\newcommand{\DPhiSquarePsiSquareB}{{(D\varphi^2\psi^2)_2}}
\newcommand{\PsiFourthA}{{(\psi^4)_1}}
\newcommand{\PsiFourthB}{{(\psi^4)_2}}

\title{Structure of two-loop SMEFT anomalous dimensions via on-shell methods}
\author{Zvi~Bern,}
\emailAdd{bern@physics.ucla.edu}

\author{Julio~Parra-Martinez,}
\emailAdd{jparra@physics.ucla.edu}

\author{and Eric Sawyer}
\emailAdd{eric.sawyer@physics.ucla.edu}

\affiliation{Mani L. Bhaumik Institute for Theoretical Physics,\\
UCLA Department of Physics and Astronomy, \\
Los Angeles, CA 90095, USA}

\abstract{We describe on-shell methods for computing one- and two-loop 
  anomalous dimensions in the context of effective field theories
  containing higher-dimension operators.  We also summarize methods for
  computing one-loop amplitudes, which are used as inputs to the computation of two-loop
  anomalous dimensions, and we explain how the structure of rational terms
  and judicious renormalization scheme choices can lead to additional
  vanishing terms in the anomalous dimension matrix at two loops.  We
  describe the two-loop implications for the Standard Model Effective
  Field Theory (SMEFT).  As a by-product of this analysis we verify a
  variety of one-loop SMEFT anomalous dimensions computed by Alonso,
  Jenkins, Manohar and Trott.  }

\begin{document}

\maketitle

\newpage
\section{Introduction}
\label{section:introduction}

Effective Field Theory (EFT) approaches have risen to
prominence in recent years as a systematic means for quantifying new physics beyond the
Standard Model.  The Standard Model Effective Field Theory (SMEFT) incorporates
the effects of new physics via higher-dimension operators built from Standard
Model fields~\cite{Buchmuller:1985jz,SMEFTReview}.  The operators are organized
according to their dimension, which gives a measure of their importance at
low-energy scales.  The SMEFT allows exploration of the effects of new physics
without requiring a complete understanding of the more fundamental high-energy
theory.  While systematic, the SMEFT involves a large number of operators and
free coefficients~\cite{Grzadkowski:2010es}, making it useful to develop
improved techniques for computing quantities of physical interest and for
understanding their structure. One such quantity is the anomalous
dimension matrix of the higher-dimension operators.  The appearance of
anomalous dimensions implies that the Wilson coefficients of operators at
scales accessible by collider experiments differ from those at the high-energy
matching scale to the more fundamental unknown theory.  These also control
operator mixing, providing important information on how experimental
constraints from one operator affect the coefficients of other operators.  This
makes evaluating the anomalous dimension matrix a crucial aspect of
interpreting results within the SMEFT. Towards this goal, here we apply
on-shell methods that greatly streamline the computation of anomalous
dimensions at one and two loops and expose hidden structure. 

A systematic and complete computation of the one-loop anomalous
dimension matrix for dimension-six operators in the SMEFT is found in
the landmark calculations of Refs.~\cite{Manohar123}.  Besides their
importance for interpreting experimental data, these calculations
reveal a remarkable structure with the appearance of nontrivial zeros
in the anomalous dimension matrix~\cite{NonrenormHolomorphicity}.
These one-loop zeros have been understood as stemming from selection
rules that arise from supersymmetry embeddings~\cite{SusyEmb},
helicity~\cite{NonrenormHelicity}, operator
lengths~\cite{NonrenormLength}, and angular
momentum~\cite{Jiang:2020sdh}.  Perhaps even more surprisingly,
nontrivial zeros in the anomalous dimension matrix of the SMEFT appear
at any loop order and for 
operators of any dimension~\cite{NonrenormLength}. In addition, a surprising number of the associated one-loop
scattering amplitudes vanish as
well~\cite{Craig:2019wmo, Jiang:2020sdh}, suggesting additional zeros
may appear in the anomalous dimensions at two loops. Here we apply on-shell methods to identify a
new set of vanishing terms in the two-loop anomalous dimension
matrix of the SMEFT. As a by product of our two-loop study, we also confirm many
one-loop anomalous dimensions computed in Refs.~\cite{Manohar123}, via
both the generalized unitarity method~\cite{GeneralizedUnitarity1} and
an elegant new unitarity-based method due to Caron-Huot and Wilhelm
for directly extracting anomalous
dimensions from cuts~\cite{Caron-Huot:2016cwu},  which builds
on insight developed in earlier work on $\mathcal N = 4$ super-Yang-Mills
theory~\cite{N4Dilitation}.

On-shell methods have proven to be quite useful in a variety of other settings,
including collider physics (see e.g.~Refs.~\cite{OnShellQCD}), ultraviolet
properties of (super)gravity (see e.g.~Refs.~\cite{TwoLoopGravitySimplified1,
TwoLoopGravitySimplified2, FourGraviton, GravityUVExamples}), theoretical
explorations of supersymmetric gauge and gravity theories (see
e.g.~Refs.~\cite{SusyGaugeExamples,SusyGravExamples}), cosmological observables
(see e.g.~Refs.~\cite{Cosmo1,Cosmo2}), and gravitational-wave physics~(see
e.g.~Refs.~\cite{GravityWave}).  They have also been used as a convenient means
for classifying interactions in EFTs such as the
SMEFT~\cite{OnShellEFTInteractions}. In addition, general properties of the
S-matrix, such as unitarity, causality and analyticity have been used to
constrain Wilson coefficients of EFTs \cite{NimaCausality}, including the SMEFT
\cite{RemmenSMEFT}.

In the context of anomalous dimensions and renormalization-group analyses,
unitarity cuts give us direct access to renormalization-scale dependence. After
subtracting infrared singularities, the renormalization-scale dependence can be
read off from remaining dimensional imbalances in the arguments of
logarithms~\cite{TwoLoopGravitySimplified2}.  The direct link between anomalous
dimensions at any loop order and unitarity cuts is made explicit in the
formulation of Caron-Huot and Wilhelm~\cite{Caron-Huot:2016cwu}.  In carrying
out our two-loop analysis we make extensive use of their formulation.  Very
recently the same formalism and general set of ideas was applied in
Refs.~\cite{RecentOneLoopAnomalous1, RecentOneLoopAnomalous2} to compute
certain SMEFT anomalous dimensions.

In general, two-loop unitarity cuts include both three-particle cuts between
two tree-level objects, as well as two-particle cuts between tree-level and
one-loop objects. Consequently, our exploration of two-loop anomalous
dimensions will require computing one-loop matrix elements first.  On-shell
methods, in particular generalized unitary~\cite{GeneralizedUnitarity1,
ChiralOnShell,GeneralizedUnitarity2}, are especially well suited for this task.
Because we feed one-loop matrix elements into higher-loop calculations, we find it 
convenient to use $D$-dimensional techniques which account for rational terms.
To carry out the integration, we decompose the integrands into gauge-invariant
tensors along the lines of Refs.~\cite{Projectors1, Projectors2}.  In this
form, the integrands can be straightforwardly reduced to a basis of scalar
integrals using integration by parts technology (as implemented, e.g., in
FIRE~\cite{FIRE}).  These one-loop amplitudes are among the building blocks
that feed into the two-loop anomalous dimension calculation.

Using the unitarity-based formalism, we indeed find that many potential
contributions to the two-loop anomalous dimension matrix vanish for a variety
of reasons, including the appearance of only scaleless
integrals~\cite{NonrenormLength}, color selection rules, vanishing rational
terms at one loop, as well as appropriate renormalization scheme choices at one
loop.  These vanishing contributions go beyond those identified in our previous
paper~\cite{NonrenormLength}.  Of the new vanishings, perhaps the most
surprising is the finding that additional zeros can be induced at two loops by
slightly adjusting the $\overline{\text{MS}}$ renormalization scheme at one
loop.  This is tied to the fact that two-loop anomalous dimensions and local
rational contributions to one-loop amplitudes are scheme dependent, and can
therefore be set to zero by appropriate finite shifts of operator coefficients,
or, equivalently, by a finite renormalization of the operators, or the addition
of finite local counterterms. 

For simplicity, we use a non-chiral version of the Standard Model, with zero
quark masses and Yukawa couplings and without an Abelian sector, but point out
overlap with the SMEFT. We note that although we only utilize Dirac fermions
here, on-shell methods are well suited for dealing with chiral fermions as well
(see e.g. Refs.~\cite{ChiralOnShell, OnShellQCD}).  In any case, this model is
a close enough cousin of the SMEFT that we can directly compare to a variety of
one-loop SMEFT anomalous dimensions calculated in Refs.~\cite{Manohar123} and
make some predictions about the structure of the two-loop anomalous dimension
matrix.  (See \sect{section:mapping} and in particular
Table~\ref{tab:twoloopSMEFT}.)

The paper is organized as follows. In Section~\ref{section:setup}, we
explain our conventions, list the higher-dimensional operators in our
simplified version of the SMEFT, and summarize the on-shell methods
that we use to obtain anomalous dimensions.  In
Section~\ref{section:onelooprat} we explain the use of generalized
unitarity in constructing full one-loop amplitudes, and we discuss the
appearance of numerous zeros in the rational terms of the
amplitudes. We also explain how finite counterterms can produce
additional zeros in the rational terms of many of the one-loop
amplitudes.  Examples of additional vanishing contributions to the
two-loop anomalous dimension matrix are presented in
Section~\ref{section:twoloop}, including those that arise from finite
counterterms at one-loop.  In Section~\ref{section:mapping} we discuss
the overlap between our simplified model and the full SMEFT in the
basis of operators used in Refs.~\cite{Manohar123}, and discus the
implications of our results for the latter theory. We give our
conclusions in Section~\ref{section:conclusion}. Appendix
\ref{section:amplitudesappendix} explains the projection method used
for integration in detail and lists the gauge invariant basis
tensors. The explicit $D$-dimensional forms of the full one-loop
amplitudes, as well as their four-dimensional finite remainders, are
relegated to the ancillary files~\cite{AttachedFiles} and
Appendix~\ref{section:amplitudesSH}, respectively.

\section{Setup and formalism}
\label{section:setup}

We now present our conventions and explain the on-shell formalisms
that we use for obtaining the anomalous dimensions. One procedure for
doing so is to extract them from ultraviolet divergences in
amplitudes. This procedure follows the generalized unitarity method
for assembling scattering amplitudes from their unitarity
cuts~\cite{GeneralizedUnitarity1, ChiralOnShell,
GeneralizedUnitarity1, OnShellQCD}.  While we describe the procedure
for obtaining the anomalous dimensions in the current section, we 
leave a more detailed discussion of the generalized unitarity method
for Section~\ref{section:onelooprat}, where it will be used to
construct full amplitudes.

As a second method, we apply the recent formalism of Caron-Huot and
Wilhelm~\cite{Caron-Huot:2016cwu}, which directly expresses the
anomalous dimensions in terms of unitarity cuts. This method is
particularly effective for computing anomalous dimensions, and is our
preferred method beyond one loop.  We show how this method helps
clarify the structure of the anomalous dimension matrix at two loops
and exposes new nontrivial zeros.

\subsection{Conventions and basic setup}
To illustrate our methods we will consider a model with dimension-four Lagrangian given by
\begin{equation}
\L^{(4)} = -\frac{1}{4}F_{\mu\nu}^aF^{a\mu\nu} 
 + D_\mu \varphi \, D^\mu \varphibar - \lambda \, (\varphi \varphibar)^2 
 + i \sum_{m=1}^{N_f} \,\bar{\psi}_m \slashed{D}\psi_m \,,
\label{Lagrangian4}
\end{equation}
where the gauge field strength, $F_{\mu\nu}^a$, is in the adjoint representation of
$SU(N)$, while $\psi_m$ and $\varphi$ are fundamental representation
Dirac fermions and scalars, respectively. The index $m$ on
the fermions denotes the flavor; for simplicity we take a
single flavor of scalars. The covariant derivative is given by
\begin{equation}
(D_\mu \psi_m)_i= \Bigl(\delta_{ij}\partial_\mu 
+ i g \frac{1}{\sqrt{2}} T^a_{ij} A^a_\mu\Bigr)(\psi_m)_j\,,
\end{equation}
where $T^a_{ij}$ is the $SU(N)$ generator. We normalize the generator
in the standard amplitudes convention by $\Tr[T^aT^b]=\delta^{ab}$
which differs from the usual textbook one, and we define $f^{abc} = -i
\Tr[[T^a,T^b]T^c]$ and $d^{abc}=\Tr[\{T^a,T^b\}T^c]$ for later
use.\footnote{Note that our structure constants, $f^{abc}$, carry an
  extra factor of $\sqrt{2}$ relative to standard textbook
  conventions~\cite{PeskinSchroeder}.}

This model theory has the general structure of the Standard Model, but with all
masses and Yukawa couplings set to zero, and with only one gauge group. Here we
also use Dirac instead of chiral fermions; the
basic methods apply just as well to cases which include chiral fermions in the context of Standard
Model calculations, as in Ref.~\cite{ChiralOnShell}.

To mimic the SMEFT we modify this Lagrangian by adding
dimension-six operators supressed by a high-energy scale $\Lambda$:
\begin{equation}
\L=\L^{(4)}+\frac{1}{\Lambda^2}\sum_k \opcoef^{(6)}_i \O^{(6)}_i  \,,
\end{equation} 
where the list of the operators that we consider here is given in
Table~\ref{tab:operators}\footnote{We note that $\O_{\PhiSixth}$ has
no nonzero four-point amplitudes through two-loops, and therefore cannot
renormalize any of the other operators \cite{NonrenormLength}. We still include it here for
completeness.}. Note that our simplified model contains
representatives from all of the operator classes of the basis used in
Refs.~\cite{Manohar123}, other than the classes $\psi^2 F\varphi$ and
$\psi^2\varphi^3$ ($\psi^2XH$ and $\psi^2H^3$ in the notation of
Refs.~\cite{Manohar123}), since operators in these classes must always
have one uncharged fermion. We defer a comparison to the full
SMEFT to \sect{section:mapping}.

At first order in $\opcoef_i/\Lambda^2$, renormalization induces mixing of the dimension-six operators, as parametrized by 
\begin{equation}
	\dot \opcoef_i\equiv \frac{\partial \opcoef_i}{\partial \log\mu}=\opcoef_j \gamma_{ji} \,.
	\label{eq:rgcoef}
\end{equation}
If the coefficient of operator $\O_j$ appears on the right-hand side of the RG equation for the coefficient of operator $\O_i$, as above, we say that $\O_j$ renormalizes $\O_i$, or that they mix under renormalization. Sometimes we write the corresponding anomalous dimension as $\gamma_{i\leftarrow j}$. In all tables which describe anomalous dimensions we will display $\gamma'_{ij} = \gamma_{ij}^{\rm T}$ to facilitate comparison with Refs.~\cite{Manohar123}.
The anomalous dimension matrix $\gamma_{ij}$ depends on the dimension-four couplings $g$ and $\lambda$, in the combinations
\begin{equation}
  \tilde g^2 = \frac{g^2}{(4\pi)^2}\,, \hskip 2 cm \tilde \lambda =  \frac{\lambda}{(4\pi)^2} \,,
\label{g4Def}
\end{equation}
which we sometimes refer to collectively as $g^{(4)}$.

\begin{table}[tb]
	\begin{center}
		\renewcommand{\arraystretch}{1.2}
		\begin{tabular}{lc}\hline
			Label \hskip 1.4 cm $\null$ & Operator \\
			\hline
			$\O _{\FCube}$ & $\frac{1}{3}f^{abc}F^{a}_{\mu\nu} F^{a}_{\nu\rho} F^{a}_{\rho\mu}$ \\[1mm]
			$\O _{\PhiSquareFSquareA}$& $(\varphi^\dagger\varphi)F^a_{\mu\nu}F^a_{\mu\nu}$ \\[1mm]
			$\O_{\PhiSquareFSquareB}$ & $d^{abc}(\varphi^\dagger T^a \varphi)F^b_{\mu\nu}F^c_{\mu\nu}$ \\[1mm]
			$\O _{\DSquarePhiFourthA}$& $(\varphi^\dagger D^\mu\varphi)^*(\varphi^\dagger D_\mu\varphi)$ \\[1mm]
			$\O _{\DSquarePhiFourthB}$& $(\varphi^\dagger\varphi)\Box(\varphi^\dagger\varphi)$ \\[1mm] 
			$\O_{\PhiSixth}$  & $(\varphi^\dagger\varphi)^3$ \\[1mm]
			$\O_{\DPhiSquarePsiSquareA}^{pr}$ & $i(\varphi^\dagger ( D_\mu-\overleftarrow{D}_\mu) \varphi)(\bar\psi{}_p \gamma^\mu\psi_r)$ \\[1mm]
			$\O_{\DPhiSquarePsiSquareB}^{pr}$ & $i(\varphi^\dagger (T^a D_\mu-\overleftarrow{D}_\mu T^a) \varphi)(\bar\psi{}_p T^a\gamma^\mu\psi_r)$ \\[1mm]
			$\O_{\PsiFourthA}^{mnpr}$ & $(\bar\psi{}_m\gamma^\mu\psi_n)(\bar\psi{}_p\gamma_\mu\psi_r)$ \\[1mm]
			$\O_{\PsiFourthB}^{mnpr}$ & $(\bar\psi{}_m\gamma^\mu T^a\psi_n)(\bar\psi{}_p\gamma_\mu T^a\psi_r)$ \\[1mm]
			\hline
		\end{tabular}
		\caption{List of dimension-six operators considered here.  For simplicity, we take the fermions to be Dirac.
       The labels $mnpr$ are flavor indices and $abc$ color indices. Note the operator $\O_{F^3}$ is normalized slightly differently than in Refs.~\cite{Manohar123}, as are the color matrices $T^a$ in the operators $\O_{\DPhiSquarePsiSquareB}$ and $\O_{\PsiFourthB}$. We will occasionally drop the $(\ )_1$ and $(\ )_2$ subscripts to refer to pairs of operators collectively. \label{tab:operators}
}
	\end{center}
\end{table}

We extract anomalous dimensions from both amplitudes
and form factors.  We define a form factor with an operator insertion as
\begin{equation}
  F_i(1^{h_1},...,n^{h_n};q)= \langle  k_1^{h_1},...,k_n^{h_n}|\O_i(q)|0\rangle\,,
\label{eq:FormFactor}
\end{equation}
which are matrix elements between an on-shell state $\langle
k_1,...,k_n|$, with particles of momenta $\{k_1...k_n\}$ and helicities $\{h_1...h_n\}$,
and an operator $\O_i$ that injects additional off-shell momentum
$q$. The states might also be dependent on the color and flavor of the
particles, but we leave this dependence implicit for the moment.
Form factors are especially useful when dealing with on-shell
states with fewer than four particles, where kinematics would
otherwise require the amplitude (with real momenta) to be zero. 
From the perspective of form factors, we can think of an amplitude
with an operator insertion as a form factor, but where the
higher-dimension operator injects zero momentum, $q=0$,
\begin{equation}
  A_i(1^{h_1},...,n^{h_n})= \langle k_1^{h_1},...,k_n^{h_n}|\O_i(0)|0\rangle\,.
\label{eq:ampO}
\end{equation}
When the inserted operator is the identity, we recover the usual
scattering amplitude, which depends only on the dimension-four couplings. We
denote such an amplitude as
\begin{equation}
A(1^{h_1},...,n^{h_n})= \langle k_1^{h_1},...,k_n^{h_n}|0\rangle =  
 \langle  k_1^{h_1},...,k_i^{h_i}|\M| -k_{i+1}^{-h_{i+1}},\ldots,-k_n^{-h_n}\rangle\,.
\label{eq:amp}
\end{equation}
Unless otherwise stated, we use an all outgoing convention where all the
particles are crossed to the final state. When crossing fermions there are
additional signs on the right-hand side of Eq.~\eqref{eq:amp} that we leave implicit
here.
In general we can write the form factors (and amplitudes) as color-space vectors, 
\begin{equation}
  F_i(1,\ldots, n)  = \sum_j \mathcal{C}^{[j]}F_{i\,[j]}(1,\ldots,n)\,,
\end{equation}
where the $\mathcal{C}^{[i]}$ are a set of independent color factors.
In the context of amplitudes, these correspond
to \emph{color-ordered}~\cite{OnShellReviews} or, more generally,
\emph{primitive}~\cite{PrimitiveAmplitudes} amplitudes.  The color
factors $\mathcal{C}^{[i]}$ depend on which particles of the amplitude
are in the adjoint or fundamental representation of $SU(N)$.  Here, we
only need the decomposition into a basis of color factors without
using special properties of the coefficients.  For the various
processes we consider, the tree and one-loop amplitudes are listed in
Appendix~\ref{section:amplitudesSH}.

We use the conventional dimensional regularization and $\overline{\rm MS}$-like
schemes throughout, in which the amplitudes and form factors,
$F_i$ satisfy the renormalization-group equations
\begin{equation}\label{eq:renormgroup}
  \left[\left(\mu\partial_\mu +\beta\,\partial\right)\delta_{ij} 
 + \left(\gamma^{\rm UV}-\gamma^{\rm IR}  \right)_{ij} \right]F_j=0,
\end{equation}
where $\partial_\mu:= \partial/\partial\mu$, $\partial:=\partial/\partial g^{(4)}$, $\beta:=\beta(g^{(4)})$ is the $\beta$-function of the collection of marginal
couplings, $\gamma_{ij}^{\rm UV}$ are the anomalous dimensions of the
higher-dimension operators, and $\gamma_{ij}^{\rm IR}$ are the IR
anomalous dimensions, arising from soft and/or collinear
divergences\footnote{The relative sign between UV and IR anomalous
dimensions is merely a convention.}. For later convenience, we introduce the shorthand 
\begin{equation}
  \Delta\gamma=\gamma^{\rm UV}-\gamma^{\rm IR}\,.
\end{equation}
The appearance of both kinds of anomalous dimensions stems from the
fact that there is a single dimensional-regularization parameter, $\epsilon = \epsilon_{\rm UV} =
\epsilon_{\rm IR}$, and single scale, $\mu = \mu_{\rm UV} = \mu_{\rm IR}$,
for both the UV and IR divergences.  As usual we take $\eps = (4 - D)/2$.

The perturbative expansion of the different quantities we consider is denoted by 
\begin{align}
F_i &= F_i^{(0)}+F_i^{(1)}+ F_i^{(2)}+\cdots\,, \nonumber\\
A_i &= A_i^{(0)}+A_i^{(1)}+ A_i^{(2)}+\cdots\,, \nonumber \\
\gamma_{ij} &=\gamma_{ij}^{(1)}+ \gamma_{ij}^{(2)}+\cdots\,, \nonumber \\
\beta &= \beta^{(1)}+ \beta^{(2)}+\cdots\,,
\label{eq:perturb}
\end{align}
where each order in the expansion includes an additional power of the
dimension-four couplings, $g^{(4)}$, as defined in Eq.~\eqref{g4Def},
compared to the previous order. Since the operators we consider here have a least
four fields, except for the $F^3$ case, any of the generated
four-point tree amplitudes are local, and directly correspond to the
operator. The amplitudes generated by the $F^3$ operator also contain a vertex
obtained from the dimension-four operators.  Thus, the four-point tree
amplitudes have no powers of $g^{(4)}$, with the exception of the
four-point amplitudes generated from the $F^3$ operator.


\subsection{Anomalous dimensions from UV divergences}

Anomalous dimensions are traditionally extracted from countertems associated to
UV divergences. For instance, in Refs.~\cite{Manohar123} the full one-loop
anomalous dimension matrix of the SMEFT was calculated by extracting the
$1/\epsilon$ divergences of the one-particle irreducible (1PI) diagrams that
generate the one-loop effective action in the background field method.
Alternatively, one might extract the anomalous dimensions from on-shell
amplitudes. Here, we use the full one-loop amplitudes to calculate the one-loop
anomalous dimension matrix of our model, and thereby verify a representative
set of the anomalous dimensions calculated in Refs.~\cite{Manohar123}.

An efficient way of determining UV divergences at one loop was
presented for the $\beta$-function in Ref.~\cite{onshellbeta}.  Here we
adopt this method to calculate one-loop anomalous dimensions. In general, the
renormalization of $\O_i$ by $\O_j$ at one loop is determined by
calculating the matrix element with external particles corresponding
to $\O_i$, but with an insertion of $\O_j$.  In general, one-loop
matrix elements can be expressed in terms of a basis of scalar
integrals
\begin{equation}
  A_i^{(1)} =\sum_s a^s_{4,\,i}\, I_{4,s} + \sum_s a^s_{3,\,i} I_{3,s} + \sum_s a^s_{2,\,i} I_{2,s} \,,
  \label{eq:oneloopdecomp}
\end{equation}
comprised of boxes $I_{4,s}$, triangles, $I_{3,s}$, and bubbles,
$I_{2,s}$, where the corresponding coefficients, $a^s_i, b^s_i$ and
$c_i$ are gauge invariant and generically depend on color and the
dimensional regularization parameter $\epsilon$.  The integrals can
then be expanded in $\epsilon$, producing both UV and IR poles
in $\epsilon$.  Only the scalar bubble integrals contain UV
divergences, so we write a formula for the anomalous dimensions
in terms of the bubble coefficients $a^s_{2,\,i}$, whose $\epsilon$
dependence can be ignored for this purpose. However, some care is
required because of cancellations between UV and IR divergences.  We
delay a detailed discussion of the infrared structure of the
amplitudes to Section~\ref{section:onelooprat}. For the moment, we just
recall that the $1/\epsilon$ pole in the bubble integrals in
Eq.~\eqref{eq:oneloopdecomp} does not contain the full UV divergence
of the amplitude. The reason for this is that there is an additional
$1/\epsilon$ pole which originates in bubble-on-external-leg diagrams,
which are scaleless and set to zero in dimensional regularization
because of a cancellation of UV and IR poles,
\begin{equation}
\begin{gathered}
\includegraphics[scale=0.6]{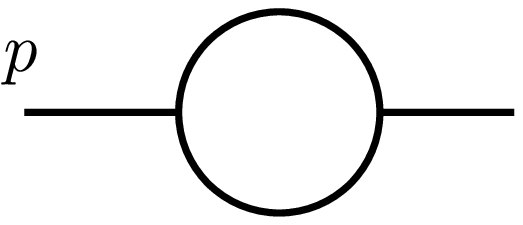} 
\end{gathered}\,\bigg\vert_{p^2=0}\hspace{0.25cm}  \propto \frac1{\epsilon_{UV}} - \frac1{\epsilon_{IR}} + \log\frac{\mu^2_{UV}}{\mu^2_{IR}}\,.
\end{equation}
Hence the bubbles on external legs give an additional UV contribution,
\begin{equation}
  - \frac{1}{2\epsilon} \gamma^{\rm IR\,(1)}_{\rm c} A_i^{(0)} : = -\frac1{2\epsilon} \sum_p  \gamma^{\rm IR}_{{\rm c},\, p} A_i^{(0)}\,,
\end{equation}
where $\gamma^c_p$ is the so-called collinear anomalous dimension of
particle $p$, and the sum is over all external states of the tree
amplitude. For the vectors, fermions and scalars in our theory the collinear anomalous dimensions are given by \cite{IRModernPapers}
\begin{equation}
\gamma^{\rm IR \,(1)}_{{\rm c},\, v} = -\tilde g^2 b_0\,,\quad 
\gamma^{\rm IR \,(1)}_{{\rm c},\, f} = -\tilde g^2 3 C_F\,,\quad
  \gamma^{\rm IR \,(1)}_{{\rm c},\, s} = -\tilde g^2 4 C_F\,,
  \label{eq:collinearanomdim}
\end{equation}
where $b_0 = (11N-2N_f-N_s/2)/3$ is the coefficient in the one-loop $\beta$-function of $g$, and $C_F= (N^2-1)/2N$ is the Casimir of the fundamental representation.
While we only consider one flavor of scalar in our model, we include the parameter $N_s$ in the $\beta$-function and elsewhere to track contributions from scalar loops.

In addition, there are contributions to the $1/\epsilon$
UV pole proportional to the one-loop $\beta$-function of the
dimension-four couplings, related to the renormalization of such
couplings
\begin{equation}
  \frac{1}{2 \epsilon} (n-L_i) \tilde\beta^{(1)} A_i^{(0)}\,,
\end{equation}
where $\tilde\beta^{(1)} =\beta^{(1)}/g^{(4)}$,  $n$ is the number of external states and $L_i$ is the length of
the operator $\O_i$, i.e., the number of fields it contains.  We
therefore conclude that the sum over bubble coefficients is related to
the UV anomalous dimensions by
\begin{equation}
\frac{1}{(4\pi)^2}\sum_s a^s_{2\,,i} = -\frac12 \bigl[ \gamma^{UV}_{ij} - \gamma^{\rm IR}_{\rm c} \delta_{ij} + (n-L_i)\tilde\beta^{(1)} \delta_{ij} \bigr]  A_j^{(0)}\,.
\end{equation}
Similar formulas have recently been used in
Ref.~\cite{RecentOneLoopAnomalous2}.  There are multiple methods by
which one might calculate these coefficients. We do so by using
generalized unitarity. For the purposes of extracting the UV
divergences, it suffices to evaluate four-dimensional cuts
\cite{onshellbeta, NonrenormHelicity}. However, we are interested in
obtaining the full amplitudes, including rational terms, as a stepping
stone towards calculating two-loop anomalous dimensions, so we use
$D$-dimensional unitarity cuts as described in
Section~\ref{section:onelooprat}.

The approach we outlined is very powerful at one loop, but at higher
loops becomes more difficult to use, because it requires two-loop
integration.  In particular, at higher loops simple decompositions of
integrals along the lines of Eq.~\eqref{eq:oneloopdecomp} do not
exist.  One might still construct the amplitudes using unitarity
methods, and then extract their UV divergences by carrying out the
loop integration, but one would like a simpler technique that avoids
much of the technical complexity.  Furthermore, to calculate two-loop
divergences, one must also keep track of evanescent one-loop
subdivergences, which contaminate the result.  By an evanescent
subdivergence we mean a subdivergence whose corresponding counterterm
vanishes in strictly four dimensions, but which cannot be ignored in
dimensional regularization (see e.g. Ref.~\cite{Evanescent, EvanescentFinite}).  While
not physical, these evanescent subdivergences greatly complicate
higher-loop calculations, and it is better to use a method that avoids
them, whenever possible. Ref.~\cite{TwoLoopGravitySimplified2} gives a
nontrivial two-loop example for Einstein gravity showing how on-shell
methods can efficiently bypass evanescent
effects~\cite{TwoLoopGravitySimplified1} to determine
renormalization-scale dependence.

\subsection{Anomalous dimensions directly from unitarity cuts}
\label{section:formalismCuts}

A much more direct way to obtain anomalous dimensions is to focus on
the renormalization-scale dependence encoded in the logarithms, and
not on the divergences. The logarithms are detectable in
four-dimensional unitarity cuts. Any dimensional imbalance in the
kinematic arguments of the logarithms must be balanced by
renormalization-scale dependence, so one can directly determine the
renormalization-scale dependence and any anomalous dimensions by
collecting the contributions from unitarity cuts.  For example, this strategy
has been used to greatly simplify the extraction of the
two-loop renormalization-scale dependence in Einstein
gravity~\cite{TwoLoopGravitySimplified2}.

The formalism of Caron-Huot and Wilhelm~\cite{Caron-Huot:2016cwu}
gives a rather neat way to carry out this strategy , allowing us to
extract the anomalous dimension at $L$-loops directly from phase-space
integrals of lower-loop on-shell form factors and amplitudes.  Among
other useful features, this makes potential zeros in the anomalous
dimension matrix much more transparent than with conventional
methods~\cite{NonrenormLength}.

By considering the analyticity of the form factors with respect to
complex shifts in momenta, along with unitarity, Caron-Huot and
Wilhelm derived the following compact equation:
\begin{equation}
e^{-i\pi D}F_i^* = S \, F_i^*\,,
\label{eq:ffabstract}
\end{equation}
which relates the phase of the S-matrix, $S$, to the dilatation
operator, $D$ (ignoring trivial overall engineering dimensions).  The
dilation operator acts on the conjugate form factor $F_i^*$.  Writing
$S=1+i\M$, Eq.~\eqref{eq:ffabstract} can be rewritten more practically
as
\begin{equation}
  (e^{-i\pi D}-1)F_i^* = i \mathcal{M}  F_i^*\,,
\label{eq:ffabstract2}
\end{equation}
where the scattering amplitude, $\mathcal M$, acts as a matrix on the
form-factor, yielding its imaginary part via the optical
theorem\footnote{In our notation the optical theorem states,
  $2\text{Im}F_i^* = \M F_i^*$ for form factors or $2\text{Im}\M = \M
  \M$ for amplitudes.}. The right-hand side of this equation is defined to be
a unitarity cut. As we discuss below, this equation precisely
captures the notion that the scale dependence of $F_i$ is encoded in the
coefficients of its logarithms. We note that the use of the complex conjugate
form factor, $F^*$, only affects the imaginary part, which do not affect our calculations
through two loops. Therefore, we drop the complex conjugation henceforth.

In dimensional regularization, the dilatation operator is related to
the single renormalization scale, $\mu$, as
$D = -\mu\partial_\mu$, reflecting the fact that $F_i$ can only
depend on dimensionless ratios $s_{ij}/ \mu$ (ignoring the overall
engineering dimensions), and that logarithms in
$s_{ij}$ kinematic variables must be balanced either by $\mu$ or by each
other. The dilatation operator then acts on the form factors as
\begin{equation}
  DF_i=-\mu\partial_\mu F_i=\left[\Delta\gamma_{ij} +\delta_{ij}\beta\, \partial  \right] F_j,
\label{eq:dilatation}
\end{equation}
where we have used the renormalization-group equation
\eqref{eq:renormgroup}.  This, together with equation
\eqref{eq:ffabstract}, gives us a powerful means to extract anomalous
dimensions.

\begin{figure}[tb]
	\centering
	\vskip -.0 cm 
		\centering
		\includegraphics[scale=0.6]{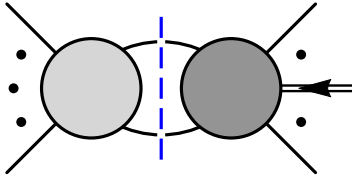} 
	\caption{
		Unitarity cut relevant for the extraction of anomalous dimensions from
		one-loop form factors. The darker blobs indicate a
		higher-dimension operator insertion. The  double-lined arrow indicates the
		insertion of additional off-shell momentum from the operator.  The dashed
		line indicates the integral over phase space of the particles crossing the
		cut.
	}
	\label{fig:formfactoroneloop}
\end{figure}

While Eqs.~\eqref{eq:ffabstract} and \eqref{eq:dilatation} are valid
non-perturbatively, we can expand in perturbation theory to obtain
order-by-order expressions for the anomalous dimensions. At one loop
the expansion yields
\begin{equation}
  \left[\Delta\gamma^{(1)}_{ij} + \delta_{ij}\beta^{(1)} \partial \right] F^{(0)}_j = -\frac{1}{\pi} (\mathcal M F_i)^{(1)} \,,
\label{eq:ffoneloop}
\end{equation}
where the superscript denotes the order in perturbation
theory.  On the right-hand side $(\M F_i)^{(1)}$ indicates 
\begin{align}
  \label{eq:oneloopsum}
  (\M F_i)^{(1)} &=\sum_{k=2}^n\sum_c \, (\M^{c}_{k\rightarrow 2})^{(0)} \otimes F_{n-k+2\,,i}^{(0)}\,,  
\end{align}
where the sums are over all kinematic channels and the $\otimes$ denotes a sum over intermediate two-particle states in the product. For a given kinematic channel this is given by the Lorentz-invariant phase-space integral
\begin{align}
  &(\M^{1\cdots k}_{k\rightarrow 2})^{(0)} \otimes F_{n-k+ 2\,,i}^{(0)} =\sum\int \dLIPStwo  \sum_{h_{1},h_{2}} \la 1\cdots k|\M| \ell_1^{h_1}\ell_2^{h_2}\ra^{(0)}\la \ell_1^{h_1}\ell_2^{h_2}\cdots n| \mathcal{O}_i|0\ra^{(0)} \nonumber \\
  &\hspace{1cm}= \sum\int \dLIPStwo  \sum_{h_{1},h_{2}} A^{(0)}(1,\cdots,k,-\ell_1^{-h_1},-\ell_2^{-h_2}) F_i^{(0)}(\ell_1^{h_1},\ell_2^{h_2},\cdots,n)\,, \label{eq:oneloopphasespace}
\end{align}
where the sum over helicities also includes a sum over different states crossing the cut.
In summary, $(\M F_i)^{(1)}$ corresponds to a sum over all one-loop two-particle unitarity cuts,
as depicted schematically in Figure~\ref{fig:formfactoroneloop}.

After rewriting the expression in terms of four-dimensional spinors,
the two-particle
phase-space integrals can be easily evaluated following the discussion 
of Ref.~\cite{Caron-Huot:2016cwu},
\begin{equation}
\left(\begin{array}{c} \lambda_1'\\ \lambda_2'\end{array}\right)=
\left(\begin{array}{cc} \cos\theta & -\sin\theta \e^{i\phi}\\\sin\theta \e^{-i\phi}&\cos\theta\end{array}\right)
\left(\begin{array}{c} \lambda_1\\ \lambda_2\end{array}\right) , \label{eq:rotation}
\end{equation}
where the $\lambda_i$ and $\tilde \lambda_i = \lambda_i^*$ spinors
depend on the momenta of the external legs and the $\lambda_i'$ and
$\tilde \lambda_i' = \lambda_i'{}^*$ spinors on the momenta of the cut
legs.  With this parametrization the integration measure is simply,
\begin{equation}
  \int \dLIPStwo \equiv \frac{1}{16\pi} \int_0^{2\pi} \frac{\d\phi}{2\pi}\int_0^{\frac{\pi}{2}}2\cos\theta\sin\theta \d\theta\,.
\end{equation}
In the definition of the phase-space measure, here we have included an
additional symmetry factor of $1/2$, relative to the usual volume of
two-particle phase space, i.e., $8\pi$. This is generally convenient
but requires some care when non-identical particles cross the cut,
where we will need to multiply by two to cancel the symmetry factor.

\begin{figure}[tb]
	\centering
	\vskip -.0 cm 
	\begin{minipage}{0.27\linewidth}
		\centering
		\includegraphics[scale=0.55]{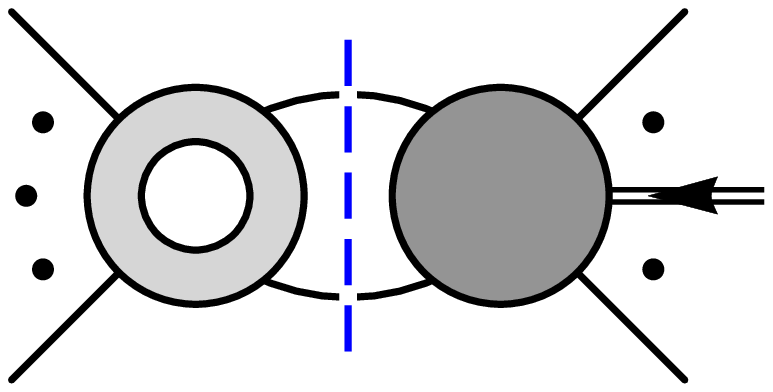}\\[-.14 cm]\textbf{(a)}
	\end{minipage}
	\hskip .7 cm
	\begin{minipage}{0.27\linewidth}
		\centering
		\includegraphics[scale=0.55]{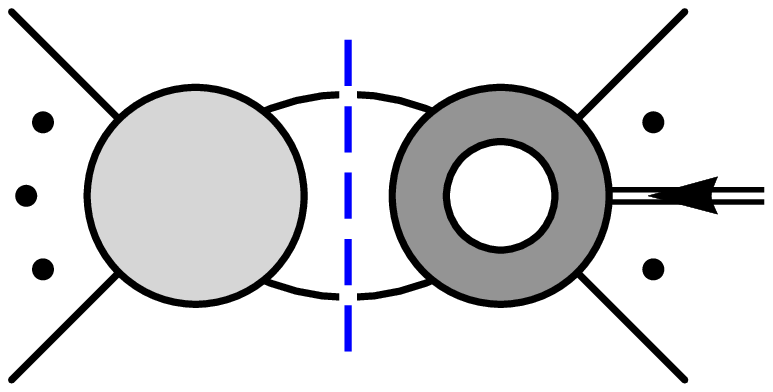}\\[-.14 cm]\textbf{(b)}
	\end{minipage}
	\hskip .7 cm 
	\begin{minipage}{.27\linewidth}
		\centering
		\includegraphics[scale=0.55]{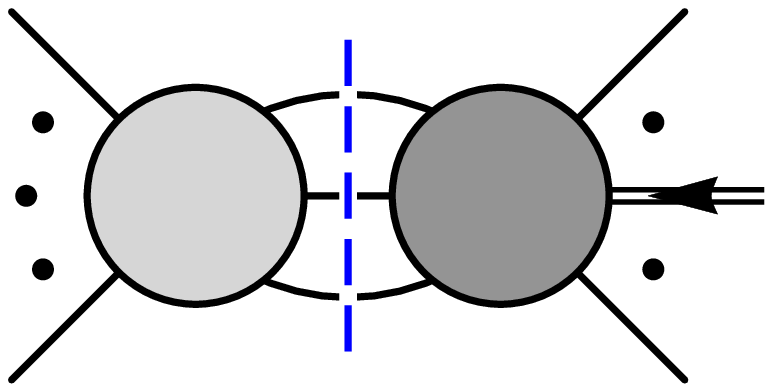}\\[-.14 cm] \textbf{(c)}
	\end{minipage}
	\caption{Unitarity cuts relevant for the extraction of
		anomalous dimensions from two-loop form factors, using the
		same notation as in Figure~\ref{fig:formfactoroneloop}.  The darker
		blobs indicate a higher-dimension operator insertion.  The
		blobs with a hole indicate a one-loop form factor or amplitude.}
	\label{fig:formfactortwoloop}
	\vspace{-0.4cm}
\end{figure}

Next consider two loops.  Expanding Eq.~\eqref{eq:ffabstract2} through
this order, we obtain
\begin{align}\label{eq:twoloopSimon}
\begin{split}
& \left[\Delta\gamma^{(1)}_{ij}+\delta_{ij}\beta^{(1)}\partial\right]F_j^{(1)}
+\left[\Delta\gamma^{(2)}_{ij}+\delta_{ij}\beta^{(2)}\partial\right]F_j^{(0)} \\
& \hskip 1.5 cm \null 
-i\pi\,\frac12\left[\Delta\gamma^{(1)}_{ik}+\delta_{ik}\beta^{(1)}\partial\right]\left[\Delta\gamma^{(1)}_{kj}
+\delta_{kj}\beta^{(1)}\partial\right]F_j^{(0)} = -\frac{1}{\pi}(\M F_i)^{(2)} \,.
\end{split}
\end{align}
On the right-hand side of this equation, $(\M F_i)^{(2)}$ denotes
collectively the three two-loop unitarity cuts displayed in
Figure~\ref{fig:formfactortwoloop},
\begin{align}
  (\M F_i)^{(2)} &=\sum_{k=2}^n\sum_c \, \left[ (\M^{c}_{k\rightarrow 2})^{(1)} \otimes F_{n-k+2\,,i}^{(0)}  + (\M^{c}_{k\rightarrow 2})^{(0)} \otimes F_{n-k+2\,,i}^{(1)} \right. \nonumber\\
  &\left. \hspace{3cm} +  (\M^{c}_{k\rightarrow 3})^{(0)} \otimes F_{n-k+3\,,i}^{(0)} \right] .
  \label{eq:twoloopsum}
\end{align}
In the first term we find two-particle cuts composed of the one-loop amplitude and the tree-level
higher-dimension form factor depicted in Figure~\ref{fig:formfactortwoloop}(a). These are
\begin{align}
  \label{eq:onezeroloopphasespaceA}
&(\M^{1\cdots k}_{k\rightarrow 2})^{(1)} \otimes F_{n-k+2\,,i}^{(0)}
  = \int \dLIPStwo  \sum_{h_{1},h_{2}} 
  \la 1 \cdots k|\M| \ell_1^{h_1}\ell_2^{h_2}\ra^{(1)} \, \la \ell_1^{h_1}\ell_2^{h_2}\cdots n| \mathcal{O}_i|0\ra^{(0)}
 \nonumber \\
 &\hspace{2cm}= \int \dLIPStwo  \sum_{h_{1},h_{2}} A^{(1)}(1,\cdots,k,-\ell_1^{-h_1},-\ell_2^{-h_2}) \, 
  F_i^{(0)}(\ell_1^{h_1},\ell_2^{h_2},\cdots,n) \,.
\end{align}
Similarly, the second term, shown in
Figure~\ref{fig:formfactortwoloop}(b), is a combination of cuts composed by the tree-level amplitude and the one-loop
higher-dimension operator, which are
\begin{align}
  \label{eq:onezeroloopphasespaceB}
&(\M^{1\cdots k}_{k\rightarrow 2})^{(0)} \otimes F_{n-k+2\,,i}^{(1)}
  = \int \dLIPStwo  \sum_{h_{1},h_{2}} 
  \la 1 \cdots k|\M| \ell_1^{h_1}\ell_2^{h_2}\ra^{(1)} \, \la \ell_1^{h_1}\ell_2^{h_2}\cdots n| \mathcal{O}_i|0\ra^{(0)}  \nonumber\\
  &\hspace{2cm}= \int \dLIPStwo  \sum_{h_{1},h_{2}} A^{(0)}(1,\cdots,k,-\ell_1^{-h_1},-\ell_2^{-h_2}) \, 
  F_i^{(1)}(\ell_1^{h_1},\ell_2^{h_2},\cdots,n) \,.
\end{align}
Finally, the third term is composed of three-particle cuts involving two tree-level objects, as in Figure~\ref{fig:formfactortwoloop}(c)
\begin{align}
  & (\M^{1\cdots k}_{k\rightarrow 3})^{(0)} \otimes F_{n-k+3\,,i}^{(1)}  = \int \dLIPSthree \!\! \sum_{h_{1},h_{2},h_{3}} \!\! \la 1\cdots k| \M| \ell_1^{h_1}\ell_2^{h_2} \ell_3^{h_3}\ra^{(0)}\la \ell_1^{h_1}\ell_2^{h_2} \ell_3^{h_3}\cdots n| \mathcal{O}_i|0\ra^{(0)} \nonumber\\
  &\hspace{0cm}= \int \dLIPSthree \!\!  \sum_{h_{1},h_{2},h_{3}} \!\! A^{(0)}(1,\cdots,k,-\ell_1^{-h_1},-\ell_2^{-h_2},-\ell_3^{-h_3}) F_i^{(0)}(\ell_1^{h_1},\ell_2^{h_2}, \ell_3^{h_3},\cdots,n) \,. 
  \label{eq:twoloopphasespace}
\end{align}
A parameterization analogous to \eqref{eq:rotation} for the three-particle cut is given in Ref.~\cite{Caron-Huot:2016cwu}. We will not evaluate any three-particle cuts in the present work, so we refer the reader to this work for more details.

\begin{figure}[tb]
	\centering
	\vskip -.0 cm 
	\begin{minipage}{.48\linewidth}
		\centering
		\hspace{10pt}\includegraphics[scale=0.6]{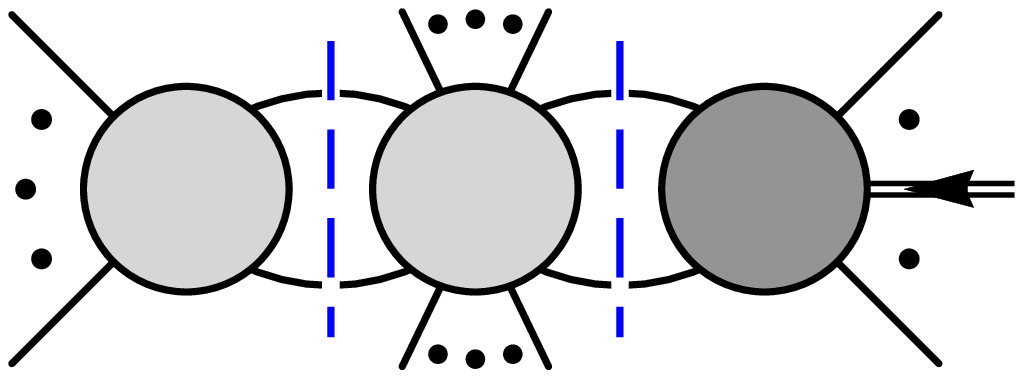}\\[-.001 cm] \textbf{(a)}
	\end{minipage}
	\begin{minipage}{.48\linewidth}
		\centering
		\includegraphics[scale=0.6]{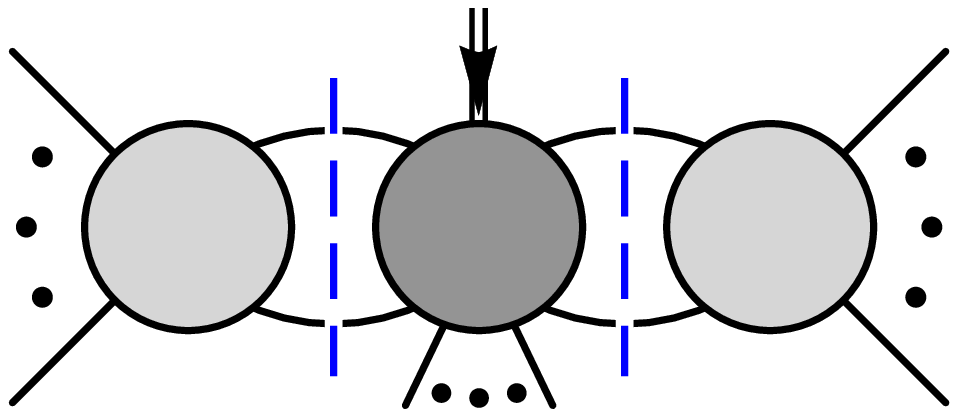}\\[-.001 cm] \textbf{(b)}
	\end{minipage}
	\caption{
	  Iterated two-particle cuts that appear on the right-hand side of Eq.~\eqref{eq:ImPart}.
	}
	\label{fig:formfactoriter}
	\vspace{-0.4cm}
\end{figure}

We can rearrange Eq.~\eqref{eq:twoloopSimon} to put it into a more convenient form for extracting two-loop anomalous dimensions.
First, note that the imaginary part of Eq.~\eqref{eq:twoloopSimon}
\begin{align}
  -i\pi\,\frac12\left[\Delta\gamma^{(1)}_{ik}+\delta_{ik}\beta^{(1)}\partial\right]\left[\Delta\gamma^{(1)}_{kj}+\delta_{kj}\beta^{(1)}\partial\right]F_j^{(0)} &= -\frac{1}{\pi}\text{Im}(\M F_i)^{(2)}\,,
  \label{eq:oneloopsquared}
\end{align}
does not feature the two-loop anomalous dimensions. Using the optical theorem again, we write its right-hand side in terms of unitarity cuts 
\begin{equation}
  \text{Im}(\M F_i)^{(2)} =  (\M \M F_i)^{(2)} \,,
\label{eq:ImPart}
\end{equation}
where the relevant cuts are the iterated two-particle cuts in Fig.~\ref{fig:formfactoriter}.  For instance $(\M \M F_i)^{(2)}$ contains terms of the form
\begin{align}
   & \!\!  \int \dLIPStwo \dLIPStwo' \sum_{h_{1},h_{2}}  \sum_{h'_{1},h'_{2}} \la \cdots| \ell_1^{h_1}\ell_2^{h_2}\ra^{(0)}  \la  \ell_1^{h_1}\ell_2^{h_2} \cdots| \ell_{1'}^{h'_1}\ell_{2'}^{h'_2}\ra^{(0)} \la \ell_{1'}^{h'_1}\ell_{2'}^{h'_2}\cdots| \O_i|0\ra^{(0)} \,,
\end{align}
which correspond to cuts of the type in Fig.~\ref{fig:formfactoriter}(a).
Note that Eq.~\eqref{eq:ImPart} does not include a factor of $1/2$ from the optical theorem because the imaginary part can arise from cutting either the one-loop amplitude or form factor, which give identical contributions.

Eq.~\eqref{eq:oneloopsquared} does not contain the two-loop anomalous dimensions but instead captures the exponentiation of one-loop anomalous dimensions and the associated logarithms. Nonetheless \eqref{eq:ImPart} can be used to simplify the real part of Eq.~\eqref{eq:twoloopSimon}, which yields 
\begin{align}\label{eq:twoloopSimon2}
  &\left[\Delta\gamma^{(1)}_{ij}+\delta_{ij}\beta^{(1)}\partial\right]\text{Re}F_j^{(1)}+\left[\Delta\gamma^{(2)}_{ij}+\delta_{ij}\beta^{(2)}\partial\right]F_j^{(0)}\\
  & \hspace{3cm} =-\frac{1}{\pi}\text{Re}(\M F_i)^{(2)}  = -\frac{1}{\pi} (\M F_i -  \M \M F_i)^{(2)}\,.  \nonumber
\end{align}
Note that the right-hand side can be rewritten using
\begin{align}
  (\M F_i -  \M \M F_i)^{(2)} =  \left[\left(\M - \frac12 \M\M\right) \left(F_i -\frac12 \M F_i\right)\right]^{(2)} =  \left[\text{Re}(\M) \text{Re}(F_i)\right]^{(2)}\,, \hskip .8 cm 
\end{align}
and with this we arrive at 
\begin{align}\label{eq:twoloopSimon3}
 & \left[\Delta\gamma^{(1)}_{ij}+\delta_{ij}\beta^{(1)}\partial\right]\text{Re}F_j^{(1)}+\left[\Delta\gamma^{(2)}_{ij}+\delta_{ij}\beta^{(2)}\partial\right]F_j^{(0)} =-\frac{1}{\pi} \left[\text{Re}(\M) \text{Re}(F_i)\right]^{(2)}\,.  
\end{align}
We use this equation to extract two-loop anomalous
dimensions.  In practice Eq.~\eqref{eq:twoloopSimon3} simply instructs
us to drop the imaginary parts of the one-loop matrix elements when
calculating the cuts in Figs.~\ref{fig:formfactortwoloop}(a) and
\ref{fig:formfactortwoloop}(b). On the left-hand side, we now see the
appearance of one-loop anomalous dimensions and the $\beta$-function, as 
well as the one-loop form factor $F_i^{(1)}$. The two-loop UV
anomalous dimension $\gamma_{ij}^{\rm UV(2)}$ contained in
$\Delta\gamma^{(2)}_{ij}$ is the object of interest, but to 
extract it we first need to remove $\gamma_{ij}^{\rm IR(2)}$,
which requires an understanding of the IR singularities, 
which we discuss below.

\subsubsection{Simplifying strategies}

A strategy that greatly simplifies the analysis is to choose an
external state with the minimal number of external legs that is
sensitive to the operator of interest, i.e. select the operator's
minimal form factor.  In this way we can avoid terms of the form
$\beta^{(n)}\partial F^{(0)}_i$ in Eqs.~\eqref{eq:twoloopSimon}
and \eqref{eq:twoloopSimon3}, since, under this choice, $F^{(0)}_i$ is
local, and thus does not depend on the dimension-four couplings,
$g^{(4)}$.  This strategy was used in Ref.~\cite{NonrenormLength} to
prove nonrenormalization theorems at the first loop order where
diagrams exist.

More generally, the $\beta$-function can no longer be eliminated by
using minimal form factors whenever the one-loop form factor
with an $\O_i$ insertion, $F^{(1)}_i$, produces a nonzero 
result with the chosen external states. In addition, the
$\beta$-function acting on the one-loop anomalous-dimension matrix
is nonzero if the matrix elements themselves are nonzero. For
example, to determine the renormalization of $\O_{F^3}$ by itself at
two loops, we would evaluate Eq.~\eqref{eq:twoloopSimon3} with
the external state $\la 1^+2^+3^+|$. In this case the term
$\beta^{(2)}\partial F_{F^3}^{(0)}$ would vanish, though the
term $\beta^{(1)}\partial F_{F^3}^{(1)}$ would remain.

Unlike the $\beta$-function, the IR anomalous dimensions are
non-trivial to eliminate. Ref.~\cite{Caron-Huot:2016cwu} removes them
by subtracting, at the integrand level, form factors of global symmetry
currents, such as the stress-tensor, which are UV finite but contain
the same IR divergences. Alternatively, one can use the same on-shell
methods to calculate them and subtract them after
integration. At one loop, the structure of infrared divergences is well
understood~\cite{IRBasicPapers, IRLaterPapers, StermanBook}, and it is
straightforward to subtract them after integration. We explain how to 
carry this out at the level of the amplitudes in the next section. Furthermore,
whenever we are interested in a leading off-diagonal element of the
anomalous dimension matrix, the IR anomalous dimensions does not appear,
since the infrared divergences are diagonal in the operators
(excluding color).

Finally, form factors are useful for operators with only two or three
external fields, since they allow nonzero results when kinematics
would otherwise set amplitudes with fewer than four external particles
to zero. Here we generally set the operator momentum insertion $q=0$
and work in terms of amplitudes whenever possible, i.e. whenever there
are four or more external states.

\subsection{Comments on evanescent operators}
When extracting anomalous dimensions from UV divergences in
dimensional regularization one must carefully keep track of evanescent
operators~\cite{Evanescent,EvanescentFinite}.  These operators are non-trivial in
$D$-dimensions, but whose matrix elements vanish for any choice of
external four-dimensional states.  In the context of the SMEFT an
example of an evanescent operator would be the Lorentz--Fierz
identities
\begin{align}
  \O_{\rm Fierz, L} &= (\bar \psi{}^m_L \gamma^\mu \psi^n_L)(\bar \psi{}^p_L \gamma_\mu \psi^r_L)+
(\bar \psi{}^p_L \gamma^\mu \psi^n_L)(\bar \psi{}^m_L \gamma_\mu \psi^r_L)\,,\nonumber\\
  \O_{\rm Fierz, R} &= (\bar \psi{}^m_R \gamma^\mu \psi^n_R)(\bar \psi{}^p_R \gamma_\mu \psi^r_R)+
(\bar \psi{}^p_R \gamma^\mu \psi^n_R)(\bar \psi{}^m_R \gamma_\mu \psi^r_R)\,,
\end{align}
(where we raised the flavor indices for convenience) which are identically zero in four but not in arbitrary
dimensions. More generally one can easily construct such operators by
antisymmetrizing over five or more Lorentz indices. In
the context of our model, an example of such an evanescent operator is
\begin{equation}
  (\bar\psi \gamma_{[\alpha} \gamma_{\mu} \gamma_{\nu} \gamma_{\sigma} \gamma_{\rho]} \psi) (\bar\psi \gamma^{[\alpha} \gamma^{\mu} \gamma^{\nu} \gamma^{\sigma} \gamma^{\rho]} \psi) \,.
\end{equation}

One-loop diagrams might contain a $1/\epsilon$ divergence proportional
to the matrix element of an evanescent operator.  While this does not
affect one-loop anomalous dimensions because we can take the external
states to be four-dimensional, when inserted in a higher-loop diagram
in the context of dimensional regularization such evanescent operators
are activated and can generate both UV divergent and finite
contributions.  In fact, they are needed to properly subtract
subdivergences. These effects must be taken into account in order to
correctly extract two-loop UV divergences and their associated
anomalous dimension. In practice we can deal with the effects of
evanescent operators~\cite{Evanescent,EvanescentFinite}, but the
number of them grows with dimension and loop order (especially in the
presence of fermions).  For this reason it would be desirable to avoid
them when possible, since they are a technical complication due to the
use of dimensional regularization, and ultimately we would expect that
they do not affect the physics~\cite{TwoLoopGravitySimplified1}.

We expect the on-shell methods presented above to completely sidestep
the issue of evanescent operators when obtaining anomalous dimension,
at least through two loops. Ref.~\cite{TwoLoopGravitySimplified2}
provides a nontrivial demonstration that complications from evanescent
operators can be completely sidestepped using on-shell methods and by
focusing on renormalization-scale dependence instead of
divergences. In the two-loop formulas used here, anomalous dimensions and
associated logarithms are given directly in terms of four-dimensional
unitarity cuts of tree and one-loop objects.  This automatically
eliminates most of the evanescent dependence, except for finite
shifts in one-loop matrix elements with evanescent operator
insertions.  We expect that any remaining evanescent dependence in
the one-loop amplitudes or form factors to be eliminated by finite
renormalizations~\cite{EvanescentFinite}.  Given the usual subtleties
of dealing with evanescent operators, it would, of course, be
important to explicitly verify that including or not including
evanescent operators in the one- and two-loop anomalous dimension matrix amounts to
a scheme choice.

\subsection{Anomalous dimensions and non-interference}

As noted in Ref.~\cite{NonInterference} helicity selection rules imply
the non-interference of SMEFT tree-level matrix elements when
constructing cross sections.  This has important consequences in the
context of the SMEFT, where the possibility of measuring the
coefficient of higher-dimension operators at colliders can be impacted
by the vanishings in the interference of the Standard-Model tree
amplitudes and those of higher-dimension operators, when computing
cross sections.  A connection between one-loop anomalous dimension and
interference terms can be seen in Eq.~\eqref{eq:ffoneloop}, where, upon
setting $q=0$, the form factors become amplitudes and the right-hand
side directly captures the interference of tree-level dimension-four
and dimension-six amplitudes.  Note that this holds even when the
anomalous dimension is not zero, in which case this equation relates
the interference terms to simpler objects, namely the one-loop
anomalous dimensions and tree-level matrix elements.  Of course, in a
realistic cross-section calculation one would not integrate over the
full phase space, due to experimental cuts.

At two loops the connection between zeros in the anomalous dimensions
and non-interference is not as direct, since it requires cancellations
between both sides of
Eq.~\eqref{eq:twoloopSimon}. Eq.~\eqref{eq:oneloopsquared} shows that,
in general, the imaginary part of the interference term is given by
the square of one-loop anomalous dimensions times tree-level matrix
elements.  Instead of non-interference, Eq.~\eqref{eq:twoloopSimon2}
shows that a vanishing two-loop anomalous dimension would imply that
the real part of interference term is simply is related to the product
of one-loop anomalous dimensions and one-loop matrix elements.  It
would be interesting to further investigate the consequences stemming
from these observations, even in the presence of experimental cuts.

\section{One-loop amplitudes and anomalous dimensions}
\label{section:onelooprat}

In this section we describe our generalized unitarity calculation of the
one-loop amplitudes with an insertion of a higher-dimensional operator in our
simplified model.  We then extract the one-loop anomalous dimension matrix of
this theory. Finally, we comment on the structure of rational terms in the
amplitudes and on the ability to set some of them to zero with a judicious
scheme choice. The results in this section are building blocks needed for the
two-loop analysis in the next section. In addition, they provide one-loop
anomalous dimensions that can be cross-checked against those in
Refs.~\cite{Manohar123}. 

\subsection{One-loop amplitudes from generalized unitarity}

The generalized unitarity method~\cite{GeneralizedUnitarity1, ChiralOnShell,
GeneralizedUnitarity2} for constructing one-loop amplitudes can be found in
various reviews, for example see Ref.~\cite{UnitarityReviews}, but here we
briefly review the procedure for the one-loop case.  To construct the full
one-loop amplitudes to all orders in the dimensional-regularization parameter
$\epsilon$, we begin with the $D$-dimensional four-point tree-level amplitudes
with or without insertions of the dimension-6 operators (given in Appendix
\ref{section:amplitudesSH}). By using $D$-dimensional tree amplitudes, we
ensure that the cuts appropriately capture the coefficients of the
$D$-dimensional box, triangle, and bubble scalar integrals that form a basis
for the full one-loop amplitudes, as in Eq.~\eqref{eq:oneloopdecomp}.  In
general, the coefficients have $\epsilon$ dependence, and expanding in $\eps$
produces rational terms that would not automatically be included if a purely
four-dimensional approach to the cuts were used~\cite{GeneralizedUnitarity1}.
Besides $\eps$, the coefficients only depend on the Mandelstam invariants
$s=(k_1+k_2)^2$, $t=(k_2+k_3)^2$ and $u=(k_1+k_3)^2$.

We construct the cuts in the standard way. For example, the
integrand-level $s$-channel cut with an $\O_n$ operator insertion is given by
\begin{align}
\begin{split}
\sum_i& \C^{[i]}\left(\left[a^{st}_{4,n[i]}I_{4,st}+a^{su}_{4,n[i]}
I_{4,su}+a^s_{3,n[i]}I_{3,s}+a^s_{2,n[i]}I_{2,s}\right]\Big\rvert_{\ell^2=0}\, \right)\\
&=\sum_{\text{states } }\sum_j\C^{[j]}A_n^{(0)}(1,2,\ell^{h_1}_1,\ell^{h_2}_2)_{[j]}
\sum_k\C^{[k]}A^{(0)}(-\ell^{h_2}_2,-\ell^{h_1}_1,3,4)_{[k]}\\
&\quad+\sum_{\text{states } }\sum_j\C^{[j]}A^{(0)}(1,2,\ell^{h_1}_1,\ell^{h_2}_2)_{[j]}
\sum_k\C^{[k]}A_n^{(0)}(-\ell^{h_2}_2,-\ell^{h_1}_1,3,4)_{[k]} \,,
\end{split}\label{eq:cutexplicit}
\end{align}
where the sum over states includes the helicity and the color, and,
for this case, $\ell_2=-(\ell_1+k_1+k_2)$. The $\C^{[i]}$ are the
appropriate color factors for the associated amplitudes.  Since the
cut legs are on-shell, where $\ell_1^2=\ell_2^2=0$. Often, the
external particles will restrict $A_n^{(0)}$ to be nonzero only for
certain cuts or placements within the cuts, depending on the field
content of the operator inserted.

\begin{figure}
	\centering
	\begin{minipage}{0.4\columnwidth}
		\centering \textbf{(a)}\\[-.5 cm] 
		\includegraphics[scale=0.6]{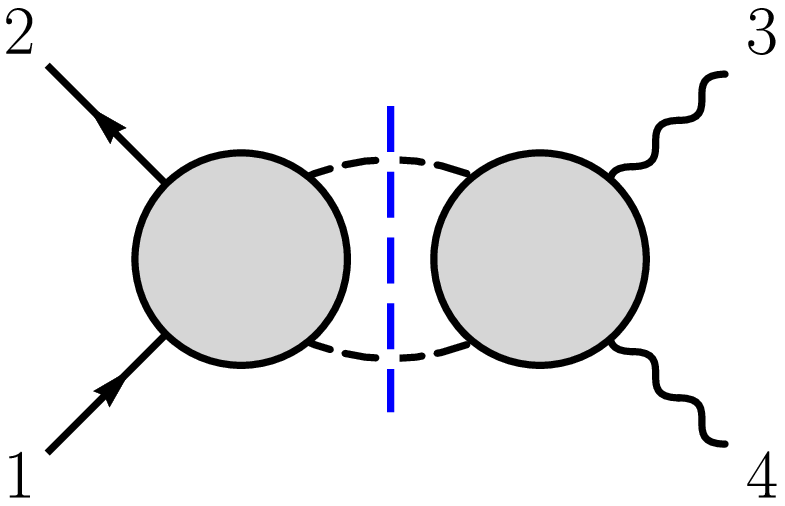}\\[.5 cm] 
	\end{minipage}
	\begin{minipage}{0.25\columnwidth}
		\centering \textbf{(b)}\\[-.5 cm] 
		\includegraphics[scale=0.6]{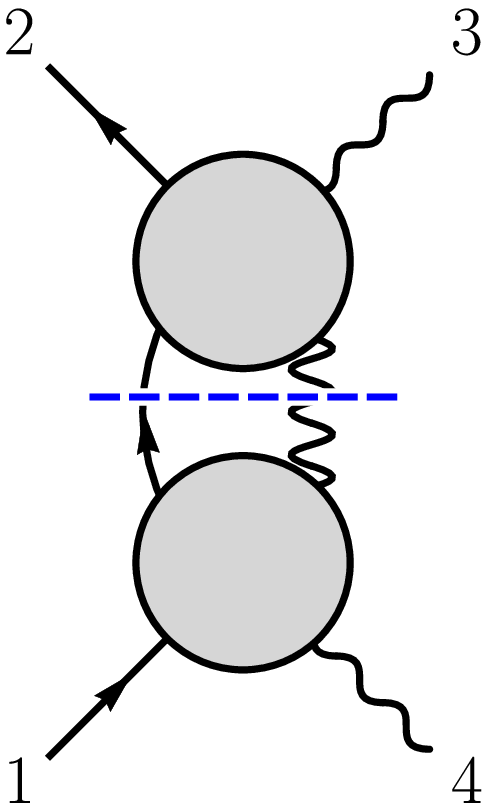}\\[.5 cm] 
	\end{minipage}
	\begin{minipage}{0.25\columnwidth}
		\centering \textbf{(c)}\\[-.5 cm]
		\includegraphics[scale=0.6]{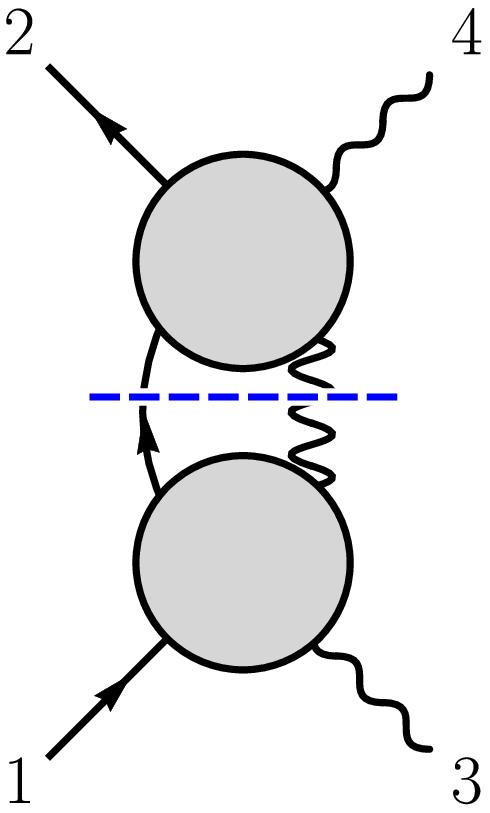}\\[-.5 cm]
	\end{minipage}
	\begin{minipage}{0.4\columnwidth}
		\centering
		\includegraphics[scale=0.6]{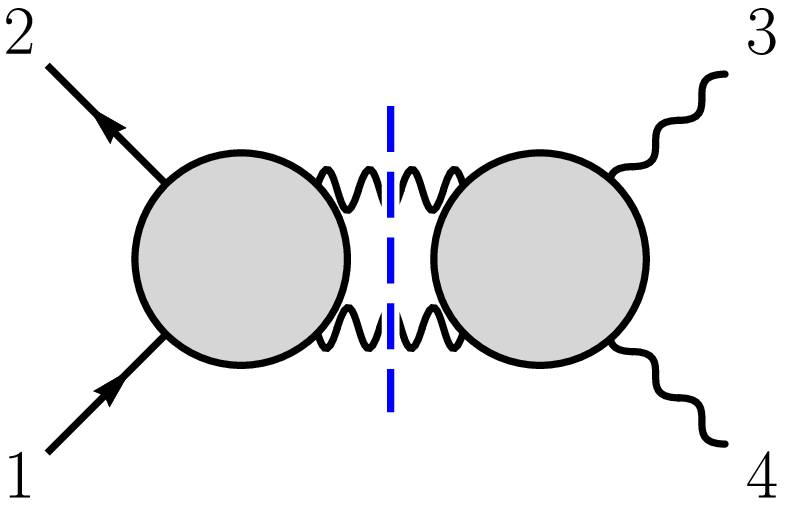}\\[-.5 cm] \textbf{(d)}
	\end{minipage}
	\begin{minipage}{0.4\columnwidth}
		\centering 
		\includegraphics[scale=0.6]{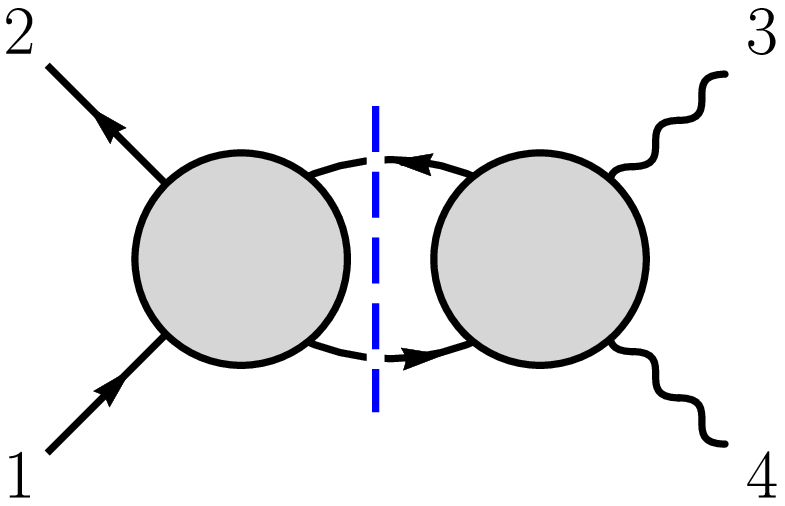}\\[-.5 cm] \textbf{(e)}
	\end{minipage}
	\caption{The necessary cut for constructing a two-fermion, two-vector
	  amplitude. For an amplitude with an insertion of a higher-dimension
	  operator, one should insert the operator into either side of the
	  diagrams when possible.  The wavy lines are vector bosons, the lines
	  with arrows fermions and the dashed lines scalars.}
	\label{fig:cuts}
\end{figure}

As an example, the cuts of the amplitude
$A^{(1)}_{\FCube}(1_{\vphantom{\bar\psi}\psi}2_{\bar\psi}3\,4)$, are
shown in Figure \ref{fig:cuts}, where the operator $\O_{\FCube}$
should be inserted on either side of the cuts, when the tree
amplitudes exist. Other amplitudes with four-point operators require
only the cuts corresponding to their correct external particles. The
color factors $\C^{[j]}\C ^{[k]}$ can be reduced to the appropriate
color basis of the full amplitude, $\C^{[i]}$, based on the external
particles. Doing so determines the contribution from each
color-decomposed cut.

We evaluate the cuts using the $D$-dimensional state sum completeness relations, 
\begin{align}\label{eq:statesum}
\begin{split}
\epsilon_i^{*\mu}\odot\epsilon_i^\nu=\sum_{\text{states }h}\epsilon_i^{*(h)\mu}\epsilon_i^{(h)\nu}&
=-g^{\mu\nu}+\frac{q^\mu k_i^\nu+k_i^\mu q^\nu}{q\cdot k_i} \,,\\[5pt]
\bar{u}_i\odot u_i=\sum_{\text{states }h}&\bar{u}^{(h)}_i u^{(h)}_i=\slashed{k}_i \,,
\end{split}
\end{align}
where $q$ is an arbitrary reference vector with $q^2=0$.

The next task is to merge the cuts and to integrate. One can merge the
cuts at the level of the integrand to find a single integrand that has
the correct cuts in all channels. However, is it is generally simpler
to merge the integrated results from each cut, treating each cut as an
off-shell object, but dropping contributions that do not have a cut in
the given channel. Integration is done by projecting each cut for a
given process onto a basis of gauge-invariant tensors, as described in
more detail in Appendix~\ref{section:amplitudesappendix}. Although the
methods we use to extract anomalous dimensions do not require us to
keep track of evanescent divergences, because the projection technique
is fully $D$ dimensional, we track them and confirm that they do not
enter our calculations of various entries in the two-loop
anomalous-dimension matrix. An alternative is to use spinor-helicity
methods~\cite{SpinorHelicity} which are much more powerful when the
number of external legs increases.  These have been
successfully used for both chiral~\cite{ChiralOnShell} and higher-loop
calculations~\cite{TwoLoopHelicityExample}, but then additional care
is needed to deal with subtleties that arise from using dimensional
regularization.

After projection, the cut integrand is rewritten in terms of inverse
propagators. We reduce the remaining integrals to
the basis of scalar integrals in Eq.~\eqref{eq:oneloopdecomp} using
integration by parts relations as implemented in FIRE~\cite{FIRE}. Cut
merging is then straightforward, as the coefficients of integrals in
the merged amplitude can be read directly off the results from each
cut, summed over the possible particles crossing the cut.  For
example, the $s$-channel cut in Eq.~\eqref{eq:cutexplicit} yields the
coefficients of the $s$-channel bubble and triangle, as well as those
of the $(s,t)$ and $(s,u)$ boxes in Eq.~\eqref{eq:oneloopdecomp}.

The full set of $D$-dimensional four-point one-loop amplitudes for the
dimension-six operators in our model are given in the ancillary
file~\cite{AttachedFiles}. These expressions are valid to all orders
in $\epsilon$, but to obtain the finite, renormalized expressions
needed to feed into our calculation of two-loop anomalous
dimensions, we need to subtract the UV poles.

The one-loop amplitudes are IR divergent. The IR singularities of gauge theories are well understood~\cite{IRBasicPapers,
IRLaterPapers, IRModernPapers, StermanBook}, and can be expressed in terms of
lower-loop amplitudes involving the same operator insertion and external
particles. The explicit form of the one-loop infrared singularity, for example,
is given by 
\begin{align}\label{eq:IRexplicit} A_i^{(1)}= \null & \boldS
  I^{(1)}A_i^{(0)} \,,
\end{align} 
where the IR operator $\boldS I^{(1)}$ is given by
\cite{IRBasicPapers,IRLaterPapers,IRModernPapers}\footnote{The difference with
the formulas in those references is due to our normalization of the $SU(N)$
generators.}
\begin{align}
\boldS I^{(1)} 
&= \frac{e^{\epsilon\gamma_E}}{\Gamma(1-\epsilon)}
\sum_{p=1}^n\sum_{q\ne p} \frac{ \boldS T_p \cdot \boldS T_q}{2}\left[\frac{\gamma^{\rm IR\, (1)}_{\rm cusp}}{\epsilon^2}
-\frac{\gamma^{\rm IR\, (1)}_{{\rm c},\, p}}{\boldS T_p^2}\frac{1}{\epsilon}\right]\left(\frac{-\mu^2}{2k_p{\cdot}k_q}\right)^\epsilon \,,
\end{align}
where the sums are over external particles.  The color charge $\boldS
T_p=\{T^a_p\}$ is a vector with respect to the generator label $a$ and a
SU($N$) matrix with respect to the outgoing particle $p$. The infrared
divergence includes a $1/\epsilon^2$ pole, with coefficient given by the cusp
anomalous dimension $\gamma^{\rm IR\, (1)}_{\rm cusp} = 4\tilde g^2$, and
$1/\epsilon$ poles, with coefficient given by the collinear anomalous dimension
of particle $p$ given in Eq.~\eqref{eq:collinearanomdim}.  By
obtaining the IR dependence of the one-loop amplitudes from
Eq.~\eqref{eq:IRexplicit}, we can subtract it from the full one-loop
amplitudes.  As always, the definition of the IR-divergent parts carries with
it some arbitrariness as to which finite pieces are included.\footnote{In
  physical quantities this arbitrariness cancels between real emission and
virtual contributions.}

The remaining poles in $\epsilon$ are UV poles, which we then match to
the appropriate tree-level counterterm amplitude containing an
insertion of the operator $\O_j$. A complication is that there can be
multiple operators corresponding to the same external particle
content, but with different color structures. Therefore, in these cases
the coefficient of a single color factor in the loop amplitude is
insufficient for the purpose of determining the anomalous dimensions,
and in principle all the color factors for the given process and
operator insertion must be considered simultaneously. For example, the
one-loop amplitude with an insertion of the $\O_{(D\varphi^2\psi^2)_2}$
operator and four external scalars determines the renormalization of
both the $\O_{(D^2\varphi^4)_1}$ and the $\O_{(D^2\varphi^4)_2}$ operators,
where the operators are given in Table~\ref{tab:operators}.

In some cases the IR structure is trivial, e.g. when the IR anomalous
dimensions are zero simply because there are no lower-loop amplitudes
for a given operator and given external state.  Our examples in
Section~\ref{section:twoloop} follow this pattern. For instance, in
the example of $\O_{(D^2\varphi^4)_1}$ renormalizing $\O_{(\psi^4)_1}$ at two
loops, there is no tree level or one-loop amplitude with an insertion
of $\O_{(D^2\varphi^4)_1}$ which has an external state of four fermions,
simply due to the lack of Feynman diagrams. Since the full IR
dependence is proportional to lower-loop amplitudes, this implies
there cannot be an IR divergence at two loops. This same reasoning
underpinned the non-renormalization theorem in Ref.~\cite{NonrenormLength}.
More generally, one needs to account for the infrared singularities.

\subsection{One-loop UV anomalous dimensions}
\label{section:oneloopuv}

After subtracting the IR singularities, the only remaining $1/\epsilon$ poles in the
amplitudes correspond to the desired one-loop anomalous dimensions,
\begin{align}
  \dot \opcoef_{\FCube}=\null &\tilde g^2 (12 N-3b_0)\opcoef_{\FCube}\,, \nonumber \\[1mm]
  \dot{\opcoef}_{\PhiSquareFSquareA}= \null & \tilde g^2 \biggl( -5 \opcoef_{F_3} - \frac{(3N^2- 7)+2 Nb_0}{N} \opcoef_{\PhiSquareFSquareA} + \frac{N^2-4}{N^2} \opcoef_{\PhiSquareFSquareB} \biggr) \nonumber \\
&+\tilde \lambda\, 4(1+N) \opcoef_{\PhiSquareFSquareA} \,,
\nonumber \\[1mm]
  \dot{\opcoef}_{\PhiSquareFSquareB}= \null & \tilde g^2 \left(-N\opcoef_{\FCube} + 2\opcoef_{\PhiSquareFSquareA} +\frac{2N^2-5 -2Nb_0}{N}\opcoef_{\PhiSquareFSquareB} \right) 
+ \tilde\lambda\, 4 \opcoef_{\PhiSquareFSquareB},
\nonumber \\[1mm]
  \dot{\opcoef}_{\DSquarePhiFourthA}=\null & \tilde g^2 \left(\frac{3  (N+1)}{N}\opcoef_{\DSquarePhiFourthA}  + \frac{2  (N-2) (N_s+9)}{3 N}\opcoef_{\DSquarePhiFourthB}+\frac{4}{3}\frac{N-2}{N}\opcoef_{\DPhiSquarePsiSquareB}^{ww} \right)\nonumber \\
& + \lambda\, 12 \opcoef_{\DSquarePhiFourthA} \,,
\nonumber \\[1mm]
  \dot \opcoef_{\DSquarePhiFourthB}= \null & \tilde g^2\left(\frac{36NC_F-(2 N-1) (N_s+9)}{3 N}\opcoef_{\DSquarePhiFourthB}
 +  \frac{ 3 (N-2) (N+1) }{2 N} \opcoef_{\DSquarePhiFourthA}
   \right.\nonumber \\
  &\left.
 \hskip .5 cm \null 
+\frac{2(2N-1)}{3N} \opcoef_{\DPhiSquarePsiSquareB}^{ww} \right) 
+ \tilde\lambda\,\left(2 (N-2) \opcoef_{\DSquarePhiFourthA} + 8 (N+1) \opcoef_{\DSquarePhiFourthB}\right) ,
 \nonumber \\
\dot \opcoef_{\DPhiSquarePsiSquareA}^{pr}= \null & 0 \,,  \nonumber
 \\  
  \dot \opcoef_{\DPhiSquarePsiSquareB}^{pr}= \null &\, \tilde g^2\left(\frac{1}{3} N_s\opcoef_{\DSquarePhiFourthB}\delta_{pr}  +   \frac{1}{3}(-9 N+N_s) \opcoef_{\DPhiSquarePsiSquareB}^{pr} +\frac{4}{3} N_f\opcoef_{\DPhiSquarePsiSquareB}^{ww}\delta_{pr} \right.\nonumber \\
  & \left. \null 
 -\frac{2}{3}N_f\opcoef_{\PsiFourthA}^{pwwr} -  \frac{2}{3} N_f \left(2  \opcoef_{\PsiFourthB}^{prww}-\frac{1}{N} \opcoef_{\PsiFourthB}^{pwwr}\right) \right) ,
\nonumber \\
\dot \opcoef_{\PsiFourthA}^{mnpr}= \null  &  \tilde g^2\frac{6 \left(N^2-1\right) }{N^2} \opcoef_{\PsiFourthB}^{mnpr}\,,  \nonumber \\
\dot \opcoef_{\PsiFourthB}^{mnpr}= \null&\tilde g^2\left(
-\frac{ N_s}{3}(\opcoef_{\DPhiSquarePsiSquareB}^{mn}\delta_{pr}+\opcoef_{\DPhiSquarePsiSquareB}^{pr}\delta_{mn})\right.
\nonumber\\&\left.
+\frac{2}{3} N_f (\delta_{mn} \opcoef_{\PsiFourthA}^{pwwr}+\delta_{pr} \opcoef_{\PsiFourthA}^{mwwn})+6 \opcoef_{\PsiFourthA}^{mnpr}  -\frac{3}{N}\opcoef_{\PsiFourthB}^{mnpr}
\nonumber\nonumber\right. \\&\left.+\frac{2N_f}{3 N}(2 N (\delta_{pr} \opcoef_{\PsiFourthB}^{mnww}+\delta_{mn} \opcoef_{\PsiFourthB}^{prww})- (\delta_{pr} \opcoef_{\PsiFourthB}^{mwwn}+\delta_{mn} \opcoef_{\PsiFourthB}^{pwwr}))  \right)
 \,.
\end{align}
Here $N_s$ is left as a parameter to track contributions from scalar loops. In our model it should be set to unity.
These anomalous dimensions have been extracted directly from the scattering amplitudes,
and, as a cross-check, we also used the unitarity cut method explained
in the previous section~\cite{Caron-Huot:2016cwu} for computing
directly the anomalous dimensions. The structure of the anomalous
dimension matrix is summarized in Table~\ref{tab:b}.  It is worth
pointing out the simplicity in the renormalization and mixing of
$\DPhiSquarePsiSquareA$ and $\PsiFourthA$, which is due to these
operators being a product of global symmetry currents, which heavily
constrains the kind of states they can overlap with. This is special
in our model, which does not contain an Abelian gauge field. In the
presence of the latter, the operators would be a product of gauge
symmetry currents (just like $\DPhiSquarePsiSquareB$ and
$\PsiFourthB$) which are renormalized \cite{Collins:2005nj}, so the
anomalous dimension matrix will receive contributions proportional to
the Abelian gauge coupling.

We use these results to verify a representative set of the
one-loop anomalous dimension calculated in Ref.~\cite{Manohar123},
including entries from nearly all classes of operators. Additional
details about this verification is given in
Section~\ref{section:mapping}.  This provides a nontrivial check on
our one-loop results, which we then feed into the two-loop anomalous
dimension calculations.

\begin{table}[tb]
	\begin{center}
		\renewcommand{\arraystretch}{1.2}
		\setlength{\tabcolsep}{4pt}
		\begin{tabular}{c|ccccccccc|}
			&\small{$\FCube$} &\small {$\PhiSquareFSquareA$} & \small{$\PhiSquareFSquareB$}& \small${\DSquarePhiFourthA}$ & \small${\DSquarePhiFourthB}$ & \small${\DPhiSquarePsiSquareA}$ &\small ${\DPhiSquarePsiSquareB}$ &\small ${\PsiFourthA}$ & \small${\PsiFourthB}$ \\
			\hline
			\small${\FCube}$ & &0  &0 &$\slashed0$ &$\slashed0$ &$\slashed0$ &$\slashed0$& $\slashed0$ & $\slashed0$ \\[1mm]\hline
			\small${\PhiSquareFSquareA}$ & & & &0 &0 &0 &0 &$\slashed0$ &$\slashed0$\\[1mm]\hline
			\small${\PhiSquareFSquareB}$ & & & &0 &0 &0 &0 &$\slashed0$ &$\slashed0$\\[1mm]\hline
			\small${\DSquarePhiFourthA}$  &0 &0 &0 & & &0 & &$\slashed0$ &$\slashed0$\\[1mm]\hline
			\small${\DSquarePhiFourthB}$ &0 &0  &0 & & &0 & &$\slashed 0$ &$\slashed0$\\[1mm]\hline
			\small${\DPhiSquarePsiSquareA}$  &0 &0 &0 & 0&0 &0 &0 &0 &0\\[1mm]\hline
			\small${\DPhiSquarePsiSquareB}$ &0 &0 &0 & 0& &0 & & &\\[1mm]\hline
			\small${\PsiFourthA}$ &0 &$\slashed0$  &$\slashed0$ &$\slashed0$ &$\slashed0$ &0 & 0 &0 &\\[1mm]\hline
			\small${\PsiFourthB}$ &0 &$\slashed0$  &$\slashed0$ &$\slashed0$ &$\slashed0$ &0 & & &\\[1mm]\hline
		\end{tabular}
\vskip .2 cm 
\caption{Structure of the zeros in the one-loop anomalous dimension
  matrix. The $\slashed0$ entries indicate there are no
  contributing one-loop diagrams, whereas a $0$ alone indicates that
  there are one-loop diagrams that could contribute, but actually give
  a vanishing result. The operators labeling the rows are renormalized by the
  operators labeling the columns.}\label{tab:b}
	\end{center}
\end{table}

\subsection{Structure of one-loop amplitudes and rational terms}

\begin{table}[tbh]
	\begin{center}
		\vspace{2mm}
		\renewcommand{\arraystretch}{1.2}
		\begin{tabular}{c|cccccccccccccc|}
		  &\rotp{$V^+V^+V^+V^+$}&\rotp{$V^+V^+V^+V^-$}&\rotp{$V^+V^+V^-V^-$}&\rotp{$\varphi\,\varphi\, V^+ V^+$}&\rotp{$\varphi\,\varphi\, V^+ V^-$}&
		  \rotp{$\varphi\,\varphi\,\varphi\,\varphi$}&\rotp{$\psi^-\psi^+ V^+V^+$}&\rotp{$\psi^-\psi^+ V^+V^-$}&\rotp{$\psi^-\psi^+ V^-V^+$}&\rotp{$\psi^+\psi^- V^-V^-$}&\rotp{$\psi^+\psi^-\varphi\,\varphi$}&\rotp{$\psi^+\psi^-\psi^+\psi^-$}&\rotp{$\psi^+\psi^-\psi^-\psi^+$}&\rotp{$\psi^+\psi^+\psi^-\psi^-$} \\
			\hline
			${\FCube}$ &L&L&R&L&R&\cR 0&L&R&R&L&\cR 0&\cR 0&\cR 0&\cR 0\\[1mm]\hline
			${\PhiSquareFSquareA}$ &R&0&R&L&R&\cR 0&0&0&0&0&0&$\slashed0$&$\slashed0$&$\slashed0$\\[1mm]\hline
			${\PhiSquareFSquareB}$ &R&0&R&L&L&\cR 0&0&0&0&0&0&$\slashed0$&$\slashed0$&$\slashed0$\\[1mm]\hline
			${\DSquarePhiFourthA}$ &$\slashed0$&$\slashed0$&$\slashed0$&\cR 0&0&\cR L$_0$&$\slashed0$&$\slashed0$&$\slashed0$&$\slashed0$&\tblue 0&$\slashed0$&$\slashed0$&$\slashed0$\\[1mm]\hline
			${\DSquarePhiFourthB}$ &$\slashed0$&$\slashed0$&$\slashed0$&R$$&0&\cR L$_0$&$\slashed0$&$\slashed0$&$\slashed0$&$\slashed0$&\cR L$_0$&$\slashed0$&$\slashed0$&$\slashed0$\\[1mm]\hline
			${\DPhiSquarePsiSquareA}$ &$\slashed0$&$\slashed0$&$\slashed0$&\tblue0&0&\tblue0&\tblue0&0&0&\tblue0&\cR L$_0$&\tblue0&\tblue0&\tblue0\\[1mm]\hline
			${\DPhiSquarePsiSquareB}$&$\slashed0$&$\slashed0$&$\slashed0$&R$$&0&\cR L$_0$&R$$&0&0&R$$&\cR L$_0$&\cR L$_0$&\cR L$_0$&\cR L$_0$\\[1mm]\hline
			${\PsiFourthA}$  &$\slashed0$&$\slashed0$&$\slashed0$&$\slashed0$&$\slashed0$&$\slashed0$&R$$&0&0&R$$&\cR L$_0$&L&L&L\\[1mm]\hline 
			${\PsiFourthB}$ &$\slashed0$&$\slashed0$&$\slashed0$&$\slashed0$&$\slashed0$&$\slashed0$&R$$&0&0&R$$&\cR L$_0$&L&L&L\\[1mm]\hline  
		\end{tabular}
\vskip .2 cm 
\begin{tabular}{rl}
  R: & rational amplitude\\
  L: & amplitude with both logarithms and rational terms\\
  $\slashed 0$: & trivial zero, no
  contributing one-loop diagrams \\
  $0$: & zero explained by angular momentum selection rules \cite{Jiang:2020sdh}  \\
  {$\tblue 0$}: &  zeros ``accidental'' to our model \\
  {\cR$0$}: &  zero from an appropriate local counterterm  \\
  {\cR L$_0$} & zero rational term from an appropriate local counterterm, logarithmic terms remain.  \\
\end{tabular}
\vskip .2 cm 
\caption{Structure of the zeros, rational terms, and logarithms in the full
  one-loop helicity amplitudes. In this table each entry indicates whether the
  operator of its row produces the amplitude with external state
corresponding to its column. $V$ denotes a vector boson, $\psi$ a fermion and $\varphi$ a scalar.}\label{tab:rationalzeros}
	\end{center}
\end{table}

After subtracting the infrared singularities and renormalization, the
amplitudes are finite. The full set of results for our renormalized
and IR-subtracted amplitudes is given in Appendix
\ref{section:amplitudesSH}. The renormalized helicity amplitudes
include a large number of zeros, including those which would otherwise
be rational contributions. A number of these zeros were pointed out in
Ref.~\cite{Craig:2019wmo}, and explained using angular-momentum
selection rules in Ref.~\cite{Jiang:2020sdh}. These selection rules
explain most, but not all, of the observed zeros, leaving some
``accidental'' zeros, displayed as a blue 0 in Table
\ref{tab:rationalzeros}.  These zeros can be considered an accident of
the simplicity of our model. In each case, the entry directly below
the blue zero shows that while the accident holds for that
particular operator, another operator with identical particle content,
but different color structure, produces a nonzero result in $\bar{\rm
  MS}$. The reason for this is that only the first of each pair of operators
is a product of global symmetry currents in our model (c.f.  our
discussion in Section~\ref{section:oneloopuv}).  In a more general
theory with an Abelian gauge field, one would expect that such zeros
would not occur.

More interesting is the surprisingly large number of amplitudes---with shaded
(red) rectangles around 0 entries in Table~\ref{tab:rationalzeros}---which do not
evaluate to zero in the standard $\bar{\rm MS}$ renormalization scheme, but
which are proportional to a linear combination of the tree-level amplitudes of
the dimension-six operators. These amplitudes can therefore be set to zero by
an appropriate choice of finite counterterms. This corresponds to a scheme
change, showing that these amplitudes are scheme dependent. Explicit examples
of how these rational shifts are related to the scheme dependence of the
two-loop anomalous dimensions is discussed at length in the next section.

Similarly, for a number of amplitudes (marked L$_0$ and in a shaded
red rectangle in Table~\ref{tab:rationalzeros}), all rational terms in the
amplitude can be removed with an appropriate choice of finite
counterterms, leaving behind logarithmic terms which cannot be
subtracted in this way. These logarithmic terms do not appear to be of the right
form to produce local results, so we may expect that they also do not
produce contributions to the two-loop anomalous dimensions via
Eq.~\eqref{eq:twoloopSimon3}.  It would be interesting to investigate
this, but we refrain from doing so here.  Remarkably, only a small
number of the one-loop amplitudes contain rational terms that cannot
be removed via finite counterterms.

As expected, however, some amplitudes do contain non-local rational
amplitudes, prohibiting such a simple subtraction by a local
counterterm.  It is interesting to note that all the nonzero rational
amplitudes of $\DSquarePhiFourthB, \DPhiSquarePsiSquareB, \PsiFourthA$
and $\PsiFourthB$ are non-local but can be individually set to zero by
the introduction of an $\FCube$ finite counterterm. This
procedure, however, will always introduce new diagrams which make
other $\slashed0$ entries in the same row nonzero. For example, since
the $\FCube$ tree contains nonzero four-vector tree amplitudes,
entries in these columns will no longer be zero. Another interesting
observation is that the UV divergence in the only nonzero amplitude of
$\DPhiSquarePsiSquareA$ cancels between terms, but the logarithms
remain. 

The vanishing one-loop amplitudes strongly suggests that many
contributions to the two-loop anomalous dimension matrix should
vanish, beyond those identified in Ref.~\cite{NonrenormLength}. For
many of the two-loop anomalous dimensions, these zeros imply that the
only contribution to the final result comes from the three-particle
cut, making their evaluation much simpler than expected, since only
four-dimensional tree-level objects are involved. In a number of
cases, including multiple examples in
Section~\ref{section:twoloop}, the three-particle cut also
vanishes, thereby immediately implying that the corresponding two-loop
anomalous dimension is zero. Of course, the amplitudes corresponding
to the entries of Table \ref{tab:rationalzeros} with shaded (red)
rectangles are not zero when working strictly in $\bar{\rm MS}$, so
one would need to evaluate the two-particle cuts in order to determine
the corresponding anomalous dimensions in this scheme.

Finally, the appearance of many zeros in Table~\ref{tab:rationalzeros}
suggests that even more zeros in the two-loop anomalous dimension
might be found by using the helicity selection rules of
Ref.~\cite{NonrenormHelicity} or the angular momentum conservation
rules of Ref.~\cite{Jiang:2020sdh}, given that the remaining
three-particle cut only involves four-dimensional tree amplitudes, which
are often restricted by these selection rules.

\section{Two-loop zeros in the anomalous dimension matrix}
\label{section:twoloop}
 
In this section we use the results of the previous
section and the tools in Section~\ref{section:formalismCuts} to obtain
two-loop anomalous dimensions in our simplified theory. These
calculations will unveil a number of mechanisms that give rise to a
wealth of new zeros in the two-loop anomalous dimension matrix. As
mentioned in the previous section, two-loop anomalous dimensions are
scheme dependent\footnote{This is in contrast to the $\beta$-function,
  which is scheme dependent starting at three
  loops~\cite{Politzer:1974fr,PeskinSchroeder}.} This makes the question of whether a
two-loop anomalous dimension is zero somewhat ill-defined. We will
show explicit examples of anomalous dimensions that are nonzero in
the $\bar{\rm MS}$ scheme, but for which we can find a scheme in which
they are zero. In addition, we demonstrate the cancellation of
logarithms in the evaluation of Eq.~\eqref{eq:twoloopSimon3} when they
appear. For simplicitly, throughout this section, we assume the case
of a single flavor of fermion, drop the flavor indices, and set $N_f=N_s=1$.  In all the
cases we consider here, the one-loop amplitudes required for the
two-loop computation are infrared finite, simplifying the discussion.

\subsection{Zeros from length selection rules}
\label{section:nonrenormlength}
First we summarize the results of our previous paper, which points out
a set of nontrivial zeros in the two-loop anomalous dimension matrix
of generic EFTs~\cite{NonrenormLength}: operators with longer
length---those with more field insertions---are often restricted
from renormalizing operators with shorter length, even if Feynman
diagrams exist. Specifically, for operators $\O_l$ and $\O_s$, with
lengths $l(\O_l)$ and $l(\O_s)$, $\O_l$ can renormalize $\O_s$ at $L$
loops only if the inequality $L>l(\O_l)-l(\O_s)$ is satisfied.  This implies, for
example, that the operator $\O_{\varphi^6}$ cannot renormalize any of the
other operators in our model (Table~\ref{tab:operators}) at two
loops. This is due to the fact that any two-loop diagram with an
insertion of $\O_{\varphi^6}$ and four external particles must contain a
scaleless integral, which evaluates to zero in dimensional
regularization. This implies that the anomalous dimensions vanish, if
there are no IR divergences.  In this case the lack of infrared
singularities follows from the fact that they are proportional to the
corresponding lower-loop amplitudes, which vanish due to the lack of diagrams when the bound is
not satisfied.

In addition, as shown in Ref.~\cite{NonrenormLength}, in a theory with
multiple types of fields, such as the SMEFT, additional vanishing can
occur at loop orders higher than indicated by the above bound.  In
general, whenever the only diagrams one can draw with an insertion of
$\O_l$ and the external particles of $\O_s$ always involve scaleless
integrals, then there will be no renormalization of $\O_s$ by
$\O_l$. In the language of Section~\ref{section:formalismCuts}, this
happens because there are no nonzero cuts on the right-hand side
of Eq.~\eqref{eq:twoloopSimon3} or the higher loop analog. Iteration pieces on the left-hand-side of Eq. \eqref{eq:twoloopSimon3}---terms other than $\gamma^{(L)}_{s\leftarrow l}F^{(0)}_s$---are also set to zero by the presence of scaleless integrals. Examples of this form of the rule in
effect include the lack of two-loop renormalization of $\O_{F^3}$ by
$\O_{D\varphi^2\psi^2}, \O_{D^2\varphi^4}$, or $\O_{\psi^4}$.

Another important consequence of the length selection rule is that, at
loop order $L=l(\O_l)-l(\O_s)+1$, only the $(L+1)$-particle cut can
contribute~\cite{NonrenormLength}.  For example, the three-particle
cut depicted in Figure~\ref{fig:twoloopexamples4}(a) is the only cut
that can contribute to $\gamma^{\rm
  UV(2)}_{\FCube\leftarrow\PhiSquareFSquareA}$. The $(L+1)$-particle cut
can then be evaluated using a four-dimensional tree-level amplitudes,
making the calculation much simpler than that of a generic $L$-loop
anomalous dimension matrix element. This observation, noted in
Ref.~\cite{NonrenormLength}, makes it straightforward to evaluate
certain two-loop SMEFT anomalous dimensions solely from three-particle
cuts~\cite{RecentOneLoopAnomalous1}.

\subsection{Zeros from vanishing one-loop rational terms}

Next, we show that the vanishing of many one-loop amplitudes and
rational terms found in Section~\ref{section:onelooprat} yields
additional zeros in the two-loop anomalous-dimension matrix of our
theory.  Somewhat surprisingly, this sometimes involves a
cancelation between different contributions to the logarithms
from one-loop terms in the cut. We will explain how this relates to
the scheme dependence of two-loop anomalous dimensions.

\subsubsection{$\O_{\psi^4}\leftarrow\O_{D^2\varphi^4}$}
\label{section:D2phi4topsi4}

We begin by determining the renormalization of $\O_{\PsiFourthA}$ and
$\O_{\PsiFourthB}$ by $\O_{\DSquarePhiFourthA}$, which we denote by
$\O_{\PsiFourthA} \leftarrow \O_{\DSquarePhiFourthA}$ and $\O_{\PsiFourthB}
\leftarrow \O_{\DSquarePhiFourthA}$.  To extract the anomalous dimensions, we
examine cuts of amplitudes with four external quarks. We can readily prove
that these anomalous dimension matrix elements are zero at two loops
in our model.  The contributing cuts would be
\begin{enumerate}
\item the three-particle cut between the five-point dimension-four
  tree amplitude and the five-point $\DSquarePhiFourthA$ amplitude,
\item the two-particle cut between
the four-point dimension-four one-loop amplitude and the four-point
$\DSquarePhiFourthA$ tree, and 
\item the two-particle cut between the
four-point dimension-four tree and the four-point $\DSquarePhiFourthA$
one-loop amplitude. 
\end{enumerate}
In all cases the external particles must be four fermions to 
match the desired operator.

In case (1), the five point amplitude containing the operator
$\DSquarePhiFourthA$ must have two external fermions, but since the
Yukawa couplings are set to zero in our simplified model, the
$\DSquarePhiFourthA$ tree must have at least four scalars, prohibiting
the required three-scalar two-fermion amplitude. For case (2), the
$\DSquarePhiFourthA$ tree must again have two fermions, so that 
there are no valid diagram and the cut vanishes.

\begin{figure}[tb]
	\centering
	\vskip -.0 cm 
	\begin{minipage}{0.49\columnwidth}
		\centering
		\includegraphics[scale=0.60]{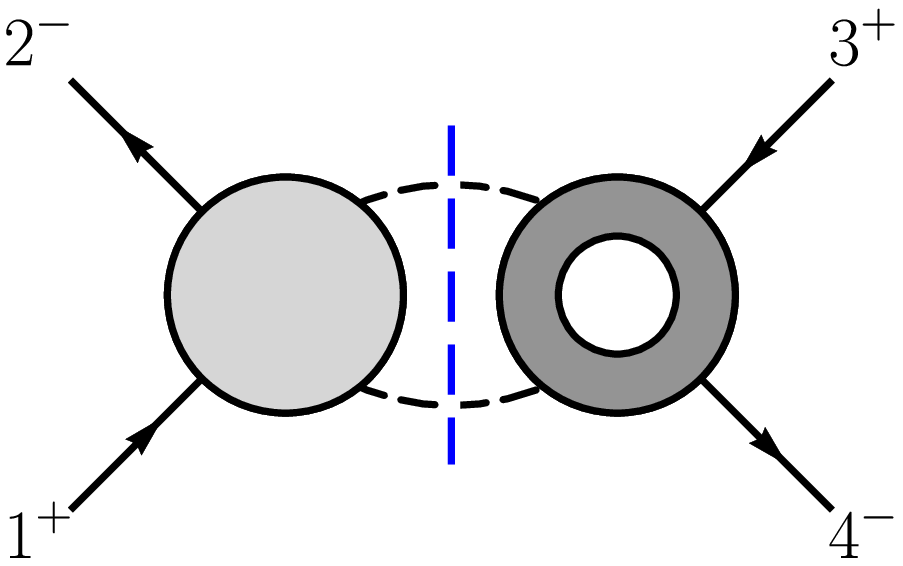}\\[-.15 cm] \textbf{(a)}
	\end{minipage}
	\begin{minipage}{0.49\columnwidth}
		\centering
		\includegraphics[scale=0.60]{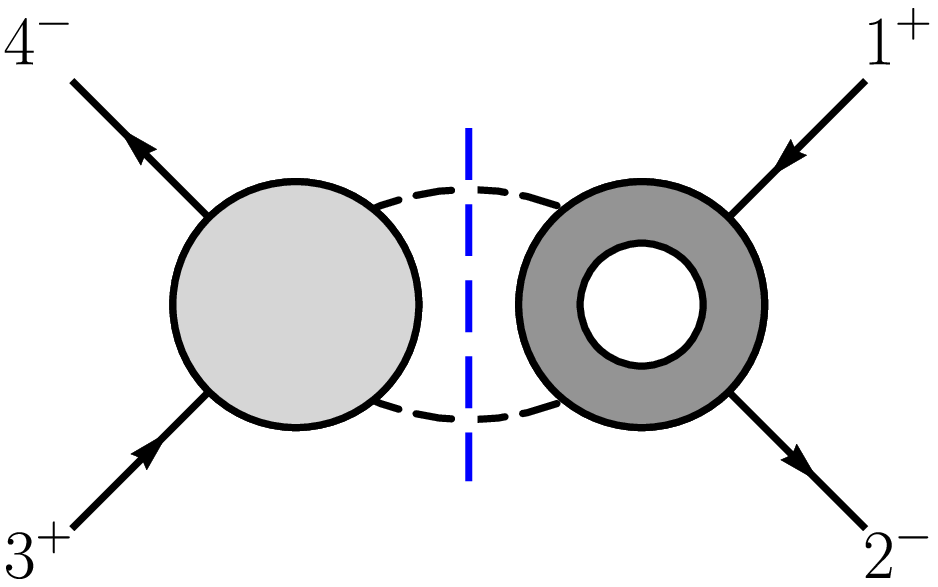}\\[-.15 cm] \textbf{(b)}
	\end{minipage}
	\caption{
		The (12)-channel (a) and (34)-channel (b) unitary cuts which determine the renormalization of $\O_{\PsiFourthA}$ by $\O_{\DSquarePhiFourthA}$ or $\O_{\DSquarePhiFourthB}$. The (23)- and (14)-channel cuts are given by the exchange of legs 2 and 4. In each, the darker blobs indicate a higher-dimension operator insertion, and the vertical (blue) dashed line indicates the integral over phase space of the particles crossing the cut.
	}
	\label{fig:examples}
	\vspace{-0.4cm}
\end{figure}

The vanishing of case (3) relies on our knowledge of the one-loop
amplitudes with an operator insertion$\DSquarePhiFourthA$, given in
Appendix~\ref{section:amplitudesSH}. In this case, the only
$\O_{\DSquarePhiFourthA}$ one-loop amplitude that can be inserted into
the cut is the two-scalar two-fermion amplitude---as in
Figure~\ref{fig:examples}---which is zero for this
operator. Therefore, all possible contributing cuts evaluate to
zero. Since $\O_{\DSquarePhiFourthA}$ does not renormalize
$\O_{\varphi^2\psi^2D}$ or $\O_{\psi^4}$ at one loop, which otherwise produce terms on the left-hand-side of Eq.~\eqref{eq:twoloopSimon3},  the vanishing of the three types of cuts implies
that the two-loop anomalous-dimension matrix element is also zero.

\label{section:Boxtopsi4}
Next, consider the case $\O_{\PsiFourthA} \leftarrow
\O_{\DSquarePhiFourthB}$, which we also show has a zero entry in the anomalous dimension
matrix of our simplified model. We organize the calculation into 
the three types of cuts as in the previous case, with the only difference being that, 
in case (3), the one-loop amplitude
with an insertion of $\O_{\DSquarePhiFourthB}$, and with two scalars and two
fermions as external particles is nonzero, and in fact has a UV
divergence. While the presence of nonzero cuts, shown
diagrammatically in Figure~\ref{fig:examples}, might seem to imply 
that the two-loop anomalous dimension must be nonzero, we will show that it
actually evaluates to zero as well.

Using the external state $\la \psim{1}{+} \psibarm{2}{-} \psim{3}{+}
\psibarm{4}{-}|$ and setting $\O_i = \O_{\DSquarePhiFourthB}$,
Eq.~\eqref{eq:twoloopSimon3} reduces to
\begin{align}\label{eq:twoloopBoxtoPsi4}
& \gamma^{\rm UV(2)}_{\psi^4 \leftarrow \DSquarePhiFourthB}           F_{\psi^4}^{(0)} 
+ \gamma^{\rm UV(1)}_{\DPhiSquarePsiSquareB \leftarrow \DSquarePhiFourthB} F_{\DPhiSquarePsiSquareB}^{(1)}
 \\
 &\hspace{2cm}=  -\frac1\pi  (\M_{2\rightarrow 2}^{12} + \M_{2\rightarrow 2}^{14} + \M_{2\rightarrow 2}^{23} + \M_{2\rightarrow 2}^{34})^{(0)} \otimes \re F_{\DSquarePhiFourthB}^{(1)}\,, \nonumber
\end{align}
where on the right-hand side we only find cuts of the form in
Figure~\ref{fig:examples} with an $\O_{\DSquarePhiFourthB}$
insertion, and the (13) and (24) channels are not allowed.  For
instance the (12)-channel cut is
\begin{align}
 &(\M_{2\rightarrow 2}^{12})^{(0)}  \otimes \re F_{\DSquarePhiFourthB}^{(1)} \\
 &\hspace{1cm}=2\int \dLIPStwo \,    
 \la \psim{1}{+} \psibarm{2}{-} | \M|\phim{\ell_{1}} \phibarm{\ell_{2}}\ra^{(0)} \,   \re
\la \phim{\ell_{1}} \phibarm{\ell_{2}} \psim{3}{+} \psibarm{4}{-}| \O_{\DSquarePhiFourthB}|0\ra^{(1)}\,. \nonumber
\end{align}
The factor of 2 is required to cancel the symmetry factor of 1/2 in
our definition of the phase-space measure.  Other terms in
Eq.~\eqref{eq:twoloopSimon3} drop out because
$\O_{\DSquarePhiFourthB}$ does not have either a one-loop or
tree-level form factor with a four-fermion external state, and does
not renormalize $\O_{\DPhiSquarePsiSquareA}$ or the $\O_{\psi^4}$
operators at one loop. In particular, the $\beta$-function also does not
appear.

For simplicity, we set the off-shell momentum $q$ to zero, and Eq.~\eqref{eq:twoloopBoxtoPsi4} then reduces to  
\begin{align} \label{eq:twoloopBoxtoPsi4amp}
  & \gamma^{\rm UV(2)}_{\psi^4\leftarrow\DSquarePhiFourthB}           A_{\psi^4}^{(0)}(\psim{1}{+} \psibarm{2}{-}  \psim{3}{+} \psibarm{4}{-}) 
  + \gamma^{\rm UV(1)}_{\DPhiSquarePsiSquareB \leftarrow \DSquarePhiFourthB} A_{\DPhiSquarePsiSquareB}^{(1)}(\psim{1}{+} \psibarm{2}{-}  \psim{3}{+} \psibarm{4}{-})
 \\
&\hspace{1cm} =- \frac{2}{\pi}\sum\int \dLIPStwo \, 
A^{(0)}(\psim{1}{+} \psibarm{2}{-}   \phim{-\!\ell_2} \phibarm{-\!\ell_1} ) \text{Re}A_{\DSquarePhiFourthB}^{(1)}(\phim{\ell_1} \phibarm{\ell_2}   \psim{3}{+} \psibarm{4}{-}) \,, \nonumber
\end{align}
where the sum is over the available channels. The relevant tree and
renormalized one-loop amplitudes needed to construct the cut are (including the
color factors):
\begin{align}
  A^{(0)}(\psim{1}{+} \psibarm{2}{-}  \phim{3} \phibarm{4}) 
  &= T^a_{i_2i_1} T^a_{i_4i_3} \, g^2 \frac{\ab{23} \spb{13} }{s} \,,
\label{eq:oneloopFFexample} \\  
  A^{(1)}_{\DSquarePhiFourthB}(\psim{1}{+} \psibarm{2}{-}  \phim{3} \phibarm{4})
  &= T^a_{i_2i_1} T^a_{i_4i_3} \, \frac{\tilde g^2 }{9}     \ab{23}  \spb{13} (3 \log (-s/\mu^2)+8) \,,
  \label{eq:oneloopFFexample2}
\end{align}
where again the flavor indices have been dropped for simplicity. Note the form
of Eq.~\eqref{eq:twoloopBoxtoPsi4amp} provides a nontrivial check on the phase
space integral on the right-hand side: $A_{\DSquarePhiFourthB}^{(1)}$ contains
terms proportional to $\log(-s/\mu^2)$, which, after the phase-space integral,
must cancel against terms in  $A_{\DPhiSquarePsiSquareB}^{(1)}$.

We can readily evaluate the cut by relabeling the amplitudes
\eqref{eq:oneloopFFexample}--\eqref{eq:oneloopFFexample2} and applying the
spinor parametrization  \eqref{eq:rotation} to the scalars crossing the cut.
This yields an integral with no poles in $z=\e^{i\phi}$, other than the pole at zero. This
can be seen by the fact that all spinor products in $A^{(0)}$ are either
proportional to $\e^{\pm i\phi}$ or else have no $\phi$ dependence under our
parametrization, whereas $A_{\DSquarePhiFourthB}^{(1)}$ only has a pole in $s$.
This makes the $\phi$ integral trivial to evaluate, resulting in:
\begin{align}  \label{eq:schannelcutexample}
  &\int_0^{\frac{\pi}{2}} d\theta \,\frac{\tilde g^4  }{18} \ab{24}  \spb{1 3} \sin ^3(2 \theta) (3 \log (-s/\mu^2)+8) T^a_{i_2i_1} T^a_{i_4i_3} 
  \nonumber \\
   &\hspace{6cm} = \frac{\tilde g^4 }{27} \ab{2 4}\spb{1 3}(3 \log (-s/\mu^2)+8)  T^a_{i_2i_1} T^a_{i_4i_3} \,,
\end{align}
for the (12)-channel cut. The (34)-channel cut gives the same result,
while the other cuts yield the same result with legs two and four
exchanged. Summing over the three other channels, we exactly match the
second term on the left-hand side of Eq.~\eqref{eq:twoloopBoxtoPsi4},
since $\gamma^{\rm
  UV(1)}_{\DPhiSquarePsiSquareB\leftarrow\DSquarePhiFourthB}=\tilde
g^2/3$ and
\begin{align}\label{eq:oneloopFFexample3}
A_{\DPhiSquarePsiSquareB}^{(1)}&=
\frac{ 2\tilde g^2  }{9} \ab{2 4}\spb{1 3} (3 \log (-s/\mu^2)+8)T^a_{i_2i_1} T^a_{i_4i_3} - (2 \leftrightarrow 4)\,.
\end{align}
Therefore the cuts exactly cancel all terms on the left-hand side of Eq.~\eqref{eq:twoloopBoxtoPsi4} involving the one-loop anomalous dimensions and
form-factors, leaving $ \gamma^{\rm
UV(2)}_{\psi^4\leftarrow\DSquarePhiFourthB}F^{(0)}_{\psi^4}=0$. Thus the
two-loop anomalous dimension $\gamma^{\rm
UV(2)}_{\psi^4\leftarrow\DSquarePhiFourthB}$ is zero.

In fact, we could have come to this conclusion simply by examining the
form of the one-loop amplitudes in Eqs.~ \eqref{eq:oneloopFFexample2}
and \eqref{eq:oneloopFFexample3}. First, note the two-loop anomalous
dimension must be $\tilde g^4$ times a number (i.e., it does not have
any kinematic dependence). Logarithmic terms resulting from the cut on
the right-hand side of \eqref{eq:twoloopBoxtoPsi4} must therefore
cancel against logarithmic terms in $A^{(1)}_{\DPhiSquarePsiSquareB}$.
Since both one-loop form factors are proportional to the factor
$(3\log(-s/\mu^2)+8)$, and since this term can be pulled out of the
phase-space integral on the right-hand side of
Eq.~\eqref{eq:twoloopBoxtoPsi4}, the cancellation of the logarithmic
terms implies cancellation of the rational term as well.  Thus, even
though there are nonzero cuts, there can be no remaining rational term
that leads to a nonzero two-loop anomalous dimension.

At this point, the vanishing of the two-loop anomalous dimensions due
to the cancellation of one-loop rational terms might seem
accidental. However, one must remember that such \emph{local} rational
pieces are scheme dependent and can be adjusted by adding
finite local counter-terms. As described in
Section~\ref{section:onelooprat}, the rational terms of both one-loop
amplitudes in \eqref{eq:oneloopFFexample2} and
\eqref{eq:oneloopFFexample3} can be set to zero by such finite
counterterms, which would also result in $\gamma^{\rm UV(2)}_{\psi^4
  \leftarrow \DSquarePhiFourthB}=0$.  For this particular example, it
just so happened that the naive $\bar{\rm MS}$ scheme has zero
anomalous dimension, but next we will see that this is not always the
case.

As a cross-check, we have verified that the Eq.~\eqref{eq:oneloopsquared}
is also satisfied.  The crucial substitution 
$\log(-s/\mu^2)\rightarrow\log(-s/\mu^2)-i\pi$, is required in the right-hand side of
that equation, coming from the analytic continuation of the amplitude from
the Euclidean region to the correct physical region, which must be carried out
for use in Eqs.~\eqref{eq:onezeroloopphasespaceA}--\eqref{eq:twoloopphasespace}.

\subsubsection{$\O_{D^2\varphi^4} \leftarrow \O_{\PsiFourthA}$}

\begin{figure}[tb]
	\centering
	\vskip -.0 cm 
	\begin{minipage}{0.49\columnwidth}
		\centering
		\includegraphics[scale=0.60]{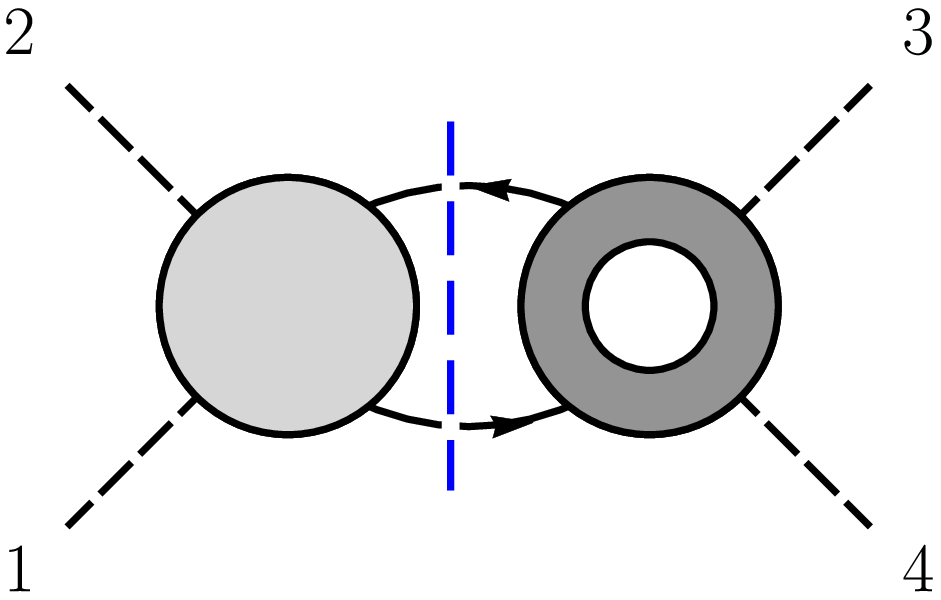}\\[-.15 cm] \textbf{(a)}
	\end{minipage}
	\begin{minipage}{0.49\columnwidth}
		\centering
		\includegraphics[scale=0.60]{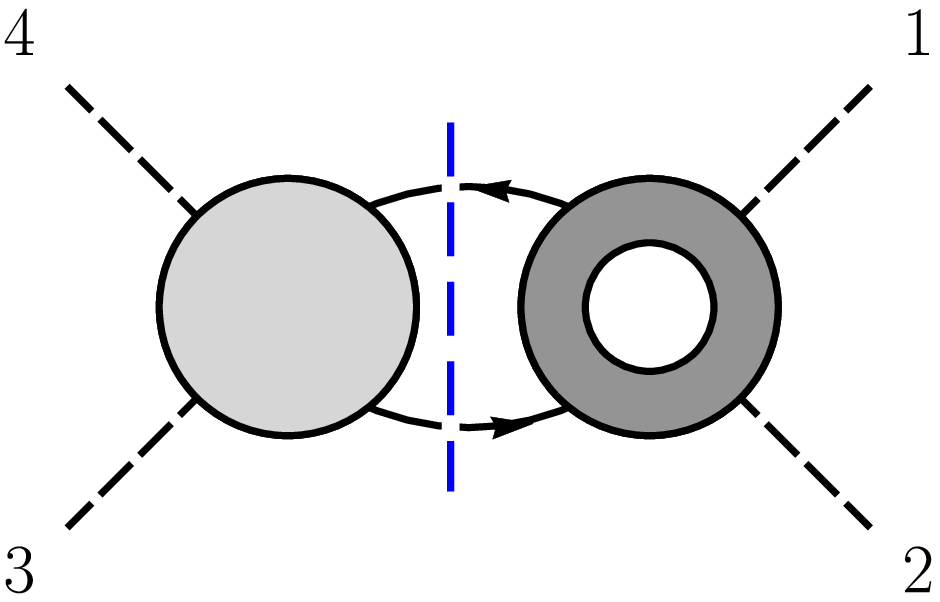}\\[-.15 cm] \textbf{(b)}
	\end{minipage}
	\caption{The (a) (12)-channel and (b) (34)-channel unitary
          cuts which determine the renormalization of
          $\O_{\DSquarePhiFourthA}$ and $\O_{\DSquarePhiFourthB}$ by
          $\O_{\PsiFourthA}$ or $\O_{\PsiFourthB}$. The (23)- and
          (14)-channel cuts are given by exchanging legs 2 and
          4. In each, the darker blobs indicate a higher-dimension
          operator insertion, and the vertical (blue) dashed line indicates the
          integral over phase space of the particles crossing the cut.
        }
	\label{fig:twoloopexamples2}
	\vspace{-0.4cm}
\end{figure}

This section will provide our first example of nonzero two-loop anomalous
dimension matrix elements in $\bar{\rm MS}$, while demonstrating how an
appropriate choice of scheme, i.e. choice of finite local counterterms, can
eliminate the two-loop anomalous dimensions of this example.

We will begin with the calculation in $\bar{\rm MS}$. Again, there is
no three-particle cut, due to the particle content of the two types of
operators in question.  Using the external state $\la
\phim{1}\phibarm{2}\phim{3}\phibarm{4}|$ and setting
$\O_i\rightarrow\O_{\PsiFourthA}$, Eq.~\eqref{eq:twoloopSimon3}
becomes
\begin{align}
&  \gamma^{\rm UV(2)}_{\DSquarePhiFourthA    \leftarrow \PsiFourthA}  F^{(0)}_{\DSquarePhiFourthA}
 + \gamma^{\rm UV(2)}_{\DSquarePhiFourthB    \leftarrow \PsiFourthA}  F^{(0)}_{\DSquarePhiFourthB}
 + \gamma^{\rm UV(1)}_{\DPhiSquarePsiSquareB \leftarrow \PsiFourthA}F^{(1)}_{\DPhiSquarePsiSquareB} \label{eq:SimonPsi4toD2phi4}\\
 &\hspace{4cm}=  -\frac1\pi  (\M_{2\rightarrow 2}^{12} + \M_{2\rightarrow 2}^{14} + \M_{2\rightarrow 2}^{23} + \M_{2\rightarrow 2}^{34})^{(0)} \otimes \re F_{\PsiFourthA}^{(1)}\,.
\nonumber
\end{align}
As for the previous example, the logarithmic terms in
the cuts must cancel against terms in the amplitude
$F^{(1)}_{\DPhiSquarePsiSquareB}$ on the left-hand side of the equation. Since we are
dealing with four-point matrix elements we will again set $q=0$. 
Then the one-loop amplitudes required for this example are
\begin{align}
A^{(1)}_{\PsiFourthA}(\psim{1}{+}\psibarm{2}{-}\phim{3}\phibarm{4})&=  \frac{2 \tilde g^2  }{9} \ab{2 3} \spb{1 3} (3 \log (-s/\mu^2)-2)T^a_{i_2i_1} T^a_{i_4i_3}\,, \label{eq:oneloopFFexample4}\\
A^{(1)}_{\DPhiSquarePsiSquareB}(\phim{1}\phibarm{2}\phim{3}\phibarm{4})&= \frac{2\tilde g^2}{9}   (t-u)(3  \log (-s/\mu^2)-5) T^a_{i_2i_1} T^a_{i_4i_3} 
 + (2\leftrightarrow4) \,, 
\label{eq:oneloopFFexample5} 
\end{align}
and the tree-level amplitudes needed are  in Eq.~\eqref{eq:oneloopFFexample} along with
\begin{align}
A_{\DSquarePhiFourthA}^{(0)}(\phim{1}\phibarm{2}\phim{3}\phibarm{4})&=\;t\delta_{i_2i_1}\delta_{i_4i_3}+s\delta_{i_4i_1}\delta_{i_2i_3} \,, \\
A_{\DSquarePhiFourthB}^{(0)}(\phim{1}\phibarm{2}\phim{3}\phibarm{4})&=\;2s\delta_{i_2i_1}\delta_{i_4i_3}+2t\delta_{i_4i_1}\delta_{i_2i_3} \,,
\end{align}
which are shown in a slightly different basis of color factors than those shown
in the appendix.  The phase-space integral is evaluated in the same manner as
the previous examples, with the result of the (12)-channel cut being
\begin{align}
  -\frac 1 \pi&\int\dLIPStwo\, \sum_{h_{1},h_{2}}A^{(0)}( \phim{1}\phibarm{2}\psim{-\ell_1}{h_1}\, \psibarm{-\ell_2}{h_2}) A^{(1)}_{\PsiFourthA}( \psim{\ell_2}{h_2}\, \psibarm{\ell_1}{h_1}\phim{3}\phibarm{4}) \nonumber \\
  &\hspace{3cm}  =  -\frac{2}{27} \tilde g^4  (t-u) (3 \log (-s/\mu^2)-2)T^a_{i_2i_1} T^a_{i_4i_3}    \,.
\end{align}
After summing over all channels and subtracting the contribution of $\gamma^{\rm UV(1)}_{\DPhiSquarePsiSquareB \leftarrow \PsiFourthA}F^{(1)}_{\DPhiSquarePsiSquareB}$ in Eq.~\eqref{eq:SimonPsi4toD2phi4}---thus canceling the logarithmic terms---the two-loop anomalous dimensions are given by 
\begin{align}
 \gamma^{\rm UV(2)}_{\DSquarePhiFourthA \leftarrow \PsiFourthA}(&t\delta_{i_2i_1}\delta_{i_4i_3}+s\delta_{41}\delta_{23})
+\gamma^{\rm UV(2)}_{\DSquarePhiFourthB \leftarrow \PsiFourthA}(2s\delta_{i_2i_1}\delta_{i_4i_3}+2t\delta_{i_4i_1}\delta_{i_2i_3})\\
=& -\frac{4}{9} \tilde g^4  (t-u)   T^a_{i_2i_1} T^a_{i_4i_3} +  (2\leftrightarrow 4)  \nonumber \,.
\end{align}  
Applying the color Fierz identity,
\begin{equation}
T^a_{ij}T^a_{kl}=\delta_{il}\delta_{kj}-\frac1N \delta_{ij}\delta_{kl} \,,
\label{eq:suNfierz}
\end{equation}
and solving for the two-loop anomalous dimensions, we find
\begin{align}
\gamma^{\rm UV(2)}_{\DSquarePhiFourthA \leftarrow \PsiFourthA}=&\null  -\frac{4 \tilde g^4  (N-2) }{9 N}  \,, \nonumber\\
\gamma^{\rm UV(2)}_{\DSquarePhiFourthA \leftarrow \PsiFourthB}=&\null \frac{2 \tilde g^4 (2 N-1) }{9 N} \,,  
\end{align}
in the $\bar{\rm MS}$ scheme. Although these anomalous dimension
matrix elements are nonzero in the $\bar{\rm MS}$ scheme, a simple
rational shift of the coefficients $\opcoef_{\DSquarePhiFourthA},
\opcoef_{\DSquarePhiFourthB}$, and $\opcoef_{\DPhiSquarePsiSquareB}$
can set them to zero. This is accomplished by the following shifts in
the coefficients:
\begin{align}
  \begin{split}
\opcoef_{\DSquarePhiFourthA}\longrightarrow&\,\tilde\opcoef_{\DSquarePhiFourthA}=\opcoef_{\DSquarePhiFourthA}+  \frac{10 \tilde g^2 (N-2) }{9 N}\opcoef_{\DPhiSquarePsiSquareB} \,,\\
\opcoef_{\DSquarePhiFourthB}\longrightarrow&\,\tilde\opcoef_{\DSquarePhiFourthB}=\opcoef_{\DSquarePhiFourthB}+ \frac{5 \tilde g^2 (2 N-1) }{9 N}\opcoef_{\DPhiSquarePsiSquareB} \,,\\
\opcoef_{\DPhiSquarePsiSquareB}\longrightarrow&\,\tilde\opcoef_{\DPhiSquarePsiSquareB}=\opcoef_{\DPhiSquarePsiSquareB}  -\frac{2 \tilde g^2 }{9}\opcoef_{\PsiFourthA}  \,,
\label{eq:coeffredef}
\end{split}
\end{align}
which yields
\begin{align}
  \tilde\gamma^{\rm UV(2)}_{\DSquarePhiFourthA \leftarrow \PsiFourthA}= 0 \,, \hskip 2 cm 
  \tilde\gamma^{\rm UV(2)}_{\DSquarePhiFourthB \leftarrow \PsiFourthA} = 0\,,
\end{align}
where the tilde indicates the modified scheme.
The shifts above are equivalent to a finite renormalization of the operator at
one loop.  Generally this can be achieved by choosing the rational terms in $
\gamma^{\rm
UV(1)}_{\DPhiSquarePsiSquareB \leftarrow \PsiFourthA}F^{(1)}_{\DPhiSquarePsiSquareB}$
to match those of the cuts. In our particular example  we set the rational
terms of both  \eqref{eq:oneloopFFexample4} and  \eqref{eq:oneloopFFexample5}
to zero. We briefly comment below on the consequences of this redefinition for the two-loop RG running of the operators involved.

While we do not present the analogous calculation for $\O_{\PsiFourthB}$ here, by inspecting Table~\ref{tab:rationalzeros}, we can deduce that the two-loop anomalous dimensions $\gamma^{\rm UV(2)}_{\DSquarePhiFourthA \leftarrow \PsiFourthB}$ and $\gamma^{\rm UV(2)}_{\DSquarePhiFourthB \leftarrow \PsiFourthB}$ can also be set to zero with the appropriate choice of finite counterterms.
 
\subsubsection{General comments about scheme redefinition}
As mentioned above, the scheme choice that sets some two-loop
anomalous dimensions to zero is equivalent to a finite renormalization
of the operators
\begin{equation}
  \tilde\O_i = Z^{\rm fin}_{ij} O_j\,, \quad \text{where} \quad Z^{\rm fin}_{ij} = \delta_{ij} + f_{ij}(g^{(4)})\,,
\end{equation}
and the quantity $f_{ij}$ is finite and has a perturbative expansion
starting at one loop, $f_{ij}(g^{(4)}) = f_{ij}^{(1)} + \cdots$.  As
usual, the redefinition of the coefficients, $\tilde c_i =Z^{{\rm fin}
  \, (c)}_{ij} c_j$ is given by the inverse, $Z^{{\rm fin}\, (c)}_{ij} =
(Z^{\rm fin}_{ij})^{-1}$.  The effect of such a scheme redefinition
can be easily analyzed using the unitarity-based formalism employed in
this paper.  Since the coupling dependence of $f_{ij}$ starts at one
loop we have that
\begin{align}
  \tilde F^{(0)}_{i} &= F^{(0)}_{i}\,, \label{eq:FFredef0} \\
  \tilde F^{(1)}_{i} &= F^{(1)}_{i} + f^{(1)}_{ij} F^{(0)}_{j}\,, \label{eq:FFredef1}
\end{align}
where the tilde indicates a form factor of the redefined operator $\tilde\O_i$.
From Eqs.~\eqref{eq:FFredef0} and~\eqref{eq:ffoneloop} we conclude the one-loop anomalous dimensions are unaffected by the finite renormalization, i.e., $\tilde\Delta\gamma^{(1)}_{ij}=\Delta\gamma^{(1)}_{ij}$. Similarly, writing Eq.~\eqref{eq:twoloopSimon3} for the redefined operator 
\begin{align}
  \left[\Delta\tilde\gamma^{(1)}_{ij}+\delta_{ij}\beta^{(1)}\partial\right]\text{Re}\tilde F_j^{(1)}&+\left[\Delta\tilde\gamma^{(2)}_{ij}+\delta_{ij}\beta^{(2)}\partial\right]\tilde F_j^{(0)} =-\frac{1}{\pi} \left[\text{Re}(\M) \text{Re}(\tilde F_i)\right]^{(2)}.  
\end{align}
and using Eqs.~\eqref{eq:FFredef0} and~\eqref{eq:FFredef1} together with Eqs.~\eqref{eq:ffoneloop} and ~\eqref{eq:twoloopSimon3}, while keeping in mind that the infrared anomalous dimensions are not changed by redefining the scheme, we find the relation between the two-loop anomalous dimensions in the two schemes,
\begin{equation}
  \tilde\gamma^{\rm UV (2)}_{ij} = \gamma^{\rm UV (2)}_{ij} +  f^{(1)}_{ik}\gamma^{\rm UV (1)}_{kj} - \gamma^{\rm UV (1)}_{ik}f^{(1)}_{kj} -  \beta^{(1)} \partial f^{(1)}_{ij}\,.
  \label{eq:twoloopanomscheme}
\end{equation}
In general, one would like to solve this equation for $f^{(1)}_{ik}$ to get as many vanishing entries as possible in $\tilde\gamma^{\rm UV (2)}_{ij}$. 

We have explicitly verified Eq.~\eqref{eq:twoloopanomscheme} in the
examples above, where we set the anomalous dimensions of the form
$\tilde\gamma^{\rm UV(2)}_{D^2 \varphi^4 \leftarrow \psi^4}$ to zero
by appropriately choosing $f^{(1)}_{D \varphi^2 \psi^2 \leftarrow
  \psi^4}$ and $f^{(1)}_{D^2 \varphi^4 \leftarrow D \varphi^2
  \psi^4}$.  In addition, $f^{(1)}_{D^2 \varphi^4 \leftarrow \psi^4}$
vanished, which from Eq.~\eqref{eq:twoloopanomscheme} implies the the
absence of a term induced by the $\beta$-function in the new two-loop
anomalous dimension. On the other hand, it is clear from
Eq.~\eqref{eq:twoloopanomscheme} that the finite renormalizations will
induce some additional running in the two-loop anomalous dimensions
$\tilde\gamma^{\rm UV(2)}_{D \varphi^2 \psi^2 \leftarrow \psi^4}$ and
$\tilde\gamma^{\rm UV(2)}_{D^2 \varphi^4 \leftarrow D \varphi^2
  \psi^4}$, proportional to the one-loop beta function and $\partial
f^{(1)}$.  However, this additional running is harmless, since those
operators already mix at one loop. Furthermore, the corresponding
entries in the two-loop anomalous-dimension matrix receive
contributions from both two- and three-particle cuts that have no a
priori reason to vanish, so we expect them in any case to run. In
summary, our scheme choice prevents certain operators from mixing at
two-loops at the expense of modifying the running of operators that,
in any case, mix at one loop in the original scheme.

\subsection{Zeros from color selection rules}

\begin{figure}[tb]
	\centering
	\vskip -.0 cm 
	\begin{minipage}{0.49\columnwidth}
		\centering
		\includegraphics[scale=0.60]{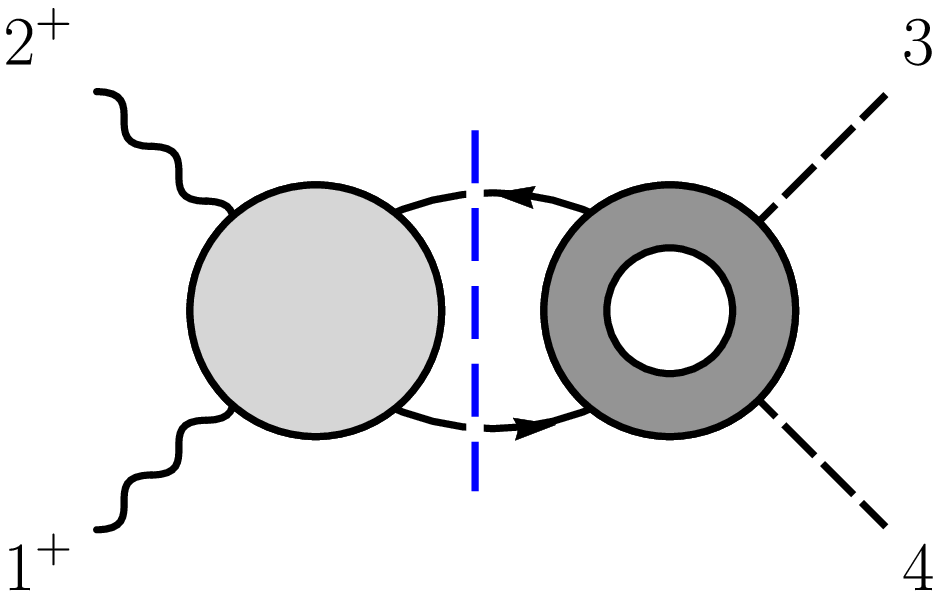}\\[-.15 cm] \textbf{(a)}
	\end{minipage}
	\begin{minipage}{0.49\columnwidth}
		\centering
		\includegraphics[scale=0.60]{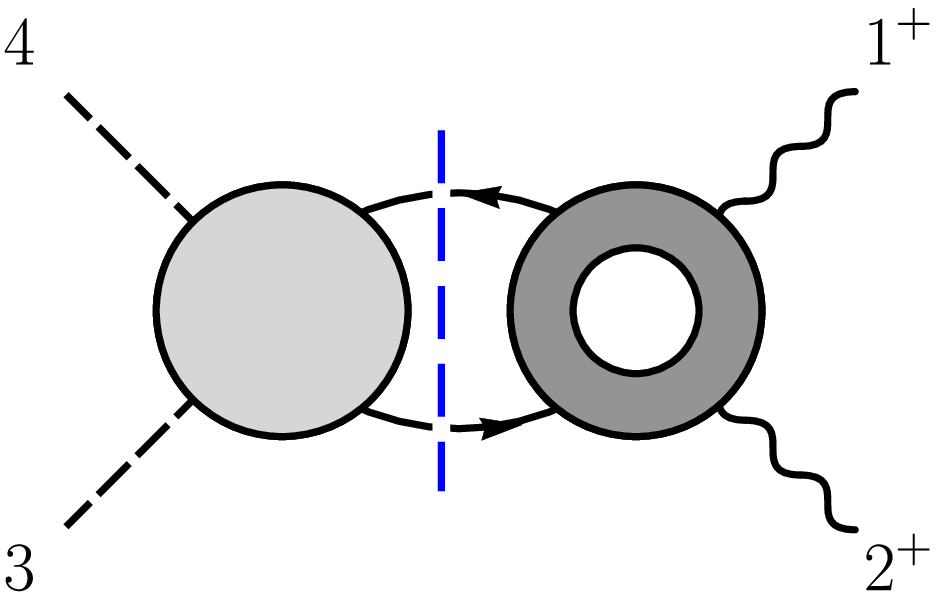}\\[-.15 cm] \textbf{(b)}
	\end{minipage}
	\caption{
		(12)-channel (a) and (34)-channel (b) unitary cuts which determine the renormalization of  $\O_{\PhiSquareFSquareA}$ and $\O_{\PhiSquareFSquareB}$ by $\O_{\PsiFourthA}$ or $\O_{\PsiFourthB}$. There are no t-channel cuts for this process. In each diagram, the darker blobs indicate a higher-dimension operator insertion, and the dashed line indicates the integral over phase space of the particles crossing the cut.
	}
	\label{fig:twoloopexamples3}
	\vspace{-0.4cm}
\end{figure}

This section will provide an example of another type of selection rule, wherein a mismatch between the color of the cuts and the color of the target operators prevents renormalization at two loops. 

\subsubsection{$\O_{\varphi^2F^2}\leftarrow\O_{\psi^4}$}
For this example we choose the external state to be $\la \phim{1}\phibarm{2} 3^+4^+|$, under which both $\O_{\PhiSquareFSquareA}$ and $\O_{\PhiSquareFSquareB}$ are nonzero. Using this state and setting $\O_i\rightarrow\O_{\PsiFourthA}$, Eq. \eqref{eq:twoloopSimon3} reduces to
\begin{align}\label{eq:SimonPsi4tophi2F2}
   \gamma^{\rm UV(2)}_{\PhiSquareFSquareA   \leftarrow \PsiFourthA}&F^{(0)}_{\PhiSquareFSquareA}
 + \gamma^{\rm UV(2)}_{\PhiSquareFSquareB   \leftarrow \PsiFourthA}F^{(0)}_{\PhiSquareFSquareB} 
 + \gamma^{\rm UV(1)}_{\DPhiSquarePsiSquareB\leftarrow \PsiFourthA}F^{(1)}_{\DPhiSquarePsiSquareB}\\
 &\hspace{1cm}=  -\frac1\pi  \Big((\M_{2\rightarrow 2}^{12})^{(0)} \otimes \re F_{\PsiFourthA}^{(1)}    + (\M_{2\rightarrow 2}^{34})^{(0)}\otimes \re F_{\PsiFourthA}^{(1)}    \Big) \nonumber\,.
\end{align}
Naively there would be the additional term $\gamma^{\rm UV(2)}_{\FCube \leftarrow \PsiFourthA}F^{(0)}_{\FCube}$ on the left-hand-side of the equation, since $\O_{F^3}$ produces a nonzero tree amplitude with the state $\la \phim{1}\phibarm{2} 3^+4^+|$. However, as was discussed in Section~\ref{section:nonrenormlength}, the length and particle content of $\O_{\PsiFourthA}$ requires $\gamma^{\rm UV(2)}_{\FCube \leftarrow \PsiFourthA}=0$. 
Setting $q=0$, the (12)-channel cut of the above equation is 
\begin{align}
  &(\M_{2\rightarrow 2}^{12})^{(0)} \otimes \re F_{\PsiFourthA}^{(1)} = \int\dLIPStwo\,\sum_{h_{1},h_{2}} A^{(0)}(\phim{1}\phibarm{2} \psim{-\ell_1}{h_1}\, \psibarm{-\ell_2}{-h_2})A^{(1)}_{\PsiFourthA}(\psim{\ell_2}{h_2}\, \psibarm{\ell_1}{h_1}3^+4^+ )\,,
\end{align}
and the (34)-channel cut is
\begin{align}
&(\M_{2\rightarrow 2}^{34})^{(0)}\otimes \re F_{\PsiFourthA}^{(1)} = \int\dLIPStwo\,\sum_{h_{1},h_{2}}A^{(0)}(3^+4^+\psim{-\ell_1}{h_1}\, \psibarm{-\ell_2}{h_2}) A^{(1)}_{\PsiFourthA}(\psim{\ell_2}{h_2}\, \psibarm{\ell_1}{h_1}\phim{1}\phibarm{2})  \,.
\end{align}
The (34)-channel cut vanishes, because the amplitude $A^{(0)}(3^+4^+\psim{-\ell_1}{}\, \psibarm{-\ell_2}{})$ is zero for all helicities of the fermions crossing the cut. This vanishing is required for the consistency of the logarithmic terms: $A^{(1)}_{\PsiFourthA}(\psim{\ell_2}{h_2}\, \psibarm{\ell_1}{h_1}\phim{1}\phibarm{2})$ includes a  term proportional to $\log(-s/\mu^2)$, but there is no term on the left-hand side that can cancel it, since $F^{(1)}_{\DPhiSquarePsiSquareB}(\phim{1}\phibarm{2} 3^+4^+)$ is purely rational. The one-loop amplitudes needed for this calculation are
\begin{align}
A_{\PsiFourthA}^{(1)}(\psim{1}{+}\, \psibarm{2}{-}3^+4^+)&=-\frac{ \tilde g^2  s [14]  \la 2 4\ra  \,
  [T^{a_3},T^{a_4}]_{i_2 i_1} } {3 \la 3 4\ra ^2} \label{eq:oneloopFFexample6}\,, \\
A_{\PsiFourthA}^{(1)}(\psim{1}{-}\, \psibarm{2}{+}3^+4^+)&=-\frac{ \tilde g^2  \la 1 2\ra  
    [2 3] [2 4]\,[T^{a_3},T^{a_4}]_{i_2 i_1}}{3 \la 3 4\ra } \,,\\
A_{\DPhiSquarePsiSquareB}^{(1)}( \phim{1}\phibarm{2} 3^+4^+)&=\frac{ \tilde g^2  s (t-u) \,[T^{a_3},T^{a_4}]_{i_2 i_1}} {3 \la 3 4\ra ^2}\label{eq:oneloopFFexample7}\,,
\end{align}
while the tree-level amplitudes needed for the cut calculation are  \eqref{eq:oneloopFFexample} and its conjugate. The phase-space integrals are carried out in the same manner as the previous example, with the simplification that the functions are now entirely rational. The result of the phase-space integral is 
\begin{align}
  -\frac 1 \pi\int&\dLIPStwo\, \sum_{h_{1},h_{2}} A^{(0)}( \phim{1}\phibarm{2} \psim{-\ell_1}{-h_1}\, \psibarm{-\ell_2}{-h_2}) A^{(1)}_{\PsiFourthA}( \psim{\ell_2}{h_2}\, \psibarm{\ell_1}{h_1}3^+4^+)\nonumber\\
&=-\frac{2\tilde g^4  s (t-u) \,[T^{a_3},T^{a_4}]_{i_2i_1}}{9  (\la 3 4\ra )^2}
= \gamma^{\rm UV(1)}_{\DPhiSquarePsiSquareB \leftarrow \PsiFourthA}A^{(1)}_{\DPhiSquarePsiSquareB} \,.
\end{align}
Thus the phase-space integral exactly cancels against this term from the left-hand-side of Eq.~\eqref{eq:SimonPsi4tophi2F2}, meaning the two-loop anomalous dimension is again zero. 

Interestingly, this can also be seen without looking at the kinematic
content of the cuts on the right-hand side of
Eq.~\eqref{eq:SimonPsi4tophi2F2}. Since the color of both
$\O_{\PhiSquareFSquareA}$ and $\O_{\PhiSquareFSquareB}$ are symmetric
in $T^3$ and $T^4$, no combination of the two can produce the color
factor $[T^3,T^4]_{i_2i_1}$. Since this is the color of
$A_{\DPhiSquarePsiSquareB}^{(1)}( \phim{1}\phibarm{2} 3^+4^+)$, and
the color of $A_{\PsiFourthA}^{(1)}(\psim{1}{\pm}\,
\psibarm{2}{\mp}3^+4^+)$ is also anti-symmetric under the exchange of
3 and 4, we can see directly from the color that neither of these
terms can contribute to the two-loop anomalous dimension, and
therefore must cancel. As in the previous example,
we can extend this argument trivially to the operator
$\O_{\PsiFourthB}$, since its two-fermion two-vector-boson amplitude
is proportional to that of $\O_{\PsiFourthA}$. In this case, the only
difference on the left-hand side would being the value of $\gamma^{\rm
  UV(1)}_{\DPhiSquarePsiSquareB\leftarrow\PsiFourthB}$ versus
$\gamma^{\rm UV(1)}_{\DPhiSquarePsiSquareB\leftarrow\PsiFourthA}$, but
the color again ensures all terms must cancel, leaving
\begin{gather}
\gamma^{\rm UV(2)}_{\PhiSquareFSquareA\leftarrow\PsiFourthA} = 
\gamma^{\rm UV(2)}_{\PhiSquareFSquareB\leftarrow\PsiFourthA}= 0 \,,\nonumber\\
\gamma^{\rm UV(2)}_{\PhiSquareFSquareA\leftarrow\PsiFourthB} =
\gamma^{\rm UV(2)}_{\PhiSquareFSquareB\leftarrow\PsiFourthB}=0 \,.
\end{gather}

Here we focused on a simple example in which the color can preclude
renormalization. In more general cases, one can directly inspect the
color of the amplitudes that compose the cuts contributing to a given
anomalous dimension and determine whether a given operator can yield a
nonzero contribution. Note that this is more efficient than
studying the color of individual Feynman diagrams, since the color
decomposed amplitudes have fewer color structures.

It is worth noting that, as mentioned in
Section~\ref{section:onelooprat}, the nonzero rational amplitudes
\eqref{eq:oneloopFFexample6}--\eqref{eq:oneloopFFexample7} can be
set to zero by introducing finite counterterms proportional to
$\opcoef_{\PsiFourthA}\O_{F^3}$ and
$\opcoef_{\DPhiSquarePsiSquareB}\O_{F^3}$, respectively. However,
since these are non-local amplitudes, doing so introduces nonzero
terms for other amplitudes, in particular any amplitudes where
$\O_{F^3}$ produces a nonzero tree-level amplitude. This would
introduce a great deal of confusion---for example, if we were to
introduce a counterterm to cancel \eqref{eq:oneloopFFexample6}, we
would then need to include additional cuts on the right-hand side of
Eq.~\eqref{eq:SimonPsi4tophi2F2}, including three-particle cuts and
cuts with nontrivial IR dependence. Canceling either
Eq.~\eqref{eq:oneloopFFexample6} or Eq.~\eqref{eq:oneloopFFexample7} with such
a counterterm would also spoil the argument of
Section~\ref{section:nonrenormlength}, as the $\O_{F^3}$
self-renormalization would contribute in a nontrivial way. Therefore
we would have to include the term $\gamma^{\rm
  UV(2)}_{\FCube\leftarrow\PsiFourthA}F^{(0)}_{\FCube}$ on the
left-hand side of Eq.~\eqref{eq:SimonPsi4tophi2F2} as well. For all of the
above reasons, we choose not to implement these finite shifts. It is
interesting however, that even though the rational terms remain in
this example, the structure of the color precludes renormalization at
two loops.

\begin{figure}[tb]
	\centering
	\vskip -.0 cm 
	\begin{minipage}{0.4\columnwidth}
		\centering
		\includegraphics[scale=0.55]{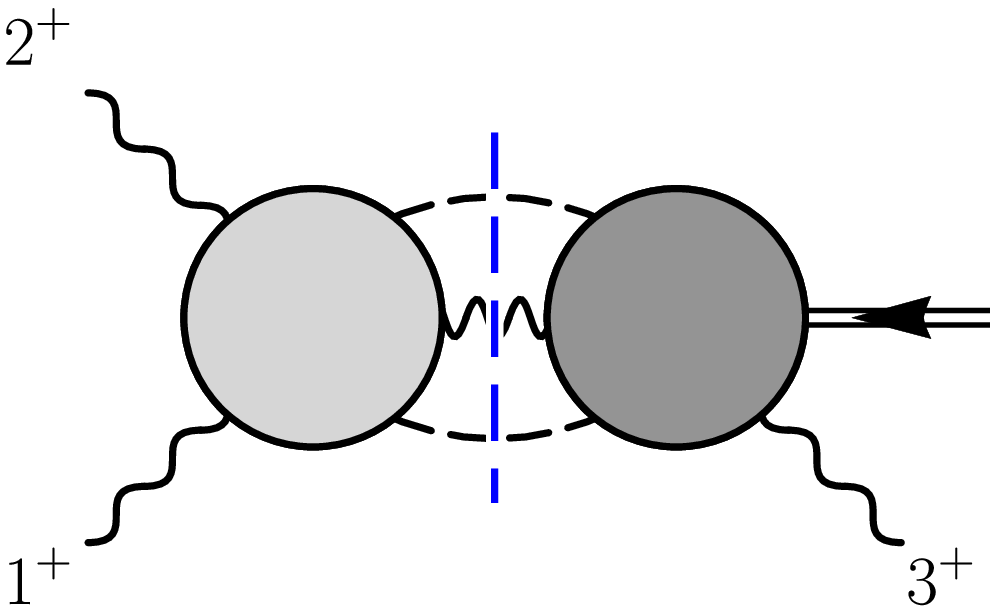}\\[-.15 cm] \textbf{(a)}
	\end{minipage}
	\begin{minipage}{0.4\columnwidth}
		\centering
		\includegraphics[scale=0.55]{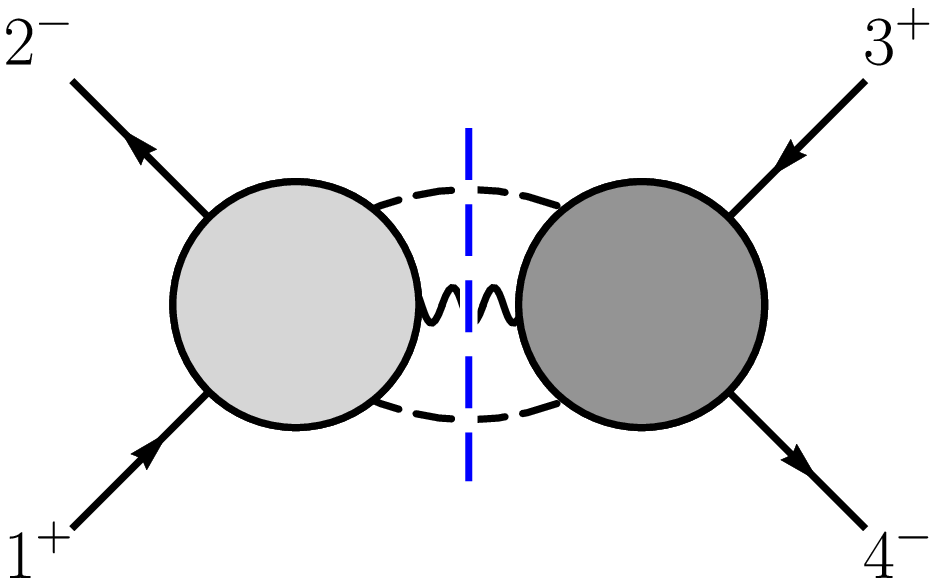}\\[-.15 cm] \textbf{(b)}
	\end{minipage}
	\caption{
		(a) Unitary cut which determines the renormalization of $\O_{F^3}$ by $\O_{\PhiSquareFSquareA}$ or $\O_{\PhiSquareFSquareB}$. Note this form factor requires $q\ne0$, and the double-lined arrow indicates this insertion of additional off-shell momentum from the operator. (b) Unitarity cut which determines the renormalization of $\O_{\PsiFourthA}$ and $\O_{\PsiFourthB}$ by $\O_{\PhiSquareFSquareA}$ or $\O_{\PhiSquareFSquareB}$. In each, the darker blobs indicate a higher-dimension operator insertion, and the dashed line indicates the integral over phase space of the particles crossing the cut.
	}
	\label{fig:twoloopexamples4}
	\vspace{-0.4cm}
\end{figure}

\subsection{Outlook on additional zeros}

\begin{table}[tb]
	\begin{center}
		\renewcommand{\arraystretch}{1.2}
		\setlength{\tabcolsep}{3pt}
		\begin{tabular}{c|cccccccccc|}	&{\small${\FCube}$}&{\small${\PhiSquareFSquareA}$}&{\small${\PhiSquareFSquareB}$}&{\small${\DSquarePhiFourthA}$}&{\small${\DSquarePhiFourthB}$}&{\small${\DPhiSquarePsiSquareA}$}&{\small${\DPhiSquarePsiSquareB}$}&{\small${\PsiFourthA}$}&{\small${\PsiFourthB}$}&{\small${\PhiSixth}$} \\
			\hline
			\small${\FCube}$ & & $\cB$ &$\cB$   &  0 &  0 &  0 &  0 &  0 &  0 & $\slashed0$\\[1mm]\hline
			\small${\PhiSquareFSquareA}$ & & & &  &  &  &  &  0 &  0 &  0 \\[1mm]\hline
			\small${\PhiSquareFSquareB}$ & & & &  &  &  &  &  0 &  0 &  0 \\[1mm]\hline
			\small${\DSquarePhiFourthA}$  &  &  &  & & &  & &  0$^*$ &  0$^*$ &  0 \\[1mm]\hline
			\small${\DSquarePhiFourthB}$ &  &   &  & & &  & &  0$^*$ &  0$^*$ &  0 \\[1mm]\hline
			\small${\DPhiSquarePsiSquareA}$  &  &  &  &  &  &  &  &  &  &$\slashed0$\\[1mm]\hline
			\small${\DPhiSquarePsiSquareB}$ &  &  &  &  & &  & & &  &$\slashed0$\\[1mm]\hline
			\small${\PsiFourthA}$ &  &$\cB$   & $\cB$ &  0 &  0 &  & & &  &$\slashed0$\\[1mm]\hline
			\small${\PsiFourthB}$ &  & $\cB$  & $\cB$ &  0 &  0 &  & & &  &$\slashed0$\\[1mm]\hline
			\small${\PhiSixth}$ &  &   &  &  &  &  & & &  &\\[1mm]\hline
		\end{tabular}
    \vskip .2 cm 
	\begin{tabular}{cl}
		$\slashed 0$ &: trivial zero, no contributing two-loop diagrams \\
		 0 &: zero predicted by the selection rules of Section \ref{section:twoloop}  \\
		{ \cB} &: only a three-particle cut is needed to evaluate $\gamma_{ij}^{\rm UV(2)}$
	\end{tabular}
		\caption{Structure of the two-loop anomalous dimension
                  matrix $\gamma_{ij}^{(2)}$ due to the collected
                  rules outlined in this section. A $\slashed0$
                  indicates there are no contributing two-loop
                  diagrams, whereas $0$ alone indicates that there are
                  one-loop diagrams that could contribute, but the
                  anomalous dimension evaluates to zero. A $0^*$
                  indicates the result is nonzero in $\bar{\rm MS}$,
                  but set to zero by introducing the appropriate
                  finite counterterms. Shading indicates the entry
                  depends only on the three-particle cut, due to
                  either the length selection rules of
                  Section~\ref{section:nonrenormlength} or the
                  vanishing of the relevant one-loop amplitudes. As
                  for Table~\ref{tab:b}, the operators labeling the
                  rows are renormalized by the operators labeling the
                  columns.  }\label{tab:twoloopzeros}
	\end{center}
\end{table}

The previous sections have demonstrated numerous zeros in the two-loop
anomalous dimension matrix, summarized in
Table~\ref{tab:twoloopzeros}. However, the previous examples are by no
means exhaustive, and more zeros may exist. The large number of zeros
in the one-loop amplitudes (Table~\ref{tab:rationalzeros}) implies
that when calculating two-loop anomalous dimensions, the
two-particle cut formed from the dimension-four tree and the
dimension-six one-loop amplitude will not contribute. In some cases,
the only contribution will come from the three-particle cut. Examples
of this include the renormalization of $\O_{F^3}$ by
$\O_{\PhiSquareFSquareA}$ or $\O_{\PhiSquareFSquareB}$, and the
renormalization of $\O_{\PsiFourthA}$ and $\O_{\PsiFourthB}$ by
$\O_{\PhiSquareFSquareA}$ or $\O_{\PhiSquareFSquareB}$. The cuts for
these examples are depicted in Figure~\ref{fig:twoloopexamples4}.
While is may seem that there no reason to expect any given
three-particle cut to evaluate to zero, it is possible that a detailed
inspection may find that helicity selection
rules~\cite{NonrenormHelicity} or angular momentum selection
rules~\cite{Jiang:2020sdh} set certain cuts to zero. For a generic
entry, the collection of these rules and the rules laid out in the
sections above greatly simplify the calculation of the
two-loop anomalous dimensions by eliminating one or more required
unitary cuts, and one might expect that overlapping rules will conspire
to eliminate all possible cuts and set additional entries in
Table~\ref{tab:twoloopzeros} to zero.

\section{Implications for the SMEFT}
\label{section:mapping}

In this section we describe the overlap of our theory with the SMEFT,
and we explain how our calculations directly confirm a large number of
the one-loop anomalous dimensions computed in Refs.~\cite{Manohar123}.
We also comment on two-loop zeros and the coupling dependence of a
subset of the two-loop anomalous-dimension matrix of the SMEFT.

\subsection{Mapping our theory to the SMEFT}

The full SMEFT is more intricate than the simplified model adopted
here, as it includes multiple gauge groups and a number of additional
operators. However, by keeping the gauge group to be a general
$SU(N)$, and by leaving the identity of the fermions unspecified, we
can access many of the entries of the anomalous-dimension matrix in
the full SMEFT basis of operators used by Refs.~\cite{Manohar123}. In
particular, since the Higgs transforms under $SU(2)$, setting $N=2$
and the number of scalars $N_s=1$ allows us to map to anomalous
dimensions or four-point amplitudes from representatives of any of the
classes of operators in Ref.~\cite{Manohar123} other than the
$\psi^2F\varphi$ class ($\psi^2XH$ in the notation of
Ref.~\cite{Manohar123}). Since the scalar is in the fundamental
representation that class necessarily involves both a left-handed
fermion charged under $SU(2)$, as well as an uncharged right-handed
fermion, which does not fit into our framework. By taking $N=3$, parts
of the anomalous dimensions in the SMEFT containing gluons can also be
obtained.  In principle, one can also compare anomalous dimensions for
additional operators using more sophisticated embeddings of the
Standard Model into $SU(N)$, including $U(1)$ charges (see e.g,
Appendix IV of Ref.~\cite{ConvertToU1}), but we do not do so here.

By specifying the flavor of the fermions, we can map to a number of
operators of the full basis used by Ref.~\cite{Manohar123} via different
choices of gauge group and helicity.  For example, by taking $N=2$ and
and left-handed helicity on the external states, we access the $SU(2)$
portions of the amplitudes involving the $q$ and $\bar q$ quark
doublets, and map onto the operators $(\bar q\gamma_\mu q)(\bar
q\gamma^\mu q)$ and $(\bar q\gamma_\mu \tau^I q)(\bar
q\gamma^\mu\tau^I q)$. One remaining difference in our approach
compared to the full SMEFT is that we treat the fermions as Dirac
instead of Weyl. This causes factor of 2 differences in the $N_f$
terms of the renormalization of $\O_{\DPhiSquarePsiSquareB}$ and
$\O_{\PsiFourthB}$ compared to Ref.~\cite{Manohar123}, which need to be
taken into account when comparing.  While our simplified model avoids
having to deal with $\gamma_5$, the generalized unitarity method has
been applied to such cases as well~\cite{ChiralOnShell}. At one loop,
the issue of Weyl versus Dirac fermions is reduced to a question of
which helicities to take in the state sum in
Eq.~\eqref{eq:oneloopphasespace}. 

Setting aside the issue of Weyl versus Dirac fermions, mapping onto the
four-fermion operators of Ref.~\cite{Manohar123}, $(\bar l\gamma_\mu l)(\bar l\gamma^\mu l)$,
$(\bar u\gamma_\mu u)(\bar u\gamma^\mu u)$, and $(\bar d\gamma_\mu
d)(\bar d\gamma^\mu d)$ is possible as well, but requires some care,
due to the presence of evanescent effects. In particular, for these
cases the operator $\O_{\PsiFourthB}$ is related to the operator
$\O_{\PsiFourthA}$ due to the $SU(N)$ Fierz identity \eqref{eq:suNfierz}
\begin{equation}
(\bar \psi{}_m \gamma^\mu T^a \psi_n)(\bar \psi{}_p \gamma_\mu T^a \psi_r)=(\bar \psi{}_m \gamma^\mu \psi_n)(\bar \psi{}_p \gamma_\mu \psi_r)\left(\delta_{i_pi_n}\delta_{i_mi_r}-\frac{\delta_{i_mi_n}\delta_{i_pi_r}}{N}\right).
\end{equation}
which, together with the Lorentz--Fierz relations for all left- or right-handed spinors 
\begin{align}
 (\bar \psi{}^m_L \gamma^\mu \psi^n_L)(\bar \psi{}^p_L \gamma_\mu \psi^r_L)&=
-(\bar \psi{}^p_L \gamma^\mu \psi^n_L)(\bar \psi{}^m_L \gamma_\mu \psi^r_L)\,,\nonumber\\
 (\bar \psi{}^m_R \gamma^\mu \psi^n_R)(\bar \psi{}^p_R \gamma_\mu \psi^r_R)&=
-(\bar \psi{}^p_R \gamma^\mu \psi^n_R)(\bar \psi{}^m_R \gamma_\mu \psi^r_R)\,,\label{eq:lorentzFierz}
\end{align}
(where we raised the flavor indices for convenience) can be applied to eliminate the need for the $\O_{\PsiFourthB}$ operator in Table~\ref{tab:operators}:
\begin{align}
 \O_{\PsiFourthB}^{mnpr} = (\bar \psi{}_m \gamma^\mu T^a \psi_n)(\bar \psi{}_p \gamma_\mu T^a \psi_r)=\O_{\PsiFourthA}^{mrpn}-\frac{1}{N}\O_{\PsiFourthA}^{mnpr} \,,
\end{align}
when there are no additional group indices preventing the particle
exchange (for example, the additional $SU(3)$ index prevents the
reduction of $(\bar q \gamma^\mu \tau^I q)(\bar q \gamma_\mu \tau^I
q)$ operator based on the $SU(2)$ Fierz identity). By choosing to
implement Eq.~\eqref{eq:lorentzFierz} or not, we can map onto either
the operators $(\bar l \gamma^\mu l)(\bar l \gamma_\mu l)$, $(\bar u
\gamma^\mu u)(\bar u \gamma_\mu u)$, or $(\bar d \gamma^\mu d)(\bar d
\gamma_\mu d)$, or onto the set of operators $(\bar q \gamma^\mu
\tau^I q)(\bar q \gamma_\mu \tau^I q)$ and $(\bar q \gamma^\mu \tau^I
q)(\bar q \gamma_\mu \tau^I q)$, respectively. Since we take all the
fermions in our operators to be charged under the same gauge group, here we
do not map onto the $(\bar LR)(\bar LR)$ or $(\bar LR)(\bar RL)$
subsets of the four-fermion operators, which require the presence of
multiple gauge groups.

It is worth noting, that there are some simplifications in the SMEFT relative
to our model with general gauge group. The symmetric color tensor $d^{abc}$ is zero
in $SU(2)$, meaning that the operator $\O_{\PhiSquareFSquareB}$ is identically
zero. In addition, this implies the color factors for the two-vector,
two-scalar or two-vector, two-fermion processes are related by
$N\{T^{a_1},T^{a_2}\}_{i_4i_3} = 2\delta^{a_1a_2}\delta_{i_4i_3}$, meaning the
number of color-ordered amplitudes is reduced for those processes in the case
of $SU(2)$.

\begin{table}[tb]
	\begin{center}
		\renewcommand{\arraystretch}{1.2} 
		\begin{tabular}{c|ccccccccc|}
			&$\O_{G}$ &$\O_{W}$ & $\O_{HW}$& \begin{tabular}{@{}c@{}}$\O_{H\Box}$ \\ $\O_{HD}$\end{tabular} &\begin{tabular}{@{}c@{}}$\O_{Hl}^{(1)}$ \\ $\O_{Hl}^{(3)}$\end{tabular} &  \begin{tabular}{@{}c@{}}$\O_{Hq}^{(1)}$ \\ $\O_{Hq}^{(3)}$\end{tabular} &$\O_{ll}$ & \begin{tabular}{@{}c@{}}$\O_{qq}^{(1)}$ \\ $\O_{qq}^{(3)}$\end{tabular} & \begin{tabular}{@{}c@{}}$\O_{uu}$ \\ $\O_{dd}$\end{tabular}\\
			\hline
			$\O_{G}$ &$ \tblue{\checkmark_3}$ &$\slashed0$  &$\slashed0$ &$\slashed0$ &$\slashed0$ &$\slashed0$ &$\slashed0$& $\slashed0$ & $\slashed0$ \\[1mm]\hline
			$\O_{W}$ &$\slashed0$ &$ \tblue{\checkmark_2}$ &$ \tblue{\checkmark_2}$ &$\slashed0$ &$\slashed0$ &$\slashed0$ &$\slashed0$ &$\slashed0$ &$\slashed0$\\[1mm]\hline
			$\O_{HW}$ &$\slashed0$ &$ \tblue{\checkmark_2}$ & $ \tblue{\checkmark_{2,\lambda}}$&$ \tblue{\checkmark_2}$ &$ \tblue{\checkmark_2}$ &$ \tblue{\checkmark_2}$ &$\slashed0$ &$\slashed0$ &$\slashed0$\\[1mm]\hline
			$\O_{H\Box},\,\O_{HD}$  &$\slashed0$ &$ \tblue{\checkmark_2}$ &$ \tblue{\checkmark_2}$ &$ \tblue{\checkmark_{2,\lambda}}$ &$ \tblue{\checkmark_2}$ &$ \tblue{\checkmark_2}$ & $\slashed0$&$\slashed0$ &$\slashed0$\\[1mm]\hline
			$\O_{Hl}^{(1)},\,\O_{Hl}^{(3)}$ &$\slashed0$ &$ \tblue{\checkmark_2}$  &$ \tblue{\checkmark_2}$ & $ \tblue{\checkmark_2}$& $ \tblue{\checkmark_{2,\lambda}}$&$ \tblue{\checkmark_2}$ &$ \tblue{\checkmark_2}$ &$\slashed 0$ &$\slashed0$\\[1mm]\hline
			$\O_{Hq}^{(1)},\,\O_{Hq}^{(3)}$  &$\slashed0$ &$ \tblue{\checkmark_2}$ &$ \tblue{\checkmark_2}$ & $ \tblue{\checkmark_2}$&$ \tblue{\checkmark_2}$ &$ \tblue{\checkmark_{2,\lambda}}$ &$\slashed0$ &$ \tblue{\checkmark_2}$ &$\slashed0$\\[1mm]\hline
			$\O_{ll}$ &$\slashed0$ &$ \tblue{\checkmark_2}$ &$\slashed0$ & $\slashed0$&$ \tblue{\checkmark_2}$ &$\slashed0$ & $ \tblue{\checkmark_2}$&$\slashed0$ &$\slashed0$\\[1mm]\hline
			$\O_{qq}^{(1)},\,\O_{qq}^{(3)}$ &$ \tblue{\checkmark_3}$ &$ \tblue{\checkmark_2}$  &$\slashed0$ &$\slashed0$ &$\slashed0$ &$ \tblue{\checkmark_2}$ & $\slashed0$&$ \tblue{\checkmark_2}$ &$\slashed0$\\[1mm]\hline
			$\O_{uu},\,\O_{dd}$ &$ \tblue{\checkmark_3}$ &$\slashed0$  &$\slashed0$ &$\slashed0$ &$\slashed0$ &$\slashed0$ &$\slashed0$ &$\slashed0$ &$ \tblue{\checkmark_3}$\\[1mm]\hline
		\end{tabular}
		\vskip .2 cm 
		\caption{Checks on the one-loop anomalous dimensions calculated
                  in Ref.~\cite{Manohar123} obtained from our
                  calculations. The $\slashed0$ entries correspond to
                  trivial cases were there are no contributing
                  diagrams. The entries $\checkmark_3$ and
                  $\checkmark_2$ are checked by setting the
                  SU(N) group to SU(3) or SU(2), respectively. In both
                  cases, only the pieces of the anomalous dimensions
                  proportional to $g_3^2$ or $g_2^2$ are accessed by
                  our amplitudes. The $\checkmark_{2,\lambda}$ cases
                  indicates that both terms proportional to $g_2^2$
                  and $\lambda$ are verified. Operators
                  have been grouped according to whether the gauge
                  dependence of the particle content is the same. As
                  for the other tables, the operators labeling the rows
                  are renormalized by the operators labeling the
                  columns.}
                 \label{tab:oneloopManohar}
	\end{center}
\end{table}

\subsection{Verification of one-loop anomalous dimensions}

From our one-loop calculations and the relations described above we
have verified entries from numerous classes of operators in the SMEFT,
as summarized in Table~\ref{tab:oneloopManohar}, following the
notation of Ref.~\cite{Manohar123}. This includes examples proportional to
$g_3^2$, $g_2^2$, and $\lambda$. In this sense our operators are a
representative sample of the full SMEFT, despite the
simplified nature of our dimension-four Lagrangian. The direct
agreement with results of Ref.~\cite{Manohar123} displayed in
Table~\ref{tab:oneloopManohar} provides a highly non-trivial check of
the validity and the effectiveness of the approach used here.

\subsection{Two-loop implications}

\begin{table}[tb]
	\begin{center}
		\renewcommand{\arraystretch}{1.2}
		\begin{tabular}{c|cccccccccc|}
			&$\O_{G}$ &$\O_{W}$ & $\O_{HW}$&$\O_{uW}$ & \begin{tabular}{@{}c@{}}$\O_{H\Box}$ \\ $\O_{HD}$\end{tabular}  &  \begin{tabular}{@{}c@{}}$\O_{Hq}^{(1)}$ \\ $\O_{Hq}^{(3)}$\end{tabular}  & \begin{tabular}{@{}c@{}}$\O_{qq}^{(1)}$ \\ $\O_{qq}^{(3)}$\end{tabular} & $\O_{uu}$& $\O_{uH}$&$\O_H$\\
			\hline
			$\O_{G}$ &  & $\slashed0$ &$\slashed0$& $\slashed0$ & $\slashed0$ &$\slashed0$ &\cRed 0 &\cRed0 &$\slashed0$ & $\slashed0$\\[1mm]\hline
			$\O_{W}$ &$\slashed0$ &  &  \cB &  \cB$0_y$ &\cRed 0 &\cRed 0 &\cRed 0 &$\slashed0$ &$\slashed0$ & $\slashed0$\\[1mm]\hline
			$\O_{HW}$ & $\slashed0$ &  &  & $0_y$ &  &  & \cRed $0$ & $\slashed0$& \cB$0_y$ &\cRed 0\\[1mm]\hline
			$\O_{uW}$ &$0_y$ & $0_y$ & $0_y$ & &$0_y$ & $0_y$ & $0_y$ & $0_y$ & \cB  &$\slashed0$\\[1mm]\hline
			$\O_{H\Box},\,\O_{HD}$  & $\slashed0$ &  & & $0_y$ &  &  & \cT  $0_y(\bcancel{g_2^4}^*)$ & \cT $0_y$& \cB $0_y$ &\cRed0\\[1mm]\hline
			$\O_{Hq}^{(1)},\,\O_{Hq}^{(3)}$  &  &  &  & $0_y$ &  &  &  &$0_y$ & \cB$0_y(\bcancel{y\lambda})$ &$\slashed0$\\[1mm]\hline
			$\O_{qq}^{(1)},\,\O_{qq}^{(3)}$ &   &  & $\cB $ & $0_y$ & \cT $0_y(\bcancel{g_2^4})$ &  &  & \cT &$\slashed0$ &$\slashed0$\\[1mm]\hline
			$\O_{uu}$ &  &  $\slashed0$ & $\slashed0$ & $0_y$ & \cT$0_y$ & \cT$0_y$ & \cT & &$\slashed0$ &$\slashed0$\\[1mm]\hline
			$\O_{uH}$ & $0_y$ & $0_y$  & $0_y$ &  & $0_y$ & $0_y$ & $0_y$ & $0_y$& &$0_y$ \cB\\[1mm]\hline
			$\O_{H}$ & $\slashed0$ &   &  & $0_y$ &  &  &\cT  & \cT&$0_y$ &\\[1mm]\hline
		\end{tabular}
\vskip .3 cm 
		\begin{tabular}{cl}
			$\slashed 0$ &: trivial zero, no contributing two-loop diagrams \\
			\cRed 0 &: zero predicted by the selection rules of Section \ref{section:twoloop}  \\
			{ \cB} &: only a three-particle cut is needed to evaluate $\gamma_{ij}^{\rm UV(2)}$  \\
			{\cT} &:  only two-particle cuts available for the relevant diagrams  \\
			$0(\bcancel{y\lambda})$, etc.&: the selection rules of Section~\ref{section:twoloop} forbid the stated coupling dependence\\
			$0_y$ &: $\gamma_{ij}^{\rm UV(2)}$ vanishes if Yukawa couplings are set to zero 
		\end{tabular}
		\vskip .2 cm 
		\caption{Predictions for the zeros and coupling
                  dependences of a representative selection of the
                  SMEFT two-loop anomalous-dimension matrix,
                  $\gamma_{ij}^{\rm UV(2)}$. The notation for the
                  operator labels follows that of \cite{Manohar123}.
                  The $g_2^4$ dependence of the entry labeled
                  $0_y(\bcancel{g_2^4}^*)$ vanishes using the
                  appropriate counterterms at one loop. The operators
                  labeling the rows are renormalized by the operators
                  labeling the columns.}\label{tab:twoloopSMEFT}
	\end{center}
\end{table}

Next we briefly discuss the implications of the zeros in the two-loop
anomalous dimensions of our simplifies model for the SMEFT.  The selection rules
of Section~\ref{section:twoloop} set a number of entries strictly to
zero, and restrict the coupling dependence of others. Our findings are summarized in
Table~\ref{tab:twoloopSMEFT}. The full SMEFT anomalous dimensions
include dependence on the Yukawa couplings, which are absent in our
simplified theory, so some of the zeros uncovered above may be replaced
by anomalous dimensions that depend on such couplings.  Nevertheless,
our results show that the coupling dependence of the anomalous
dimensions is simpler than one might have expected, and that some of
the entries are zero or do not have pure dependence on the gauge
couplings. Though most of the strictly zero examples rely on the length selection rule, which does not depend on the gauge group or the presence of Yukawa couplings, the anomalous-dimension matrix element $\gamma_{HW\leftarrow qq}^{(2)}$ relies solely on the color selection rules. In this case, including Yukawa and U(1) couplings will not affect this zero, as the cuts still cannot match the color of the target operator. 

In addition to the zeros, we find that many of the entries only
receive contributions from either three- or two-particle cuts, which
should greatly simplify their computation. One interesting example is
the element $\gamma_{qq\leftarrow HW}^{(2)}$, which only has a three-particle cut due to the vanishing of the one-loop amplitudes that
would contribute to the two-particle cut. For this example, we have
also checked the one-loop amplitudes with Yukawa and U(1)
couplings do not contribute. As can also be seen in Table
\ref{tab:twoloopSMEFT}, many entries vanish when the Yukawa couplings
are set to zero. Many of these zeros are trivial due to the particle
content of the operators involved, but in some cases a closer
examination of the diagrams is required to see that only diagrams with
Yukawa couplings will produce nonvanishing results.
	
Note that the operators in Table \ref{tab:twoloopSMEFT} are merely a representative set, in that all of the operators of the SMEFT are restricted by one or more of our selection rules, either in terms of which operators they can renormalize, or vice versa. In particular, the length selection rules apply independently of the gauge group or the presence of Yukawa couplings, which allows us to include operators of the classes $\psi^2F\varphi$ and $\psi^2\varphi^3$ in Table \ref{tab:twoloopSMEFT}.
We would also like to stress that our
analysis of the structure of the two-loop anomalous dimensions is
not an exhaustive study of the SMEFT anomalous dimensions.  For this
reason, we expect that there could be additional
vanishings or structures that can be uncovered under closer scrutiny.

\section{Conclusions}
\label{section:conclusion}

In this paper we applied on-shell methods to investigate the structure of the
two-loop anomalous dimension matrix of dimension-six operators, in both a
simplified model and in the SMEFT. At one loop, we used both the standard
generalized unitarity method~\cite{GeneralizedUnitarity1} and the recently
developed approach for extracting anomalous dimensions directly from unitarity
cuts~\cite{Caron-Huot:2016cwu}.  At two loops, we find the latter method to be
especially effective, with the former method providing one-loop amplitudes as
inputs. As an initial step, we reorganized the basic equation  for the
two-loop anomalous dimension in the latter approach so as to simplify one-loop
iterations.  Using this equation, we revealed a number of vanishing
contributions in the two-loop anomalous dimension matrix of the SMEFT. Our
analysis was based on a simplified model without U(1) or Yukawa interactions.
Nevertheless, as summarized in Table~\ref{tab:twoloopSMEFT}, by analyzing the
overlap of our simplified model with the SMEFT we found that a remarkable
number of SMEFT two-loop anomalous dimensions either vanish or have a simpler
dependence on the Standard Model couplings than naively expected.

The structure we uncovered has a number of origins, including length selection
rules, color selection rules, and zeros in the one-loop amplitudes with
dimension-six operator insertions. Additional zeros arise from the choice of an
$\bar{\rm MS}$-like scheme which includes additional finite renormalizations
designed to set various rational terms in one-loop amplitudes to zero. This
suggests that there exist interesting schemes that make the structure of the
renormalization-group running beyond one loop more transparent.  The full
implications of choosing such schemes clearly deserve further study.

Since one-loop amplitudes are used as input for the two-loop
calculation, we have computed the full set of four-point amplitudes
with dimension-six operator insertions in our simplified version of the
SMEFT. As a byproduct, these amplitudes have allowed us to verify a
large subset of the one-loop anomalous dimensions calculated in
Refs.~\cite{Manohar123}. 

The zeros that we found in the two-loop anomalous dimension matrix relied on
choosing examples with trivial infrared dependence, as well as a lack of a
three-particle cut. However, the methods can be applied just as well to any
generic anomalous dimension matrix element at two or higher loops. It would be
interesting to investigate whether there are additional zeros at two loops
beyond those we identified.  The large number of zeros in the one-loop
amplitudes restrict the number of cuts that can contribute, suggesting that
other mechanisms, such as helicity or angular-momentum selection rules, may set
the remaining cuts to zero in some cases. 

The presented methods are quite general, and should be applicable to general
EFTs. In addition, while we have focused on ultraviolet anomalous dimensions
here, this method could equally be applied to the evaluation of infrared
anomalous dimensions, such as the soft anomalous dimension, by the use of
ultraviolet protected operators such as the stress-tensor or global symmetry
currents. It would also be interesting to understand the implications, if any,
of the vanishing of two-loop anomalous dimensions for the interference of
Standard Model and higher-dimension operator matrix elements beyond tree level,
in the presence of experimental cuts.  Another obvious direction would be to
include dimension seven and eight operators into the analysis~\cite{Dim78}.

In summary, we have demonstrated that the on-shell methods applied here are
well suited for computing anomalous dimensions and associated scattering
amplitudes at one and two loops. We used these methods to expose new structures
in the guise of vanishing terms in the anomalous matrix of the SMEFT beyond one
loop.  Our analysis here was not exhaustive, so it is likely that further
vanishing contributions and new structures exist at two loops and beyond. Our
results also suggest that a judicious choice of renormalization scheme can help
expose such structures.

\section*{Acknowledgments}

We thank Clifford Cheung, Nathaniel Craig, Enrico Herrmann,
Chia-Hsien Shen and especially Aneesh Manohar for many helpful
discussions. Z.B. is supported by the U.S.  Department of Energy (DOE)
under award number DE-SC0009937. J.P-.M.\ is supported by the US
Department of State through a Fulbright scholarship. We are also
grateful to the Mani L. Bhaumik Institute for Theoretical Physics for
additional support.

\newpage
\begin{appendix}

\section{Integral reduction via gauge-invariant tensors}
\label{section:amplitudesappendix}

In this appendix we summarize the projection technique that we use to perform
tensor reduction of loop integrals in \sect{section:onelooprat}.  The same technique has been previously
used in Refs.~\cite{Projectors1, Projectors2} and is a convenient method for
decomposing $D$-dimensional tensor loop integrands (or cuts) into a basis of scalar
master integrals,  in a way that makes dimensional regularization, and any
associated chiral and evanescent issues relatively straightforward. In
particular this technique is well suited to deal with integrals with high-rank
numerators, which naturally arise in loop amplitudes with insertions of
higher-dimension operators. 

We start by noting that scattering amplitudes amplitudes are gauge invariant
and can therefore be decomposed into a basis of gauge-invariant tensors, $T_m$.  For a
given amplitude labeled by $i$ we have,
\begin{equation}
A^{(L)}_i=\sum_m \mathcal A^{(L)}_{i,m}(k_j) T_m(k_j,\epsilon_j, u_j,\bar u_j) \,,
\label{Decomposition}
\end{equation}
where the coefficients, $\mathcal A^{(L)}_{i,m}$, only depend on the
external momenta, and all dependence on the polarization vectors or spinors is
contained entirely within the basis tensors, $T_m$.  The basis tensors for the
various processes we consider in this paper are given below and in the
supplementary material~\cite{AttachedFiles}.  They are found by writing down
the most general polynomials built from Lorentz invariant products of external
polarizations, spinor and momenta and then demanding gauge invariance.

The desired coefficient of tensor $T_j$ can be extracted using a projector  
\begin{equation}
P_n=c_{nm}T_m^* \,,
\end{equation}
where $c_{nm}$ is the inverse of the matrix
\begin{equation}
m_{nm}=T^*_n\odot T_m \,.
\end{equation}
Here the product $\odot$ corresponds to the state sum in
Eq.~\eqref{eq:statesum}, taken over all particles.  
The coefficient of the tensor is then simply given by
\begin{equation}
  \mathcal A^{(L)}_{i,m} = P_m\odot A^{(L)}_i\,.
\end{equation}
The projectors for
all  processes consider in this paper are given explicitly in an ancillary
file~\cite{AttachedFiles}.  

Once projected, any gauge invariant quantity can be summarized as a list of the
coefficients corresponding to each basis tensor. In the case of a loop integrand or cut thereof, each coefficient is a rational function of scalar propagators and inverse
propagators (and irreducible numerators beyond one loop). The
integrals corresponding to each term in the projected quantity
are then in a form that can be reduced to a basis of 
master integrals using by integration by parts (IBP) relations.  
This can be done using by using IBP programs such as FIRE~\cite{FIRE}.

As described in \sect{section:onelooprat}, we can apply 
this procedure cut by cut to determine the coefficients of each gauge invariant 
tensor in the full amplitude. 

\subsection*{Basis tensors}

Basis tensors for the four-vector amplitudes are taken from
\cite{Projectors2}, which we reproduce here. Beginning with the
linearized field strength for each external particle:
\begin{equation}
F_{i\,\mu\nu} \equiv k_{i\,\mu} \varepsilon_{i\,\nu} - k_{i\,\nu} \varepsilon_{i\,\mu}\,,
\label{FDef}
\end{equation}
one can construct the following combinations,
\begin{equation}
\begin{aligned}
\Ff{st}   & \equiv (F_1 F_2 F_3 F_4)\,, &
\Ff{tu}   & \equiv (F_1 F_4 F_2 F_3)\,, &
\Ff{us}   & \equiv (F_1 F_3 F_4 F_2)\,,\\
\Ft{s}    & \equiv (F_1 F_2)(F_3 F_4)\,, \hskip 5mm&
\Ft{t}    & \equiv (F_1 F_4)(F_2 F_3)\,, \hskip 5mm&
\Ft{u}    & \equiv (F_1 F_3)(F_4 F_2)\,,
\end{aligned}
\label{Fcombinations}
\end{equation}
where parentheses one the right-hand side of the above equations indicate taking the trace over adjacent Lorentz indices. The four-vector basis tensors are then given by
\begin{align}
T_{vvvv}^{\rm tree} &=
-\frac{1}{2}(\Ft{s}+\Ft{t}+\Ft{u})
+2\,(\Ff{st}+\Ff{tu}+\Ff{us}) \,,
\nonumber\\
T_{vvvv}^{++++} &=
-2\,\Ff{st}+\frac{1}{2}\,(\Ft{s}+\Ft{t}+\Ft{u})
\,, \nonumber \\
T_{vvvv}^{-+++} &=
-T_{F^3}-(\Ff{tu}-\Ff{us})\,(s-t)
+(\Ff{st}-\frac{1}{4}(\Ft{s}+\Ft{t}+\Ft{u}))\,(s+t)
\,, \nonumber \\
T_{vvvv}^{--++} &=
\Ft{s}-\Ft{t}+2\,(\Ff{tu}-\Ff{us})
\,,\\
T_{vvvv}^{-+-+} &=
2\,\Ff{st}-\frac{1}{2}(\Ft{s}+\Ft{t}-\Ft{u})
\,, \nonumber \\
T_{vvvv}^{\rm ev1} &=
-(2\,\Ff{st}+\frac{3}{2}\,(\Ft{s}+\Ft{t}+\Ft{u}))\,(s+t)
+2\,(\Ff{us}\,(3\,s+t)+\Ff{tu}\,(s+3\,t))
\,, \nonumber \\
T_{vvvv}^{\rm ev2} &=
-(2\,\Ff{st}-\frac{1}{2}(\Ft{s}+\Ft{t}+\Ft{u}))\,(s-t)
+2\,(\Ff{tu}-\Ff{us})\,(s+t)
\,, \nonumber
\label{SevenTensorsCyclic}
\end{align}  
where the $v$ labels signifies that a leg is a vector boson, and $T_{F^3}$ is proportional to the $F^3$ amplitude \cite{SupersymStringAmps}:
\begin{equation}
T_{F^3}=-istA_{F^3}^{(0)}=-istu\left(\frac{(F_s^2)^2}{4s^2}+\frac{(F_t^2)^2}{4t^2}+\frac{(F_u^2)^2}{4u^2} - \frac{g_1g_2g_3g_4}{(stu)^2}    \right) \,,
\end{equation}
where 
$g_i\equiv (k_{i+1}F_ik_{i-1})$. We note that we have written this expression in an explicitly gauge-invariant form at the expense of manifest locality.
These tensors are nonzero only under the indicated (and parity conjugate) helicity
configurations, along with cyclic permutations. $T_{vvvv}^{\rm tree}$ is
nonzero for helicities $(1^-2^+3^-4^+)$, $(1^-2^-3^+4^+)$, and cyclic
permutations. $T_{vvvv}^{\rm ev1}$ and $T_{vvvv}^{\rm ev2}$ are
evanescent, i.e. zero for all helicity configurations in four
dimensions. This can be made manifest by rewriting them as
\begin{align}
  \begin{split}
    T_{vvvv}^{\rm ev1} & = \frac12 k_4^{[\alpha} F_1^{\vphantom{[}\mu\nu} F_2^{\sigma\rho]} k_2{}_{\alpha} F_4{}_{\mu\nu} F_3{}_{\sigma\rho} + \frac12 k_4^{[\alpha} F_3^{\vphantom{[}\mu\nu} F_2^{\sigma\rho]} k_2{}_{\alpha} F_4{}_{\mu\nu} F_1{}_{\sigma\rho}\,, \\
	T_{vvvv}^{\rm ev2} & = \frac12 k_2^{[\alpha} F_1^{\vphantom{[}\mu\nu} F_3^{\sigma\rho]} k_1{}_{\alpha} F_2{}_{\mu\nu} F_4{}_{\sigma\rho}\,,
\end{split}
\end{align}
where the anti-symmetrization does not include a symmetry factor.

The two-vector, two-scalar tensors are also nonzero under specific helicity combinations, and are given by 
\begin{align}
  T^{+-}_{vv ss}&=  2(k_3F_1F_2k_4) + 2(k_4F_1F_2k_3) - (k_3\cdot k_4) (F_1F_2)\,, \quad
T^{++}_{vv ss}&= -(F_1F_2) \,,
\end{align}
where the $v$ and $s$ labels specify the corresponding legs are vectors or scalars.

Similarly, the two-vector, two-fermion tensors are linear combinations of those in Ref.~\cite{Projectors1}, chosen to again be nonzero only under specific helicities:
\begin{align}
  &T^{-+++}_{ffvv}= -\frac 1 {2^4} (\bar{u}_{2}\slashed  F_4  \slashed F_3  \slashed k_2 u_{1}) \,, \qquad
   T^{-+-+}_{ffvv} = -\frac 1 {2^4} (\bar{u}_{2}\slashed  F_4  \slashed k_2  \slashed F_3 u_{1}) \,, \nonumber\\
  &T^{-++-}_{ffvv}= -\frac 1 {2^4} (\bar{u}_{2}\slashed  F_3  \slashed k_1  \slashed F_4 u_{1}) \,, \qquad
   T^{-+--}_{ffvv} = -\frac 1 {2^4} (\bar{u}_{2}\slashed  k_1  \slashed F_4  \slashed F_3 u_{1}) \,, \nonumber\\
   &\hspace{2cm}T^{\rm ev}_{ffvv}= \frac12 k_1^{[\alpha} F_3^{\mu\nu}  F_4^{\rho\sigma]}  (\bar{u}_{2} \gamma_{\alpha} \gamma_{\mu}  \gamma_{\nu} \gamma_{\rho}\gamma_{\sigma} u_{1})
\,,
\end{align}
where $f$ now indicates a leg as a fermion, $\slashed F_i = F_{i\,\mu\nu} \gamma^\mu\gamma^\nu$, and the antisymmetrization in $T^{\rm ev}$ includes a symmetry factor of $1/5!$.
As for the four-vector case, we encounter an evanescent tensor,
$T^{\rm ev}_{ffvv}$ which vanishes for all four-dimensional
helicities. For the two-fermion two-scalar case there is only a single basis
tensor:
\begin{align}
T_{ffss}=\bar{u}_{2}\slashed k_{3}u_{1}\,.
\end{align}
Finally, the four-fermion tensors are,
\begin{align}
T^1_{ffff} &= (\bar u_2\gamma^\mu u_1)(\bar u_4\gamma_\mu u_3) \,, \nonumber\\
T^2_{ffff} &= (\bar u_2\slashed k_4 u_1)(\bar u_4\slashed k_2 u_3)\,, \nonumber \\
T^3_{ffff} &= (\bar u_2\gamma^\mu\gamma^\nu \gamma^\rho  u_1)(\bar u_4\gamma_\mu\gamma_\nu\gamma_\rho u_3) -16(\bar u_2\gamma^\mu u_1)(\bar u_4\gamma_\mu u_3)\,, \nonumber \\
T^4_{ffff} &= t \,(\bar u_2\gamma^\mu\slashed k_4 \gamma^\rho  u_1)(\bar u_4\gamma_\mu\slashed k_2\gamma_\rho u_3) - 4u (\bar u_2\slashed k_4 u_1)(\bar u_4\slashed k_2 u_3)\,,
\end{align}
plus those given by the exchange of legs 2 and 4. It should be noted, however, that in practice it is unnecessary to calculate the coefficients of the exchanged tensors, since they are fixed by the symmetry of the contributing diagrams. 
$T^3_{ffff}$ and $T^4_{ffff}$ are chosen to be zero for the helicity configuration $\psim{1}{+} \psibarm{2}{-}  \psim{3}{+} \psibarm{4}{-}$ and its conjugate, 
so that these tensors are evanescent if the spinors are Weyl of the same handedness.

\section{Tree-level and one-loop amplitudes}
\label{section:amplitudesSH}

In this appendix we collect tree- and one-loop amplitudes.  
In addition to the spinor-helicity amplitudes given below, expressions that are valid to all orders in the dimensional regularization parameter $\epsilon$ are provided in a supplementary file \cite{AttachedFiles}. While we do not require one-loop amplitudes without higher-dimension operators for our specific examples in Section \ref{section:twoloop}, they would be required for the calculation of a generic two-loop anomalous dimension matrix element. These one-loop dimension-4 amplitudes can be found in various references; e.g. Refs. \cite{OneLoopSMAmplitudes} gives the relevant amplitudes which exclude scalars.

The amplitudes and form factors can be written as vectors in color space, 
\begin{equation}
A^{(L)}(\lambda_1\lambda_2\lambda_3\lambda_4)
=S_{\lambda_1\lambda_2\lambda_3\lambda_4}
\sum_i \mathcal{C}_{\lambda_1\lambda_2\lambda_3\lambda_4}^{[i]}A^{(L)}(\lambda_1\lambda_2\lambda_3\lambda_4)_{[i]} \,,
\end{equation}
where $S_{\lambda_1\lambda_2\lambda_3\lambda_4}$ is a
helicity-dependent factor which which depending 
on spinors when evaluated using four-dimensional spinor helicity.
These factors are pure phases for the amplitudes with an
even number of pairs of external fermions, and for the amplitudes with
an odd number of fermions their square is a dimensionless ratio of
$s,t$, or $u$ and powers thereof. The full list of
$S_{\lambda_1\lambda_2\lambda_3\lambda_4}$ for each process is listed below.

The IR dependence has been stripped from the amplitudes below, but can be reconstructed, if desired, using the basic IR formulas given in the text, which we reproduce here:
\begin{align}
A_i^{(1)}= \null & \boldS I^{(1)}A_i^{(0)} 
+A_i^{(1)\text{fin}}\,, \hskip 1 cm
\end{align}
where the IR operator $\boldS I^{(1)}$ is given by 
\begin{align}
\boldS I^{(1)} 
&= \frac{e^{\epsilon\gamma_E}}{\Gamma(1-\epsilon)}
\sum_{p=1}^n\sum_{q\ne p} \frac{ \boldS T_p \cdot \boldS T_q}{2}\left[\frac{\gamma^{\rm IR\, (1)}_{\rm cusp}}{\epsilon^2}-\frac{\gamma^{\rm IR\, (1)}_{{\rm c},\, p}}{\boldS T_p^2}\frac{1}{\epsilon}\right]\left(\frac{-\mu^2}{2k_p{\cdot}k_q}\right)^\epsilon ,
\end{align}
with
\begin{equation}
  \gamma^{\rm IR\, (1)}_{\rm cusp} = \tilde g^2 4\,, \quad
\gamma^{\rm IR \,(1)}_{{\rm c},\, v} = -\tilde g^2 b_0\,,\quad 
\gamma^{\rm IR \,(1)}_{{\rm c},\, f} = -\tilde g^2 3 C_F\,,\quad
\gamma^{\rm IR \,(1)}_{{\rm c},\, s} = -\tilde g^2 4 C_F\,.
\end{equation}
Explicit evaluations of $\boldS I^{(1)}$ for various processes can be found, for example, in Refs. \cite{Projectors1,TwoLoopHelicityExample}. All results below are reported in the Euclidean region
and the $\bar{\rm MS}$ scheme. As a shorthand, logarithms are given by:
\begin{align}
\begin{split}
X^2 & =\log\left(\frac{s}{t}\right)^2+\pi^2, \hskip 1.5 cm  Y^2=\log\left(\frac{s}{u}\right)^2+\pi^2, \hskip 1.5 cm  Z^2=\log\left(\frac{u}{t}\right)^2+\pi^2, \\[5pt]
& \hspace{0.9cm} X_s  =\log\left(\frac{\mu^2}{-s}\right), \hskip 1.1 cm  X_t=\log\left(\frac{\mu^2}{-t}\right), \hskip 1.1 cm X_u=\log\left(\frac{\mu^2}{-u}\right).
\end{split}
\end{align}
In general we drop the Wilson coefficients, for example $\opcoef_{F^3}$ for amplitudes with an $\O_{F^3}$ insertion, since it is in this form that the amplitudes are used in Eq. \eqref{eq:twoloopSimon3}. However we have contracted the Wilson coefficients with the amplitudes for operators which include fermions, since doing so simplifies the flavor information for these cases.

\subsection{Four-vector amplitudes}
The color factors for the four-vector amplitudes are
\begin{gather}
\mathcal{C}_{vvvv}^{[1]}=\Tr[T^1T^2T^3T^4]\,, \hskip 1 cm 
\mathcal{C}_{vvvv}^{[2]}=\Tr[T^1T^3T^2T^4]\,, \nonumber \\
\mathcal{C}_{vvvv}^{[3]}=\Tr[T^1T^2T^4T^3]\,, \hskip 1 cm 
\mathcal{C}_{vvvv}^{[4]}=\Tr[T^1T^4T^2T^3]\,, \\
\mathcal{C}_{vvvv}^{[5]}=\Tr[T^1T^3T^4T^2]\,, \hskip 1 cm 
\mathcal{C}_{vvvv}^{[6]}=\Tr[T^1T^4T^3T^2]\,,  \nonumber\\
\mathcal{C}_{vvvv}^{[7]}=\Tr[T^1T^2]\Tr[T^3T^4]\,,\quad
\mathcal{C}_{vvvv}^{[8]}=\Tr[T^1T^3]\Tr[T^2T^4]\,,\quad
\mathcal{C}_{vvvv}^{[9]}=\Tr[T^1T^4]\Tr[T^2T^3]\,, \nonumber
\end{gather}
where only two partial amplitudes---one single-trace and one double-trace---are independent in general, and the rest are given by relabelings.

We remove dimensionless prefactors from the helicity amplitudes. These are all phases except for the amplitudes involving only one pair of fermions. For the four-vector amplitudes, the spinor prefactors are are given by 
\begin{align}
S(1^+2^+3^+4^+)&=\frac{[1 2] [3 4]}{\langle 1 2\rangle  \langle 3 4\rangle } \,,\hskip 1 cm 
S(1^-2^+3^+4^+)=\frac{\langle 1 2\rangle  \langle 1 4\rangle  [2 4]}{\langle 2 3\rangle  \langle 2 4\rangle  \langle 3 4\rangle }\,, \nonumber \\
S(1^-2^-3^+4^+)&=\frac{\langle 1 2\rangle  [3 4]}{\langle 3 4\rangle  [1 2]}\,, \hskip 1.1 cm 
S(1^-2^+3^-4^+)=\frac{\langle 1 3\rangle  [2 4]}{\langle 2 4\rangle  [1 3]} \,.
\end{align}
The tree-level $D$-dimensional amplitudes are given by
\begin{align}
A^{(0)}(1234)_{[1]}&=\frac{- g^2}{st}T_{vvvv}^{\rm tree}\,,\nonumber\\
A^{(0)}_{}(1234)_{[7]}&=0\,,\nonumber\\
A^{(0)}_{F^3}(1234)_{[1]}&=\frac{  g }{2 stu}\left(4stT^{++++}_{vvvv}-2uT^{-+++}_{vvvv}+(s-t)T^{\rm ev2}_{vvvv} \right)   \,,\nonumber\\
A^{(0)}_{F^3}(1234)_{[7]}&=0\,,
\end{align}
which have four-dimensional helicity values
\begin{align}
A^{(0)}(1^-2^+3^+4^+)_{[1]}&=A^{(0)}(1^=2^+3^+4^+)_{[1]}=0  \,,\nonumber\\
A^{(0)}(1^-2^-3^+4^+)_{[1]}&=-\frac{g^2 s}{t}  \,,\nonumber\\
A^{(0)}(1^-2^+3^-4^+)_{[1]}&=-\frac{g^2 u^2}{s t}  \,,\nonumber\\
A^{(0)}(1^\pm 2^\pm 3^\pm 4^\pm)_{[7]}&=0    \,,
\end{align}
\begin{align}
A^{(0)}_{F^3}(1^+2^+3^+4^+)_{[1]}&=2 g   s   \,,\nonumber\\
A^{(0)}_{F^3}(1^-2^+3^+4^+)_{[1]}&=-gu  \,,\nonumber\\
A^{(0)}_{F^3}(1^-2^-3^+4^+)_{[1]}&=A^{(0)}_{F^3}(1^-2^+3^-4^+)_{[1]}=0   \,,\nonumber\\
A^{(0)}_{F^3}(1^\pm 2^\pm 3^\pm 4^\pm)_{[7]}&=0 \,.
\end{align}
The one-loop amplitudes with one insertion of the $F^3$ operator are
\begin{align}
  A^{(1)\rm fin}_{F^3}(1^+2^+3^+4^+)_{[1]}&= g\tilde g^2   \big((4N(t-u)+2ub_0) X_s + (4N(s-u)+2ub_0)X_t   \nonumber\\
  & \hspace{0.5cm} \null - \frac12(44 N + 2N_f - N_s)u\big) \,, \nonumber\\
  A^{(1)\rm fin}_{F^3}(1^-2^+3^+4^+)_{[1]}&=   g\tilde g^2  \Big(N \frac{u^2-s t}{u}  X^2 \nonumber\\
    & \hspace{0.5cm} \null + (2N (t - u) + b_0 u) X_s +( 2N (s - u)  + b_0 u) X_t - 12 u\Big) \,,\nonumber \\
A^{(1)}_{F^3}(1^-2^+3^-4^+)_{[1]}&=0\,, \nonumber\\
A^{(1)}_{F^3}(1^-2^-3^+4^+)_{[1]}&=\frac{g\tilde g^2   }{6} \left(4 N (u-s)-(2 N_f-N_s) (u-t)\right) \,,
\end{align}
where $\tilde g^2 = g^2/(4\pi)$ as defined in Eq.\eqref{g4Def}, and $b_0 = (11N-2N_f-N_s/2)/3$.
The double-trace amplitudes with an $\O_{F^3}$ insertion are given by the $U(1)$ decoupling identity
\begin{equation}
  A^{(1)}_{F^3}(1234)_{[7]} = \frac1N \left( A^{(1)}_{F^3}(1234)_{[1]} + A^{(1)}_{F^3}(1243)_{[1]} + A^{(1)}_{F^3}(1423)_{[1]}\right)\,.
\end{equation}
The amplitudes with one insertion of a $\varphi^2F^2$ operators are
\begin{align}
A^{(1)}_{\PhiSquareFSquareA}(1^\pm 2^\pm 3^\pm 4^\pm )_{[1]}&=0\,,\nonumber\\
A^{(1)}_{\PhiSquareFSquareA}(1^+2^+3^+4^+)_{[7]}&= 4   \tilde g^2 N_s s\,,\nonumber\\
A^{(1)}_{\PhiSquareFSquareA}(1^-2^+3^+4^+)_{[7]}&=A^{(1)}_{\PhiSquareFSquareA}(1^-2^+3^-4^+)_{[7]}=0\,,\nonumber\\
A^{(1)}_{\PhiSquareFSquareA}(1^-2^-3^+4^+)_{[7]}&=4 \tilde g^2   N_s s\,,\nonumber\\
A^{(1)}_{\PhiSquareFSquareB}(1^+2^+3^+4^+)_{[1]}&=-2 \tilde g^2   N_s u\,,\nonumber\\
A^{(1)}_{\PhiSquareFSquareB}(1^-2^+3^+4^+)_{[1]}&=A^{(1)}_{\PhiSquareFSquareB}(1^-2^+3^-4^+)_{[1]}=0\,,\nonumber\\
A^{(1)}_{\PhiSquareFSquareB}(1^-2^-3^+4^+)_{[1]}&=2  \tilde g^2  N_s s\,,\nonumber\\
A^{(1)}_{\PhiSquareFSquareB}(1^+2^+3^+4^+)_{[7]}&=A^{(1)}_{\PhiSquareFSquareB}(1^-2^-3^+4^+)_{[7]}=-\frac{4 \tilde g^2  N_s s}{N}\,,\nonumber\\
A^{(1)}_{\PhiSquareFSquareB}(1^-2^+3^+4^+)_{[7]}&=A^{(1)}_{\PhiSquareFSquareB}(1^-2^+3^-4^+)_{[7]}=0  \,.
\end{align}

\subsection{Four-fermion amplitudes}
The color structures for the four-fermion amplitudes are
\begin{align}
  \mathcal{C}_{ffff}^{[1]}=T^a_{i_2i_1}T^a_{i_4i_3}\,,\qquad
  \mathcal{C}_{ffff}^{[2]}=T^a_{i_4i_1}T^a_{i_2i_3}\,.
\end{align}
Note for any operator, due to the anti-symmetry of the amplitudes under exchange of  (anti-)fermions:
\begin{align}
A^{(L)}_{\O}(1^+_{\vphantom{\bar \psi_n}\psi_m} 2^-_{\,\bar \psi_n} 3^+_{ \vphantom{\bar \psi_n} \psi_p} 4^-_{\,\bar \psi_r})_{[2]}&= -A^{(L)}_{\O}(1^+_{\vphantom{\bar \psi_n}\psi_m} 2^-_{\,\bar \psi_r} 3^+_{ \vphantom{\bar \psi_n} \psi_p} 4^-_{\,\bar \psi_n})_{[1]}\; (s\leftrightarrow t) \,,\nonumber\\
A^{(L)}_{\O}(1^+_{\vphantom{\bar \psi_n}\psi_m} 2^-_{\,\bar \psi_n} 3^-_{ \vphantom{\bar \psi_n} \psi_p} 4^+_{\,\bar \psi_r})_{[2]}&=  -A^{(L)}_{\O}(1^+_{\vphantom{\bar \psi_n}\psi_m} 2^+_{\,\bar \psi_r} 3^-_{ \vphantom{\bar \psi_n} \psi_p} 4^-_{\,\bar \psi_n})_{[1]}\; (s\leftrightarrow t) \,,\nonumber\\
A^{(L)}_{\O}(1^+_{\vphantom{\bar \psi_n}\psi_m} 2^+_{\,\bar \psi_n} 3^-_{ \vphantom{\bar \psi_n} \psi_p} 4^-_{\,\bar \psi_r})_{[2]}&= - A^{(L)}_{\O}(1^+_{\vphantom{\bar \psi_n}\psi_m} 2^-_{\,\bar \psi_r} 3^-_{ \vphantom{\bar \psi_n} \psi_p} 4^+_{\,\bar \psi_n})_{[1]}\; (s\leftrightarrow t)  \,.
\end{align}
The overall spinor phases are
\begin{gather}\
  S(\psim{1}{+} \psibarm{2}{-}  \psim{3}{+} \psibarm{4}{-})=\frac{\langle 2 4\rangle  [1 2]}{\langle 3 4\rangle  [2 4]}\,,\qquad
  S(\psim{1}{+} \psibarm{2}{-}  \psim{3}{-} \psibarm{4}{+})=\frac{\langle 2 3\rangle  [1 2]}{\langle 3 4\rangle  [2 3]}
\,,\nonumber\\
S(\psim{1}{+} \psibarm{2}{+}  \psim{3}{-} \psibarm{4}{-})=\frac{[12]}{[34]}  \,.
\end{gather}
The tree-level $D$-dimensional amplitudes are given by 
\begin{align}
A^{(0)}(1_{\vphantom{\bar \psi_n}\psi_m} 2_{\,\bar \psi_n} 3_{\vphantom{\bar \psi_n}\psi_p} 4_{\,\bar \psi_r})_{[1]}&=  g^2\frac{u_2\gamma^\mu u_1\;\bar u_4\gamma_\mu u_3}{2s}\delta_{mn}\delta_{pr}\,,\nonumber\\
A^{(0)}_{\PsiFourthA}(1_{\vphantom{\bar \psi_n}\psi_m} 2_{\,\bar \psi_n} 3_{\vphantom{\bar \psi_n}\psi_p} 4_{\,\bar \psi_r})_{[1]}&=  \frac{N}{N^2-1}(\opcoef_{\PsiFourthA}^{nmrp} u_2\gamma^\mu u_1\;\bar u_4\gamma_\mu u_3- \opcoef_{\PsiFourthA}^{rmnp}N u_4\gamma^\mu u_1\;\bar u_2\gamma_\mu u_3)  \,,\nonumber\\
A^{(0)}_{\PsiFourthB}(1_{\vphantom{\bar \psi_n}\psi_m} 2_{\,\bar \psi_n} 3_{\vphantom{\bar \psi_n}\psi_p} 4_{\,\bar \psi_r})_{[1]}&=\opcoef_{\PsiFourthB}^{nmrp} u_2\gamma^\mu u_1\;\bar u_4\gamma_\mu u_3  \,,
\end{align}
which have four-dimensional values
\begin{align} 
A^{(0)}(1^+_{\vphantom{\bar \psi_n}\psi_m} 2^-_{\,\bar \psi_n} 3^+_{ \vphantom{\bar \psi_n} \psi_p} 4^-_{\,\bar \psi_r})_{[1]}&=  \frac{g^2 u}{s}\delta_{mn}\delta_{pr} \,,\nonumber\\
A^{(0)}(1^+_{\vphantom{\bar \psi_n}\psi_m} 2^-_{\,\bar \psi_n} 3^-_{ \vphantom{\bar \psi_n} \psi_p} 4^+_{\,\bar \psi_r})_{[1]}&=   -\frac{g^2 t}{s}\delta_{mn}\delta_{pr} \,,\nonumber\\
A^{(0)}(1^+_{\vphantom{\bar \psi_n}\psi_m} 2^+_{\,\bar \psi_n} 3^-_{ \vphantom{\bar \psi_n} \psi_p} 4^-_{\,\bar \psi_r})_{[1]}&=  0  \,,\nonumber\\
A^{(0)}_{\PsiFourthA}(1^+_{\vphantom{\bar \psi_n}\psi_m} 2^-_{\,\bar \psi_n} 3^+_{ \vphantom{\bar \psi_n} \psi_p} 4^-_{\,\bar \psi_r})_{[1]}&=  -\frac{2 N u (N \opcoef_{\PsiFourthA}^{rmnp}+\opcoef_{\PsiFourthA}^{nmrp})}{N^2-1}  \,,\nonumber\\
A^{(0)}_{\PsiFourthA}(1^+_{\vphantom{\bar \psi_n}\psi_m} 2^-_{\,\bar \psi_n} 3^-_{ \vphantom{\bar \psi_n} \psi_p} 4^+_{\,\bar \psi_r})_{[1]}&= \frac{2 N t \opcoef_{\PsiFourthA}^{nmrp}}{N^2-1}   \,,\nonumber\\
A^{(0)}_{\PsiFourthA}(1^+_{\vphantom{\bar \psi_n}\psi_m} 2^+_{\,\bar \psi_n} 3^-_{ \vphantom{\bar \psi_n} \psi_p} 4^-_{\,\bar \psi_r})_{[1]}&=  \frac{2 N^2 s \opcoef_{\PsiFourthA}^{nmrp}}{N^2-1}  \,,\nonumber\\
A^{(0)}_{\PsiFourthB}(1^+_{\vphantom{\bar \psi_n}\psi_m} 2^-_{\,\bar \psi_n} 3^+_{ \vphantom{\bar \psi_n} \psi_p} 4^-_{\,\bar \psi_r})_{[1]}&= -2 u \opcoef_{\PsiFourthB}^{nmrp}  \,,\nonumber\\
A^{(0)}_{\PsiFourthB}(1^+_{\vphantom{\bar \psi_n}\psi_m} 2^-_{\,\bar \psi_n} 3^-_{ \vphantom{\bar \psi_n} \psi_p} 4^+_{\,\bar \psi_r})_{[1]}&= 2 t \opcoef_{\PsiFourthB}^{nmrp}  \,,\nonumber\\
A^{(0)}_{\PsiFourthB}(1^+_{\vphantom{\bar \psi_n}\psi_m} 2^+_{\,\bar \psi_n} 3^-_{ \vphantom{\bar \psi_n} \psi_p} 4^-_{\,\bar \psi_r})_{[1]}&=  0    \,.
\end{align}
The amplitudes with one insertion of the $F^3$ operator are
\begin{align}
A^{(1)}_{F^3}(1^+_{\vphantom{\bar \psi_n}\psi_m} 2^-_{\,\bar \psi_n} 3^+_{ \vphantom{\bar \psi_n} \psi_p} 4^-_{\,\bar \psi_r})_{[1]}&=\frac{1}{3} g\tilde g^2  u\delta_{mn}\delta_{pr}\,,\nonumber\\
A^{(1)}_{F^3}(1^+_{\vphantom{\bar \psi_n}\psi_m} 2^-_{\,\bar \psi_n} 3^-_{ \vphantom{\bar \psi_n} \psi_p} 4^+_{\,\bar \psi_r})_{[1]}&=-\frac{1}{3} g\tilde g^2  t\delta_{mn}\delta_{pr} \,,\nonumber\\
A^{(1)}_{F^3}(1^+_{\vphantom{\bar \psi_n}\psi_m} 2^+_{\,\bar \psi_n} 3^-_{ \vphantom{\bar \psi_n} \psi_p} 4^-_{\,\bar \psi_r})_{[1]}&= 0  \,.
\end{align}
The amplitudes with one insertion of a $D\varphi^2\psi^2$ operator are
\begin{align}
A^{(1)}_{\DPhiSquarePsiSquareA}(1^\pm_{\vphantom{\bar \psi_n}\psi_m} 2^\pm_{\,\bar \psi_n} 3^\pm_{ \vphantom{\bar \psi_n} \psi_p} 4^\pm_{\,\bar \psi_r})_{[1]}&=  0   \,,\nonumber\\
A^{(1)}_{\DPhiSquarePsiSquareB}(1^+_{\vphantom{\bar \psi_n}\psi_m} 2^-_{\,\bar \psi_n} 3^+_{ \vphantom{\bar \psi_n} \psi_p} 4^-_{\,\bar \psi_r})_{[1]}&=-\frac{1}{9} \tilde g^2 N_s (3 X_s+8) u (\opcoef_{\DPhiSquarePsiSquareB}^{rp} \delta_{mn}+\opcoef_{\DPhiSquarePsiSquareB}^{nm} \delta_{pr}) \,,\nonumber\\
A^{(1)}_{\DPhiSquarePsiSquareB}(1^+_{\vphantom{\bar \psi_n}\psi_m} 2^-_{\,\bar \psi_n} 3^-_{ \vphantom{\bar \psi_n} \psi_p} 4^+_{\,\bar \psi_r})_{[1]}&= \frac{1}{9} \tilde g^2 N_s (3 X_s+8) t (\opcoef_{\DPhiSquarePsiSquareB}^{rp} \delta_{mn}+\opcoef_{\DPhiSquarePsiSquareB}^{nm} \delta_{pr})  \,,\nonumber\\
A^{(1)}_{\DPhiSquarePsiSquareB}(1^+_{\vphantom{\bar \psi_n}\psi_m} 2^+_{\,\bar \psi_n} 3^-_{ \vphantom{\bar \psi_n} \psi_p} 4^-_{\,\bar \psi_r})_{[1]}&= 0    \,.
\end{align}
The amplitudes with one insertion of a $\psi^4$ operator are
\begin{align}
A^{(1)\rm fin}_{\PsiFourthA}(1^+_{\vphantom{\bar \psi_n}\psi_m} 2^-_{\,\bar \psi_n} 3^+_{ \vphantom{\bar \psi_n} \psi_p} 4^-_{\,\bar \psi_r})_{[1]}&=  \frac{2 \tilde g^2 u}{9 t}  \Big(t \big(72 N \opcoef_{\PsiFourthA}^{rmnp}
\nonumber\\&\qquad+N_f (3 X_s+2) (\delta_{mn} \opcoef_{\PsiFourthA}^{rwwp}+\delta_{pr} \opcoef_{\PsiFourthA}^{nwwm})\big)
\nonumber\\&\qquad+9 (2 s+t (3 X_u+25)) \opcoef_{\PsiFourthA}^{nmrp}\Big)\,,\nonumber\\
A^{(1)\rm fin}_{\PsiFourthA}(1^+_{\vphantom{\bar \psi_n}\psi_m} 2^-_{\,\bar \psi_n} 3^-_{ \vphantom{\bar \psi_n} \psi_p} 4^+_{\,\bar \psi_r})_{[1]}&=   -\frac{2}{9} \tilde g^2 (N_f t (3 X_s+2) (\delta_{mn} \opcoef_{\PsiFourthA}^{rwwp}+\delta_{pr} \opcoef_{\PsiFourthA}^{nwwm})
\nonumber\\&\qquad+9 (2 s+t (5-3 X_t)) \opcoef_{\PsiFourthA}^{nmrp}) \,,\nonumber\\
A^{(1)\rm fin}_{\PsiFourthA}(1^+_{\vphantom{\bar \psi_n}\psi_m} 2^+_{\,\bar \psi_n} 3^-_{ \vphantom{\bar \psi_n} \psi_p} 4^-_{\,\bar \psi_r})_{[1]}&= -16 \tilde g^2 N s \opcoef_{\PsiFourthA}^{rmnp}  \,,\nonumber\\
A^{(1)\rm fin}_{\PsiFourthB}(1^+_{\vphantom{\bar \psi_n}\psi_m} 2^-_{\,\bar \psi_n} 3^+_{ \vphantom{\bar \psi_n} \psi_p} 4^-_{\,\bar \psi_r})_{[1]}&= \frac{2 \tilde g^2 u}{9 N s t} \Big(9 s \left(2 \left(N^2-1\right) s+t \left(13 N^2-3 X_u-25\right)\right) \opcoef_{\PsiFourthB}^{nmrp}
\nonumber\\&\qquad+t (N_f s (2 N (3 X_s+5) (\delta_{mn} \opcoef_{\PsiFourthB}^{rpww}+\delta_{pr} \opcoef_{\PsiFourthB}^{nmww})
\nonumber\\&\qquad-(3 X_s+2) (\delta_{mn} \opcoef_{\PsiFourthB}^{rwwp}+\delta_{pr} \opcoef_{\PsiFourthB}^{nwwm}))
\nonumber\\&\qquad+9 N (s (3 X_u+17)+2 t) \opcoef_{\PsiFourthB}^{rmnp})\Big)   \,,\nonumber\\
A^{(1)\rm fin}_{\PsiFourthB}(1^+_{\vphantom{\bar \psi_n}\psi_m} 2^-_{\,\bar \psi_n} 3^-_{ \vphantom{\bar \psi_n} \psi_p} 4^+_{\,\bar \psi_r})_{[1]}&=  -\frac{2 \tilde g^2}{9 N} \Big(9 \left(2 \left(N^2-1\right) s-t \left(3 (N^2-1) X_t -3N^2 +5\right)\right) \opcoef_{\PsiFourthB}^{nmrp}
\nonumber\\&\qquad+N_f t (2 N (3 X_s+5) (\delta_{mn} \opcoef_{\PsiFourthB}^{rpww}+\delta_{pr} \opcoef_{\PsiFourthB}^{nmww})
\nonumber\\&\qquad-(3 X_s+2) (\delta_{mn} \opcoef_{\PsiFourthB}^{rwwp}+\delta_{pr} \opcoef_{\PsiFourthB}^{nwwm}))\Big)  \,,\nonumber\\
A^{(1)\rm fin}_{\PsiFourthB}(1^+_{\vphantom{\bar \psi_n}\psi_m} 2^+_{\,\bar \psi_n} 3^-_{ \vphantom{\bar \psi_n} \psi_p} 4^-_{\,\bar \psi_r})_{[1]}&= 2 \tilde g^2 (3 s (X_s+1)-2 t) \opcoef_{\PsiFourthB}^{rmnp}  \,.
\end{align}

\subsection{Four-scalar amplitudes}
The color structures for this process are identical to those of the four fermion case:
\begin{align}
\mathcal{C}_{ssss}^{[1]}=T^a_{i_2i_1}T^a_{i_4i_3}\,,\qquad
\mathcal{C}_{ssss}^{[2]}=T^a_{i_4i_1}T^a_{i_2i_3}\,.
\end{align}
There is no spinor phase in this case, as the scalars do not carry helicity weight.
The tree-level amplitudes are
\begin{align}
A^{(0)}(1_{\vphantom{\bar\varphi}\varphi}2_{\bar\varphi}3_{\vphantom{\bar\varphi}\varphi}4_{\bar\varphi})_{[1]}&=  -\frac{ g^2 (t-u)}{2 s}-\frac{2 \lambda  N}{N-1} \,,\nonumber\\
A^{(0)}_{\DSquarePhiFourthA}(1_{\vphantom{\bar\varphi}\varphi}2_{\bar\varphi}3_{\vphantom{\bar\varphi}\varphi}4_{\bar\varphi})_{[1]}&=  \frac{N (N s+t)}{N^2-1}  \,,\nonumber\\
A^{(0)}_{\DSquarePhiFourthB}(1_{\vphantom{\bar\varphi}\varphi}2_{\bar\varphi}3_{\vphantom{\bar\varphi}\varphi}4_{\bar\varphi})_{[1]}&=  \frac{2 N (N t+s)}{N^2-1}     \,.
\end{align}
The one-loop amplitudes with an insertion of the $F^3$ operator are
\begin{align}
A^{(1)}_{F^3}(1_{\vphantom{\bar\varphi}\varphi}2_{\bar\varphi}3_{\vphantom{\bar\varphi}\varphi}4_{\bar\varphi})_{[1]}&=-\frac{1}{6}   g\tilde g^2 N (t-u)  \,.
\end{align}
The one-loop amplitudes with an insertion of a $\varphi^2F^2$ operator are
\begin{align}
A^{(1)}_{\PhiSquareFSquareA}(1_{\vphantom{\bar\varphi}\varphi}2_{\bar\varphi}3_{\vphantom{\bar\varphi}\varphi}4_{\bar\varphi})_{[1]}&=2   \tilde g^2 (N t+s)\,,\nonumber\\
A^{(1)}_{\PhiSquareFSquareB}(1_{\vphantom{\bar\varphi}\varphi}2_{\bar\varphi}3_{\vphantom{\bar\varphi}\varphi}4_{\bar\varphi})_{[1]}&=\frac{2   \tilde g^2 \left(N^2-4\right) s}{N}  \,.
\end{align}
The one-loop amplitudes with an insertion of a $D^2\varphi^4$ operator are
\begin{align}
A^{(1)\rm fin}_{\DSquarePhiFourthA}(1_{\vphantom{\bar\varphi}\varphi}2_{\bar\varphi}3_{\vphantom{\bar\varphi}\varphi}4_{\bar\varphi})_{[1]}&= 
\frac{\tilde g^2}{2} \Big(-4(4 N+3) s-2(3 N+5) t
\nonumber\\&\qquad-3(N-2) t X_t-3 s X_s+3 u X_u\Big)
\nonumber\\&\quad+\frac{2\tilde\lambda}{N-1}  \Big(2 N ((N-3) t-2 s)-N s X_s
\nonumber\\&\qquad+(N-2) N t X_t+N u X_u\Big)
\,,\nonumber\\
A^{(1)\rm fin}_{\DSquarePhiFourthB}(1_{\vphantom{\bar\varphi}\varphi}2_{\bar\varphi}3_{\vphantom{\bar\varphi}\varphi}4_{\bar\varphi})_{[1]}&=
\frac{\tilde g^2}{9} \Big(-2 (2 t (9 N+4 N_s+27)+(4 N_s+45) s)
\nonumber\\&\qquad+27 (2 N-1) t X_t-3 X_s ((N_s-18) s+2 N_s t)+27 u X_u)
\nonumber\\&\qquad+\frac{4\tilde\lambda N}{N-1}  (-4 (N t+s)+(1-2 N) t X_t-s X_s+u X_u)
\,.
\end{align}
The one-loop amplitudes with an insertion of a $D\varphi^2\psi^2$ operator are
\begin{align}
A^{(1)}_{\DPhiSquarePsiSquareA}(1_{\vphantom{\bar\varphi}\varphi}2_{\bar\varphi}3_{\vphantom{\bar\varphi}\varphi}4_{\bar\varphi})_{[1]}&=0\,,\nonumber\\
A^{(1)}_{\DPhiSquarePsiSquareB}(1_{\vphantom{\bar\varphi}\varphi}2_{\bar\varphi}3_{\vphantom{\bar\varphi}\varphi}4_{\bar\varphi})_{[1]}&=   -\frac{2}{9} \opcoef_{\DPhiSquarePsiSquareB}^{ww} \tilde g^2 (3 X_s+5) (t-u)  \,.
\end{align}

\subsection{Two-fermion, two-vector amplitudes}
The color factors for the two-fermion, two-vector amplitudes are
\begin{equation}
\mathcal{C}_{ffvv}^{[1]}=(T^3T^4)_{i_2i_1}\,,\hskip 1 cm
\mathcal{C}_{ffvv}^{[2]}=(T^4T^3)_{i_2i_1}\,,\hskip 1 cm
\mathcal{C}_{ffvv}^{[3]}=\Tr[T^3T^4]\delta_{i_2i_1}\,.
\end{equation}
In this case the spinor prefactors are not pure phases, but have magnitudes equal to ratios of $s$, $t$, and $u$:
\begin{align}\
  S(1^-_{\vphantom{\bar \psi_n}\psi_p} 2^+_{\,\bar \psi_r} 3^+4^+)&=\frac{\langle 1 3\rangle  [3 4]}{\langle 2 3\rangle  \langle 3 4\rangle }\,,\hskip 1.7 cm
S(1^-_{\vphantom{\bar \psi_n}\psi_p} 2^+_{\,\bar \psi_r} 3^-4^+)=\frac{\langle 1 3\rangle ^3}{\langle 1 2\rangle  \langle 3 4\rangle  \langle 4 1\rangle }\,,\nonumber\\
S(1^-_{\vphantom{\bar \psi_n}\psi_p} 2^+_{\,\bar \psi_r} 3^+4^-)&=\frac{\langle 1 4\rangle ^3}{\langle 1 2\rangle  \langle 3 1\rangle  \langle 4 3\rangle }\,,\hskip 1 cm
S(1^-_{\vphantom{\bar \psi_n}\psi_p} 2^+_{\,\bar \psi_r} 3^-4^-)=\frac{\langle 3 4\rangle ^3}{\langle 2 3\rangle  \langle 2 4\rangle  [1 2]}  \,.
\end{align}
The tree-level amplitudes for this process are
\begin{align}
A^{(0)}(1_{\vphantom{\bar \psi_n}\psi_p} 2_{\,\bar \psi_r} 34)_{[1]}&= -\frac{g^2}{s t} (2T^{-+-+}_{ffvv}-2T^{-++-}_{ffvv}+T^{\rm ev}_{ffvv})  \delta_{pr} \,,\nonumber\\
A^{(0)}_{F^3}(1_{\vphantom{\bar \psi_n}\psi_p} 2_{\,\bar \psi_r} 34)_{[1]}&= -\frac{2 g }{s} (T^{-+++}_{ffvv}+T^{-+--}_{ffvv}) \delta_{pr} \,,
\end{align}
which evaluate in four dimensions as
\begin{align}
A^{(0)}(1^-_{\vphantom{\bar \psi_n}\psi_p} 2^+_{\,\bar \psi_r} 3^+4^+)_{[1]}&=0   \,,\nonumber\\
A^{(0)}(1^-_{\vphantom{\bar \psi_n}\psi_p} 2^+_{\,\bar \psi_r} 3^+4^-)_{[1]}&=  g^2 \delta_{pr}\,,\nonumber\\
A^{(0)}(1^-_{\vphantom{\bar \psi_n}\psi_p} 2^+_{\,\bar \psi_r} 3^-4^+)_{[1]}&=   \frac{ g^2 t}{u}\delta_{pr}\,,\nonumber\\
A^{(0)}(1^-_{\vphantom{\bar \psi_n}\psi_p} 2^+_{\,\bar \psi_r} 3^-4^-)_{[1]}&=  0 \,,\nonumber\\
A^{(0)}_{F^3}(1^-_{\vphantom{\bar \psi_n}\psi_p} 2^+_{\,\bar \psi_r} 3^+4^+)_{[1]}&=  -g   t  \delta_{pr} \,,\nonumber\\
A^{(0)}_{F^3}(1^-_{\vphantom{\bar \psi_n}\psi_p} 2^+_{\,\bar \psi_r} 3^+4^-)_{[1]}&= 0    \,,\nonumber\\
A^{(0)}_{F^3}(1^-_{\vphantom{\bar \psi_n}\psi_p} 2^+_{\,\bar \psi_r} 3^-4^+)_{[1]}&=   0  \,,\nonumber\\
A^{(0)}_{F^3}(1^-_{\vphantom{\bar \psi_n}\psi_p} 2^+_{\,\bar \psi_r} 3^-4^-)_{[1]}&=  \frac{ g  t u}{s}  \delta_{pr} \,.
\end{align}
The one-loop amplitudes with an insertion of a $F^3$ operator are
\begin{align}
A^{(1)\rm fin}_{F^3}(1^-_{\vphantom{\bar \psi_n}\psi_p} 2^+_{\,\bar \psi_r} 3^+4^+)_{[1]}&=
\frac{g\tilde g^2\delta_{pr}}{36Nu} \Big(2 t u \big(34 N^2+N(5 N_f+2 N_s)-18\big)
  \nonumber\\&\qquad+ 9N t u ( (4 N_f+N_s) X_s  + 2 (N-b_0) X_t )  
\nonumber\\&\qquad+18   N^2 (t-u)t X^2 \Big) \,,\nonumber\\
A^{(1)\rm fin}_{F^3}(1^-_{\vphantom{\bar \psi_n}\psi_p} 2^+_{\,\bar \psi_r} 3^-4^+)_{[1]}&=0\,,\nonumber\\
A^{(1)}_{F^3}(1^-_{\vphantom{\bar \psi_n}\psi_p} 2^+_{\,\bar \psi_r} 3^+4^-)_{[1]}&=g\tilde g^2 \delta_{pr} N\frac{ s u}{ t}\,,\nonumber\\
A^{(1)\rm fin}_{F^3}(1^-_{\vphantom{\bar \psi_n}\psi_p} 2^+_{\,\bar \psi_r} 3^-4^-)_{[1]}&=-\frac{u}{s}A^{(1)\rm fin}_{F^3}(1^-_{\vphantom{\bar \psi_n}\psi_p} 2^+_{\,\bar \psi_r} 3^+4^+)_{[1]}\,,
\end{align}
\begin{align} 
A^{(1)\rm fin}_{F^3}(1^-_{\vphantom{\bar \psi_n}\psi_p} 2^+_{\,\bar \psi_r} 3^+4^+)_{[3]}&=g \tilde g^2\delta_{pr} \bigg(\frac{(3N+b_0)}{2 N} t (X_u-X_t)
\nonumber\\&\qquad+\frac{  (t-u) }{2 su} (st X^2 +su Y^2 + ut Z^2 ) \bigg) \,,\nonumber\\
A^{(1)}_{F^3}(1^-_{\vphantom{\bar \psi_n}\psi_p} 2^+_{\,\bar \psi_r} 3^-4^+)_{[3]}&= g\tilde g^2 \delta_{pr}2\frac{ s t}{u}    \,,\nonumber\\
A^{(1)}_{F^3}(1^-_{\vphantom{\bar \psi_n}\psi_p} 2^+_{\,\bar \psi_r} 3^+4^-)_{[3]}&=g\tilde g^2 \delta_{pr}2\frac{ s u}{t} \,,\nonumber\\
A^{(1)\rm fin}_{F^3}(1^-_{\vphantom{\bar \psi_n}\psi_p} 2^+_{\,\bar \psi_r} 3^-4^-)_{[3]}&= -\frac{u}{s}A^{(1)\rm fin}_{F^3}(1^-_{\vphantom{\bar \psi_n}\psi_p} 2^+_{\,\bar \psi_r} 3^+4^+)_{[3]}\,.
\end{align}
The one-loop amplitudes with an insertion of a $\varphi^2F^2$ operator all evaluate to zero:
\begin{align}
A^{(1)}_{\PhiSquareFSquareA}(1^\pm_{\vphantom{\bar \psi_n}\psi_p} 2^\pm_{\,\bar \psi_r} 3^\pm 4^\pm )_{[1]}&=A^{(1)}_{\PhiSquareFSquareA}(1^\pm_{\vphantom{\bar \psi_n}\psi_p} 2^\pm_{\,\bar \psi_r} 3^\pm 4^\pm )_{[3]}=0\,,\nonumber\\
A^{(1)}_{\PhiSquareFSquareB}(1^\pm_{\vphantom{\bar \psi_n}\psi_p} 2^\pm_{\,\bar \psi_r} 3^\pm 4^\pm )_{[1]}&=A^{(1)}_{\PhiSquareFSquareB}(1^\pm_{\vphantom{\bar \psi_n}\psi_p} 2^\pm_{\,\bar \psi_r} 3^\pm 4^\pm )_{[3]}=0  \,.
\end{align}
The one-loop amplitudes with an insertion of a $D\varphi^2\psi^2$ operator are
\begin{align}
A^{(1)}_{\DPhiSquarePsiSquareA}(1^\pm_{\vphantom{\bar \psi_n}\psi_p} 2^\pm_{\,\bar \psi_r} 3^\pm 4^\pm )_{[1]}&=A^{(1)}_{\DPhiSquarePsiSquareA}(1^\pm_{\vphantom{\bar \psi_n}\psi_p} 2^\pm_{\,\bar \psi_r} 3^\pm 4^\pm )_{[3]}=0\,,\nonumber\\
A^{(1)}_{\DPhiSquarePsiSquareB}(1^-_{\vphantom{\bar \psi_n}\psi_p} 2^+_{\,\bar \psi_r} 3^+4^+)_{[1]}&=\frac{1}{3} \tilde g^2\opcoef_{\DPhiSquarePsiSquareB}^{rp}N_s t\,,\nonumber\\
A^{(1)}_{\DPhiSquarePsiSquareB}(1^-_{\vphantom{\bar \psi_n}\psi_p} 2^+_{\,\bar \psi_r} 3^+4^-)_{[1]}&=A^{(1)}_{\DPhiSquarePsiSquareB}(1^-_{\vphantom{\bar \psi_n}\psi_p} 2^+_{\,\bar \psi_r} 3^-4^+)_{[1]}=0\,,\nonumber\\
A^{(1)}_{\DPhiSquarePsiSquareB}(1^-_{\vphantom{\bar \psi_n}\psi_p} 2^+_{\,\bar \psi_r} 3^-4^-)_{[1]}&=-\frac{1}{3 s} \tilde g^2 \opcoef_{\DPhiSquarePsiSquareB}^{rp}N_s t u\,,\nonumber\\
A^{(1)}_{\DPhiSquarePsiSquareB}(1^\pm_{\vphantom{\bar \psi_n}\psi_p} 2^\pm_{\,\bar \psi_r} 3^\pm 4^\pm )_{[3]}&=0  \,.
\end{align}
The one-loop amplitudes with an insertion of a $\psi^4$ operator are
\begin{align}
A^{(1)}_{\PsiFourthA}(1^\pm_{\vphantom{\bar \psi_n}\psi_p} 2^\pm_{\,\bar \psi_r} 3^\pm 4^\pm )_{[1],[3]}&=\frac{ N_f}{N_s}\frac{\opcoef_{\PsiFourthA}^{rwwp}}{\opcoef_{\DPhiSquarePsiSquareB}^{rp}}A^{(1)}_{\DPhiSquarePsiSquareB}(1^\pm_{\vphantom{\bar \psi_n}\psi_p} 2^\pm_{\,\bar \psi_r} 3^\pm 4^\pm )_{[1],[3]}\,,\nonumber\\
A^{(1)}_{\PsiFourthB}(1^\pm_{\vphantom{\bar \psi_n}\psi_p} 2^\pm_{\,\bar \psi_r} 3^\pm 4^\pm )_{[1],[3]}&=\frac{2N\opcoef_{\PsiFourthB}^{rpww}-\opcoef_{\PsiFourthB}^{rwwp}}{\opcoef_{\PsiFourthA}^{rwwp}}A^{(1)}_{\PsiFourthA}(1^\pm_{\vphantom{\bar \psi_n}\psi_p} 2^\pm_{\,\bar \psi_r} 3^\pm 4^\pm )_{[1],[3]}  \,.
\end{align}

\subsection{Two-scalar, two-vector amplitudes}
The color basis for this process is analogous to the that of the previous:
\begin{equation}
\mathcal{C}_{vvss}^{[1]}=(T^1T^2)_{i_4i_3}\,,\hskip 1 cm
\mathcal{C}_{vvss}^{[2]}=(T^2T^1)_{i_4i_3}\,,\hskip 1 cm
\mathcal{C}_{vvss}^{[3]}=\Tr[T^1T^2]\delta_{i_4i_3}\,.
\end{equation}
The spinor factors are again pure phases:
\begin{equation}
  S(1^+2^+3_{\vphantom{\bar\varphi}\varphi}4_{\bar\varphi})=\frac{[1 2]}{\langle 1 2\rangle }\,,\quad\quad
  S(1^+2^-3_{\vphantom{\bar\varphi}\varphi}4_{\bar\varphi})=\frac{\langle 2 3\rangle  \langle 2 4\rangle  [1 2] [3 4]}{\langle 1 2\rangle  \langle 3 4\rangle  [2 3] [2 4]}\,.
\end{equation}
The $D$-dimensional tree-level expressions are given by
\begin{align}
A^{(1)}(123_{\vphantom{\bar\varphi}\varphi}4_{\bar\varphi})_{[1]}&=  -\frac{g^2}{st}T^{+-}_{vvss} \,,\nonumber\\
A^{(1)}(123_{\vphantom{\bar\varphi}\varphi}4_{\bar\varphi})_{[3]}&=  0 \,,\nonumber\\
A^{(1)}_{F^3}(123_{\vphantom{\bar\varphi}\varphi}4_{\bar\varphi})_{[1]}&=  \frac{g  (t-u)}{2 s} T^{++}_{vvss} \,,\nonumber\\
A^{(1)}_{F^3}(123_{\vphantom{\bar\varphi}\varphi}4_{\bar\varphi})_{[3]}&= 0   \,,\nonumber\\
A^{(1)}_{\PhiSquareFSquareA}(123_{\vphantom{\bar\varphi}\varphi}4_{\bar\varphi})_{[1]}&= 0  \,,\nonumber\\
A^{(1)}_{\PhiSquareFSquareA}(123_{\vphantom{\bar\varphi}\varphi}4_{\bar\varphi})_{[3]}&= -2  T^{++}_{vvss}  \,,\nonumber\\
A^{(1)}_{\PhiSquareFSquareB}(123_{\vphantom{\bar\varphi}\varphi}4_{\bar\varphi})_{[1]}&= -2  T^{++}_{vvss}    \,,\nonumber\\
A^{(1)}_{\PhiSquareFSquareB}(123_{\vphantom{\bar\varphi}\varphi}4_{\bar\varphi})_{[3]}&=-\frac{4}{N}  T^{++}_{vvss}    \,,
\end{align}
with four-dimensional helicity values
\begin{align}
A^{(1)}(1^+2^+3_{\vphantom{\bar\varphi}\varphi}4_{\bar\varphi})_{[1]}&= 0  \,,\nonumber\\
A^{(1)}(1^+2^-3_{\vphantom{\bar\varphi}\varphi}4_{\bar\varphi})_{[1]}&=\frac{g^2 u}{s}   \,,\nonumber\\
A^{(1)}(1^\pm 2^\pm 3_{\vphantom{\bar\varphi}\varphi}4_{\bar\varphi})_{[3]}&=  0 \,,\nonumber\\
A^{(1)}_{F^3}(1^+2^+3_{\vphantom{\bar\varphi}\varphi}4_{\bar\varphi})_{[1]}&=  \frac{1}{2} g   (t-u)  \,,\nonumber\\
A^{(1)}_{F^3}(1^+2^-3_{\vphantom{\bar\varphi}\varphi}4_{\bar\varphi})_{[1]}&=   0 \,,\nonumber\\
A^{(1)}_{F^3}(1^\pm 2^\pm 3_{\vphantom{\bar\varphi}\varphi}4_{\bar\varphi})_{[3]}&=  0  \,,\nonumber\\
A^{(1)}_{\PhiSquareFSquareA}(1^\pm 2^\pm 3_{\vphantom{\bar\varphi}\varphi}4_{\bar\varphi})_{[1]}&= 0  \,,\nonumber\\
A^{(1)}_{\PhiSquareFSquareA}(1^+2^+3_{\vphantom{\bar\varphi}\varphi}4_{\bar\varphi})_{[3]}&= -2s   \,,\nonumber\\
A^{(1)}_{\PhiSquareFSquareA}(1^+2^-3_{\vphantom{\bar\varphi}\varphi}4_{\bar\varphi})_{[3]}&=  0 \,,\nonumber\\
A^{(1)}_{\PhiSquareFSquareB}(1^+2^+3_{\vphantom{\bar\varphi}\varphi}4_{\bar\varphi})_{[1]}&=  -2s    \,,\nonumber\\
A^{(1)}_{\PhiSquareFSquareB}(1^+2^-3_{\vphantom{\bar\varphi}\varphi}4_{\bar\varphi})_{[1]}&= 0   \,,\nonumber\\
A^{(1)}_{\PhiSquareFSquareB}(1^+2^+3_{\vphantom{\bar\varphi}\varphi}4_{\bar\varphi})_{[3]}&=  -\frac{4s}{N}    \,,\nonumber\\
A^{(1)}_{\PhiSquareFSquareB}(1^+2^-3_{\vphantom{\bar\varphi}\varphi}4_{\bar\varphi})_{[3]}&=0  \,.
\end{align}
The one-loop amplitudes with an insertion of the $F^3$ operator are
\begin{align}
  A^{(1)\rm fin}_{F^3}(1^+2^+3_{\vphantom{\bar\varphi}\varphi}4_{\bar\varphi})_{[1]}&=-\frac{g\tilde g^2}{72N} \Big( 8 ( ( 52 N^2 -18) s +  ( 77 N^2-36) t +  N (t-u) (5 N_f + 2 N_s))
    \nonumber\\&\qquad+18 N ( (2 N (5 t - 7 u) - 3 b_0 (t - u))  X_t + X_s (2 N s  + b_0 (u-t)))
\nonumber\\&\qquad-72 N^2 t X^2    \Big)\,,\nonumber\\
A^{(1)}_{F_3}(1^+2^-3_{\vphantom{\bar\varphi}\varphi}4_{\bar\varphi})_{[1]}&=\frac{1}{2} g\tilde g^2  N u\,,\nonumber\\
A^{(1)\rm fin}_{F^3}(1^+2^+3_{\vphantom{\bar\varphi}\varphi}4_{\bar\varphi})_{[3]}&=  \frac{g\tilde g^2}{4N} \Big((8 N t + b_0 (t - u)) X_t  
+ (8 N u + b_0 (u-t)) X_u -4Ns X_s
\nonumber\\&\qquad+\frac{1}{6s}(stX^2+suY^2+t u Z^2)\Big) \,,\nonumber\\
A^{(1)}_{F_3}(1^+2^-3_{\vphantom{\bar\varphi}\varphi}4_{\bar\varphi})_{[3]}&=-g\tilde g^2  s  \,.
\end{align}
The one-loop amplitudes with an insertion of a $\varphi^2F^2$ operator are
\begin{align}
A^{(1)\rm fin}_{\PhiSquareFSquareA}(1^+2^+3_{\vphantom{\bar\varphi}\varphi}4_{\bar\varphi})_{[1]}&=
-\frac{  \tilde g^2 s}{ N}( (b_0-2N) X_t +  b_0  X_u ) +4 \tilde g^2 s
\,,\\
A^{(1)}_{\PhiSquareFSquareA}(1^+2^-3_{\vphantom{\bar\varphi}\varphi}4_{\bar\varphi})_{[1]}&=  2\tilde g^2  (s+3 t) \,,\nonumber\\
A^{(1)\rm fin}_{\PhiSquareFSquareA}(1^+2^+3_{\vphantom{\bar\varphi}\varphi}4_{\bar\varphi})_{[3]}&=
\tilde g^2 s \left( 4  C_F -2(b_0+3C_F)X_s \right)
\nonumber\\&\qquad+4   \tilde\lambda  (N+1) s (X_s+2)\,,\nonumber\\
A^{(1)}_{\PhiSquareFSquareA}(1^+2^-3_{\vphantom{\bar\varphi}\varphi}4_{\bar\varphi})_{[3]}&= 0\,,\nonumber\\
A^{(1)\rm fin}_{\PhiSquareFSquareB}(1^+2^+3_{\vphantom{\bar\varphi}\varphi}4_{\bar\varphi})_{[1]}&=
\frac{\tilde g^2 s}{N^2} \Big( 6 N \left(2 N^2-3\right)
+  N   \left(3-Nb_0\right) X_s +2 b_0  X_u 
\nonumber\\&\qquad+  \big(2 N (N^2-4) -  b_0 ( N^2-2) \big) X_t\Big) 
\nonumber\\&\qquad+4   \tilde\lambda  s (X_s+2) \,,\nonumber\\ 
A^{(1)}_{\PhiSquareFSquareB}(1^+2^-3_{\vphantom{\bar\varphi}\varphi}4_{\bar\varphi})_{[1]}&=-\frac{2 \tilde g^2  }{ N} \left(N^2 u+4 t\right)\,,\nonumber\\
A^{(1)\rm fin}_{\PhiSquareFSquareB}(1^+2^+3_{\vphantom{\bar\varphi}\varphi}4_{\bar\varphi})_{[3]}&=\frac{\tilde g^2 s}{N^2} \Big( 2 \left(b_0 N-3\right) X_s  +b_0 N  (X_t+X_u) 
-3 \left(4 N^2-1\right) \Big)
\nonumber\\&\qquad -\frac{8}{N}   \tilde\lambda  s (X_s+2)
\,,\nonumber\\
A^{(1)}_{\PhiSquareFSquareB}(1^+2^-3_{\vphantom{\bar\varphi}\varphi}4_{\bar\varphi})_{[3]}&=-4\tilde g^2   s  \,.
\end{align}
The one-loop amplitudes with an insertion of a $D^2\varphi^4$ operator are
\begin{align}
A^{(1)}_{\DSquarePhiFourthA}(1^+2^+3_{\vphantom{\bar\varphi}\varphi}4_{\bar\varphi})_{[1]}&=-\frac{1}{2} \tilde g^2   N_s s\,,\nonumber\\
A^{(1)}_{\DSquarePhiFourthA}(1^+2^-3_{\vphantom{\bar\varphi}\varphi}4_{\bar\varphi})_{[1]}&=0\,,\nonumber\\
A^{(1)}_{\DSquarePhiFourthA}(1^\pm 2^\pm 3_{\vphantom{\bar\varphi}\varphi}4_{\bar\varphi})_{[3]}&=-A^{(1)}_{\DSquarePhiFourthA}(1^\pm 2^\pm 3_{\vphantom{\bar\varphi}\varphi}4_{\bar\varphi})_{[1]}\,,\nonumber\\
A^{(1)}_{\DSquarePhiFourthB}(1^+2^+3_{\vphantom{\bar\varphi}\varphi}4_{\bar\varphi})_{[1]}&=\frac{1}{3} \tilde g^2   N_s (s-t)\,,\nonumber\\
A^{(1)}_{\DSquarePhiFourthB}(1^+2^-3_{\vphantom{\bar\varphi}\varphi}4_{\bar\varphi})_{[1]}&=0\,,\nonumber\\
A^{(1)}_{\DSquarePhiFourthB}(1^\pm 2^\pm 3_{\vphantom{\bar\varphi}\varphi}4_{\bar\varphi})_{[3]}&=4 A^{(1)}_{\DSquarePhiFourthA}(1^\pm 2^\pm 3_{\vphantom{\bar\varphi}\varphi}4_{\bar\varphi})_{[1]}  \,.
\end{align}
The one-loop amplitudes with an insertion of a $D\varphi^2\psi^2$ operator are
\begin{align}
A^{(1)}_{\DPhiSquarePsiSquareA}(1^\pm 2^\pm 3_{\vphantom{\bar\varphi}\varphi}4_{\bar\varphi})_{[1]}&=A^{(1)}_{\DPhiSquarePsiSquareA}(1^\pm 2^\pm 3_{\vphantom{\bar\varphi}\varphi}4_{\bar\varphi})_{[3]}=0\,,\nonumber\\
A^{(1)}_{\DPhiSquarePsiSquareB}(1^+2^+3_{\vphantom{\bar\varphi}\varphi}4_{\bar\varphi})_{[1]}&=\frac{1}{3} \tilde g^2\opcoef_{\DPhiSquarePsiSquareB}^{ww}N_f (t-u)\,,\nonumber\\
A^{(1)}_{\DPhiSquarePsiSquareB}(1^+2^-3_{\vphantom{\bar\varphi}\varphi}4_{\bar\varphi})_{[1]}&=A^{(1)}_{\DPhiSquarePsiSquareB}(1^\pm 2^\pm 3_{\vphantom{\bar\varphi}\varphi}4_{\bar\varphi})_{[3]}=0  \,.
\end{align}

\subsection{Two-fermion, two-scalar amplitudes}
The color structures for this process are identical to those of the four fermion case:
\begin{align}
\mathcal{C}_{ffss}^{[1]}=T^a_{i_2i_1}T^a_{i_4i_3}\,,\qquad
\mathcal{C}_{ffss}^{[2]}=T^a_{i_4i_1}T^a_{i_2i_3}\,.
\end{align}
There is only one independent spinor prefactor (which again is not a pure phase for this case):
\begin{align}\
S(1_{\vphantom{\bar \psi}\psi} 2_{\,\bar \psi} 3_{\vphantom{\bar\varphi}\varphi}4_{\bar\varphi})&=\frac{\langle 2 3\rangle  [1 3]}{s}  \,.
\end{align}
The tree-level amplitudes for this process are given by
\begin{align}
A^{(0)}(1_{\vphantom{\bar \psi_n}\psi_p} 2_{\,\bar \psi_r} 3_{\vphantom{\bar\varphi}\varphi}4_{\bar\varphi})_{[1]}&= g^2\frac{\bar{u}_{2}\slashed k_{3}u_{1}}{s}\delta_{pr} \,,\nonumber\\
A^{(0)}(1_{\vphantom{\bar \psi_n}\psi_p} 2_{\,\bar \psi_r} 3_{\vphantom{\bar\varphi}\varphi}4_{\bar\varphi})_{[2]}&= 0  \,,\nonumber\\
A^{(0)}_{\DPhiSquarePsiSquareA}(1_{\vphantom{\bar \psi_n}\psi_p} 2_{\,\bar \psi_r} 3_{\vphantom{\bar\varphi}\varphi}4_{\bar\varphi})_{[1]}&=  -\frac{2 \opcoef_{\DPhiSquarePsiSquareA}^{rp} N (\bar{u}_{2}\slashed  k_{3}u_{1})}{N^2-1}    \,,\nonumber\\
A^{(0)}_{\DPhiSquarePsiSquareA}(1_{\vphantom{\bar \psi_n}\psi_p} 2_{\,\bar \psi_r} 3_{\vphantom{\bar\varphi}\varphi}4_{\bar\varphi})_{[2]}&= -\frac{2\opcoef_{\DPhiSquarePsiSquareA}^{rp} N^2 (\bar{u}_{2}\slashed  k_{3}u_{1})}{N^2-1}  \,,\nonumber\\
A^{(0)}_{\DPhiSquarePsiSquareB}(1_{\vphantom{\bar \psi_n}\psi_p} 2_{\,\bar \psi_r} 3_{\vphantom{\bar\varphi}\varphi}4_{\bar\varphi})_{[1]}&= -2 \opcoef_{\DPhiSquarePsiSquareB}^{rp}(\bar{u}_{2}\slashed  k_{3}u_{1})  \,,\nonumber\\
A^{(0)}_{\DPhiSquarePsiSquareB}(1_{\vphantom{\bar \psi_n}\psi_p} 2_{\,\bar \psi_r} 3_{\vphantom{\bar\varphi}\varphi}4_{\bar\varphi})_{[2]}&=  0  \,,
\end{align}
with four-dimensional helicity values
\begin{align}
A^{(0)}(1^+_{\vphantom{\bar \psi_n}\psi_p} 2^-_{\,\bar \psi_r} 3_{\vphantom{\bar\varphi}\varphi}4_{\bar\varphi})_{[1]}&=  g^2 \delta_{pr}  \,,\nonumber\\
A^{(0)}(1^+_{\vphantom{\bar \psi_n}\psi_p} 2^-_{\,\bar \psi_r} 3_{\vphantom{\bar\varphi}\varphi}4_{\bar\varphi})_{[2]}&=  0 \,,\nonumber\\
A^{(0)}_{\DPhiSquarePsiSquareA}(1^+_{\vphantom{\bar \psi_n}\psi_p} 2^-_{\,\bar \psi_r} 3_{\vphantom{\bar\varphi}\varphi}4_{\bar\varphi})_{[1]}&=   -\frac{4 \opcoef_{\DPhiSquarePsiSquareA}^{rp} N s}{N^2-1}  \,,\nonumber\\
A^{(0)}_{\DPhiSquarePsiSquareA}(1^+_{\vphantom{\bar \psi_n}\psi_p} 2^-_{\,\bar \psi_r} 3_{\vphantom{\bar\varphi}\varphi}4_{\bar\varphi})_{[2]}&= -\frac{4 \opcoef_{\DPhiSquarePsiSquareA}^{rp} N^2 s}{N^2-1}  \,,\nonumber\\
A^{(0)}_{\DPhiSquarePsiSquareB}(1^+_{\vphantom{\bar \psi_n}\psi_p} 2^-_{\,\bar \psi_r} 3_{\vphantom{\bar\varphi}\varphi}4_{\bar\varphi})_{[1]}&= -4 \opcoef_{\DPhiSquarePsiSquareB}^{rp} s  \,,\nonumber\\
A^{(0)}_{\DPhiSquarePsiSquareB}(1^+_{\vphantom{\bar \psi_n}\psi_p} 2^-_{\,\bar \psi_r} 3_{\vphantom{\bar\varphi}\varphi}4_{\bar\varphi})_{[2]}&=  0  \,.
\end{align}
The one-loop amplitudes with an insertion of the $F^3$ operator are
\begin{align}
A^{(1)}_{F^3}(1^+_{\vphantom{\bar \psi_n}\psi_p} 2^-_{\,\bar \psi_r} 3_{\vphantom{\bar\varphi}\varphi}4_{\bar\varphi})_{[1]}&=\frac{1}{6}   g\tilde g^2 N s \delta_{pr}\,,\nonumber\\
A^{(1)}_{F^3}(1^+_{\vphantom{\bar \psi_n}\psi_p} 2^-_{\,\bar \psi_r} 3_{\vphantom{\bar\varphi}\varphi}4_{\bar\varphi})_{[2]}&=0  \,.
\end{align}
The one-loop amplitudes with an insertion of a $\varphi^2F^2$ operator all evaluate to zero:
\begin{align}
A^{(1)}_{\PhiSquareFSquareA}(1^+_{\vphantom{\bar \psi_n}\psi_p} 2^-_{\,\bar \psi_r} 3_{\vphantom{\bar\varphi}\varphi}4_{\bar\varphi})_{[1]}&=A^{(1)}_{\PhiSquareFSquareA}(1^+_{\vphantom{\bar \psi_n}\psi_p} 2^-_{\,\bar \psi_r} 3_{\vphantom{\bar\varphi}\varphi}4_{\bar\varphi})_{[2]}=0\,,\nonumber\\
A^{(1)}_{\PhiSquareFSquareB}(1^+_{\vphantom{\bar \psi_n}\psi_p} 2^-_{\,\bar \psi_r} 3_{\vphantom{\bar\varphi}\varphi}4_{\bar\varphi})_{[1]}&=A^{(1)}_{\PhiSquareFSquareB}(1^+_{\vphantom{\bar \psi_n}\psi_p} 2^-_{\,\bar \psi_r} 3_{\vphantom{\bar\varphi}\varphi}4_{\bar\varphi})_{[2]}=0  \,.
\end{align}
The one-loop amplitudes with an insertion of a $D^2\varphi^4$ operator are
\begin{align}
A^{(1)}_{\DSquarePhiFourthA}(1^+_{\vphantom{\bar \psi_n}\psi_p} 2^-_{\,\bar \psi_r} 3_{\vphantom{\bar\varphi}\varphi}4_{\bar\varphi})_{[1]}&=A^{(1)}_{\DSquarePhiFourthA}(1^+_{\vphantom{\bar \psi_n}\psi_p} 2^-_{\,\bar \psi_r} 3_{\vphantom{\bar\varphi}\varphi}4_{\bar\varphi})_{[2]}=0\,,\nonumber\\
A^{(1)}_{\DSquarePhiFourthB}(1^+_{\vphantom{\bar \psi_n}\psi_p} 2^-_{\,\bar \psi_r} 3_{\vphantom{\bar\varphi}\varphi}4_{\bar\varphi})_{[1]}&=\frac{1}{9}   \tilde g^2 N_s s (3 X_s+8) \delta_{pr}\,,\nonumber\\
A^{(1)}_{\DSquarePhiFourthB}(1^+_{\vphantom{\bar \psi_n}\psi_p} 2^-_{\,\bar \psi_r} 3_{\vphantom{\bar\varphi}\varphi}4_{\bar\varphi})_{[2]}&=0  \,.
\end{align}
The one-loop amplitudes with an insertion of a $D\psi^2\varphi^2$ operator are
\begin{align}
A^{(1)\rm fin}_{\DPhiSquarePsiSquareA}(1^+_{\vphantom{\bar \psi_n}\psi_p} 2^-_{\,\bar \psi_r} 3_{\vphantom{\bar\varphi}\varphi}4_{\bar\varphi})_{[1]}&=  -\tilde g^2 s (3 X_t-3 X_u-16) \opcoef_{\DPhiSquarePsiSquareA}^{rp} \,,\nonumber\\
A^{(1)\rm fin}_{\DPhiSquarePsiSquareA}(1^+_{\vphantom{\bar \psi_n}\psi_p} 2^-_{\,\bar \psi_r} 3_{\vphantom{\bar\varphi}\varphi}4_{\bar\varphi})_{[2]}&= 16 \tilde g^2 N s \opcoef_{\DPhiSquarePsiSquareA}^{rp} \,,\nonumber\\
A^{(1)\rm fin}_{\DPhiSquarePsiSquareB}(1^+_{\vphantom{\bar \psi_n}\psi_p} 2^-_{\,\bar \psi_r} 3_{\vphantom{\bar\varphi}\varphi}4_{\bar\varphi})_{[1]}&= \frac{\tilde g^2 s}{9 N}
\opcoef_{\DPhiSquarePsiSquareB}^{rp}\Big(8 \left(9 N^2+N N_s-18\right)
\nonumber\\&\qquad-27 \left(N^2-1\right) X_t+3 N N_s X_s-27 X_u\Big)
\nonumber\\&\quad+\frac{4}{9} \tilde g^2 N_f s (3 X_s+5) \opcoef_{\DPhiSquarePsiSquareB}^{ww} \delta_{pr}\,,\nonumber\\
A^{(1)\rm fin}_{\DPhiSquarePsiSquareB}(1^+_{\vphantom{\bar \psi_n}\psi_p} 2^-_{\,\bar \psi_r} 3_{\vphantom{\bar\varphi}\varphi}4_{\bar\varphi})_{[2]}&=-3 \tilde g^2 s (X_t-X_u) \opcoef_{\DPhiSquarePsiSquareB}^{rp}   \,.
\end{align}
The one-loop amplitudes with an insertion of a $\psi^4$ operator are
\begin{align}
A^{(1)}_{\PsiFourthA}(1^+_{\vphantom{\bar \psi_n}\psi_p} 2^-_{\,\bar \psi_r} 3_{\vphantom{\bar\varphi}\varphi}4_{\bar\varphi})_{[1]}&=-\frac{2}{9} \tilde g^2 N_f s (3 X_s+2) \opcoef_{\PsiFourthA}^{rwwp}\,,\nonumber\\
A^{(1)}_{\PsiFourthA}(1^+_{\vphantom{\bar \psi_n}\psi_p} 2^-_{\,\bar \psi_r}  3_{\vphantom{\bar\varphi}\varphi}4_{\bar\varphi})_{[2]}&=0\,,\nonumber\\
A^{(1)}_{\PsiFourthB}(1^+_{\vphantom{\bar \psi_n}\psi_p} 2^-_{\,\bar \psi_r} 3_{\vphantom{\bar\varphi}\varphi}4_{\bar\varphi})_{[1]}&= \frac{2 \tilde g^2 N_f s}{9 N} ((3 X_s+2) \opcoef_{\PsiFourthB}^{rwwp}-2 N (3 X_s+5) \opcoef_{\PsiFourthB}^{rpww}) \,,\nonumber\\
A^{(1)}_{\PsiFourthB}(1^+_{\vphantom{\bar \psi_n}\psi_p} 2^-_{\,\bar \psi_r} 3_{\vphantom{\bar\varphi}\varphi}4_{\bar\varphi})_{[2]}&=   0   \,.
\end{align}

\end{appendix}



\end{document}